\documentclass[aps,eqsecnum,twocolumn,showpacs,preprintnumbers,nofootinbib,prd,superscriptaddress,groupedaddress,10pt]{revtex4-2}

\makeatletter
\def\l@subsubsection#1#2{}
\def\l@subsubsubsection#1#2{}
\makeatother

\usepackage{empheq}
\usepackage{graphicx,amssymb,amsmath,amsthm,amsfonts,epsfig,epsf,fixmath}
\usepackage[usenames]{color}
\usepackage{epstopdf}
\usepackage{mathtools}
\usepackage{mathrsfs}
\usepackage{relsize}
\usepackage{graphics}
\usepackage{placeins} 
\usepackage{booktabs} 
\usepackage{siunitx} 
\DeclareSIUnit \parsec {pc} 
\usepackage{aas_macros}
\usepackage{bm}
\usepackage{dcolumn}
\usepackage{latexsym}
\usepackage{rotating}
\usepackage{longtable} 

\usepackage{enumerate}
\usepackage{tensor,multirow}
\usepackage{upgreek}
\usepackage{url}
\usepackage{float}
\usepackage[pdfencoding=auto, psdextra,linktocpage]{hyperref}
\usepackage{xcolor}
\definecolor{darkblue}{rgb}{0, 0.3, 0.6}
\hypersetup{
    colorlinks=true,
    linkcolor=darkblue,
    filecolor=darkblue,      
    urlcolor=darkblue,
    citecolor = darkblue,
    pdftitle={Piovano et al.: Spinning particles near Kerr black holes: Orbits and gravitational-wave fluxes through the Hamilton-Jacobi formalism},
    pdfpagemode=FullScreen,
    }
\usepackage{orcidlink}

\def\nn{\nonumber}

\renewcommand{\Re}{\mathrm{Re}\,} 
\renewcommand{\Im}{\mathrm{Im}\,} 
\newcommand{\osc}{\underline{\Delta}}

\newcommand{\dd}{\mathrm{d}}
\newcommand{\wmean}{\vec{\mathsf{w}}}
\newcommand{\parder}[2]{\frac{\partial #1}{\partial #2}}
\newcommand{\totder}[2]{\frac{\mathrm{d} #1}{\mathrm{d} #2}}
\newcommand{\totderhigh}[3]{\frac{\mathrm{d}^{#3} #1}{\mathrm{d}{#2}^{#3}}}
\newcommand{\Lzrd}{L_{z\text{rd}}}
\newcommand{\nswsh}{\!\prescript{}{-2}{S^{a \omega}_{\ell m}}} 
\newcommand{\rswsh}{S^c_{\ell m}} 
\newcommand{\Rin}{R^{\text{in}}_{\ell m \omega}}
\newcommand{\Rup}{R^{\text{up}}_{\ell m\omega}} 

\newcommand{\bea}{\begin{eqnarray}}
\newcommand{\eea}{\end{eqnarray}}
\newcommand{\be}{\begin{equation}}
\newcommand{\ee}{\end{equation}}
\newcommand{\ba}{\begin{align}}
\newcommand{\ea}{\end{align}}
\newcommand{\sgn}{\mathrm{sgn}}

\begin{document}
\title{
Spinning particles near Kerr black holes: Orbits and gravitational-wave fluxes through the Hamilton-Jacobi formalism
}
\author{
Gabriel Andres Piovano$^1$\orcidlink{0000-0003-1782-6813},
Christiana Pantelidou$^1$\orcidlink{0000-0002-7983-8636},
Jake Mac Uilliam$^1$\orcidlink{0009-0003-5172-2126},
Vojt{\v e}ch Witzany$^2\orcidlink{0000-0002-9209-5355}$
}
\affiliation{$^{1}$ School of Mathematics and Statistics, University College Dublin, Belfield, Dublin 4, Ireland}
\affiliation{$^{2}$ Institute of Theoretical Physics, Faculty of Mathematics and Physics, Charles University, CZ-180 00 Prague, Czech Republic}

\begin{abstract} 
Extreme mass ratio inspirals (EMRIs) are among the key sources of gravitational waves for the LISA space-based gravitational wave detector. Achieving sufficient accuracy in the gravitational wave template for these binaries requires modeling the effects of the spin of the comparably light secondary compact object. In this work, we employ the solution of the Hamilton-Jacobi equations for the motion of spinning bodies in Kerr space-time for the first time to obtain general bound orbits. Specifically, we implement a new solver for the Mathisson-Papatrou-Dixon equations of motion reduced to first-order form. Our approach provide novel semi-analytical expressions for the spin corrections to the orbital motion and frequencies, valid for any choice of referential geodesics, and new analytic expressions for the constants of motion shifts.
Then, using the Teukolsky formalism we compute gravitational wave energy and angular momentum fluxes sourced by these orbits valid to linear order in secondary spin, and provide waveform snapshots corresponding to the motion. The solver and the novel method we have developed substantially improve on previous studies in terms of speed and accuracy. Additionally, we include the full effect of a general precessing secondary spin in the waveform for the first time. As such, it provides a breakthrough building block for the modeling of waveforms of precessing compact binaries at large mass ratios. 
\end{abstract}
\maketitle
\tableofcontents

\section{Introduction and summary}
\subsection{The promise of next-generation gravitational-wave detectors}

We are nearing a decade since the first direct detection of gravitational waves (GWs) by LIGO \cite{LIGOScientific:2016aoc}. Now, with other detectors having joined the network and several rounds of upgrades, detections of GWs from compact binary mergers are currently being announced by the LIGO-Virgo-Kagra collaboration on a weekly basis \cite{KAGRA:2013rdx,gracedbO4}. In the meantime, proposals for next-generation terrestrial detectors such as the Einstein Telescope \cite{Maggiore:2019uih} or the Cosmic Explorer \cite{Reitze:2019iox} promise to reach deeper in the $\sim$ 10s-100s Hz sky and broaden the sensitivity even down to $\sim$ few Hz. As a result, these new detectors should probe novel types of sources with different masses, mass ratios, and dynamical states. 

In addition, the space-based detector LISA \cite{LISA:2017pwj,Colpi:2024xhw} has recently been adopted by the European Space Agency. It is planned to be launched and begin science operation in the 2030s. LISA will probe the $\sim$ 0.1s-100s mHz frequency range, thus capturing relativistic orbital dynamics involving massive black holes in the mass range $10^5 - 10^8 M_\odot$. Apart from massive black hole binaries \cite{Klein:2015hvg}, LISA is also expected to detect GWs from extreme mass ratio inspirals (EMRIs) \cite{Babak:2017tow}, where solar-mass compact objects of mass $\mu \sim 1 - 100 M_\odot$ spiral into massive black holes in the centers of galaxies. Most EMRIs are expected to be formed dynamically by $N$-body scattering in a dense nuclear star cluster around the massive black hole \cite{Amaro-Seoane:2012lgq,Babak:2017tow}, and are thus expected to have significant eccentricities and generic inclinations between the spin and orbital degrees of freedom throughout the entire LISA band. The case of EMRIs is particularly complicated, as they will spend $\sim 10^4$ orbital cycles in the LISA band. This poses significant modeling challenges for accurate LISA waveform generation, which is needed for the detection and parameter estimation of the signal superposed with strong detector noise \cite{LISAConsortiumWaveformWorkingGroup:2023arg}. 

\subsection{The two body problem at extreme mass ratios}

The relativistic two-body problem has no exact solution and needs to be treated within various approximation schemes. In the weak field, one can employ Post-Newtonian or Post-Minkowski methods, which iterate from flat space-time dynamics and possibly also slow motion (in the case of Post-Newtonian theory) \cite{Blanchet:2013haa}. Fully strong-field setups can then be treated by numerical relativity at significant costs that grow with every simulated orbital cycle \cite{baumgarte2010numerical,LISAConsortiumWaveformWorkingGroup:2023arg}. Nevertheless, EMRIs spend many cycles both in band and in the strong field. Which approach should be used for these inspirals? The answer is that they are best modeled by a third approach, gravitational self-force \cite{Barack:2018yvs,vandeMeent:2020xgc,Pound:2021qin}.

In gravitational self-force, one iterates the assumption that the compact object of mass $\mu$ moves on an orbit locally close to geodesic motion in the spacetime of the massive black hole of mass $M$ (the Kerr spacetime). One then employs black hole perturbation theory to find corrections to the equations of motion in powers of the mass ratio $q \equiv \mu/M$. A necessary ingredient in this procedure is a two-timescale approximation, where one systematically employs the fact that the radiation reaction acts on a much slower timescale than the orbital motion \cite{Hinderer:2008dm}. When evolving the binary through a finite frequency band such as that of LISA, the phase elapsed during the inspiral can then be decomposed as
\begin{align}
    \phi_{\rm insp} = \frac{1}{q} \phi_{\rm 0 PA} + \phi_{\rm 1 PA} + \mathcal{O}(q)\,,
\end{align}
where $\phi_{n\mathrm{PA}}$ are unknown $\mathcal{O}(1)$ coefficients. 
The leading \textit{adiabatic} order (0PA) of the phase is obtained by computing linear curvature perturbations sourced by a point mass moving along a geodesic, finding the rate at which GW energy and angular momentum is radiated from the space-time by these perturbations, and then taking orbital energy and angular momentum from the point mass at a matching rate  \cite{PhysRevD.67.084027,Sago:2005fn,Hughes:2005qb,Hughes:2021exa}. The next-to-leading, post-1-adiabatic (1PA) order then consists of a number of different contributions (for a more complete summary of the state of the art, see Section 4.3 of Ref. \cite{LISAConsortiumWaveformWorkingGroup:2023arg}). These include local self-force corrections to the equations of motion that go beyond the flux averaging \cite{Mino:1996nk,Barack:2011ed,vandeMeent:2017bcc}, second-order fluxes of energy and angular momentum \cite{Pound:2012nt,Warburton:2021kwk}, and a number of contributions appearing due to the spin of the secondary compact object (for detailed references, see the next section). 

By Lindblom's criterion \cite{Lindblom:2008cm}, modeling the inspiral and the outgoing waveform phase up to 1PA \textit{guarantees} that the waveform is indistinguishable from the true waveform for signals with signal-to-noise ratios up to $\sim \mathcal{O}(1/q)$. On the other hand, neglecting the 1PA contribution produces waveforms that are potentially distinguishable from true waveforms for \textit{any} detectable signal (signal-to-noise ratios $\sim \mathcal{O}(1)$).\footnote{Note, however, that because of degeneracies in the parameter space, 1PA waveforms are close to 0PA waveforms with shifted parameters~\cite{Piovano:2021iwv,Burke:2023lno}, at least for the (relatively simple) cases for which we fully understand 1PA waveforms.} Obtaining the full 1PA piece of the phasing and the waveform thus constitutes a clear goal for EMRI modeling. Furthermore, obtaining the 1PA piece can be a valuable source of information for the modeling of binaries even at comparable mass ratios within formalisms such as the Effective one-body models, see, e.g., Refs. \cite{Damour:2009sm,LeTiec:2011dp,Bini:2013zaa,Damour:2016gwp,Antonelli:2019fmq,Albertini:2022rfe,vandeMeent:2023ols}.

\subsection{Spinning particles near black holes}

Consider an isolated body that moves in a curved spacetime background such that its physical radius is much smaller than the ambient curvature radius. Starting only from the conservation of the stress-energy tensor and a multipole expansion, one can derive the celebrated equations for the evolution of the linear and angular momentum of this system known as the Mathisson-Papapetrou-Dixon (MPD) equations \cite{Mathisson:1937zz,Papapetrou:1951pa,Dixon:1970I,Dixon:1970II,DixonIII}. An analogous set of equations can also be derived for the evolution of momenta of black holes in ambient curved space-time \cite{1975PhRvD..11.1387D,Thorne:1984mz}.  When wrangled correctly, the MPD equations provide a set of second-order differential equations for the motion of the body and the evolution of its multipoles. The leading correction in these equations is the relativistic dipole, which couples the internal angular momentum of the body to the motion of its center of mass through a \textit{spin-curvature} term. 

How does this expansion fit into the $n$PA counting? Since the physical radius is of the order of the gravitational radius for compact objects, the multipole expansion of the equations can be viewed as a mass-ratio expansion for EMRIs. The 1PA order then only requires the inclusion of the effects of the leading dipole correction. This is also the focus of our paper. 

The evolution of the pole-dipole MPD system in black hole spacetimes, often dubbed as the ``motion of a spinning particle'', has a rich history (see Refs. \cite{Semerak:1999qc,Kyrian:2007zz} for a literature review). The MPD equations in Kerr spacetime are integrable only to linear perturbative order in the small body spin \cite{Rudiger:1981,Rudiger:1983,Gibbons:1993ap,Kubiznak:2011ay,Witzany:2019nml}, and will break integrability when general spin-squared terms are allowed \cite{Suzuki:1996gm,Suzuki:1999si,Zelenka:2019nyp,Ramond:2024ozy}. Only when the small body is a Kerr black hole, integrability is preserved up to quadratic order in the spin~\cite{Compere:2023alp}.
Even under the linear approximations, the MPD equations have no known closed-form  solutions in the generic case (for results on special parameters/configurations, see, e.g., \cite{Hackmann:2014tga,Witzany:2023bmq}). However, the computation of the 1PA terms requires a careful extraction of a number of linearized spin corrections of interest.  Specifically one needs to compute the spin corrections to fundamental frequencies of motion as a function of the energy and total angular momentum of the binary, and the GW fluxes of energy and angular momenta of the binary. 

Both of these tasks are best handled when one first semi-analytically decomposes solutions of the MPD equations in Kerr spacetime in the frequency domain. The first work in this direction was carried out by Ruangsri et al.~\cite{Ruangsri:2015cvg} by decomposing the MPD equations by brute force, which was then refined by Drummond \& Hughes~\cite{Drummond:2022xej,Drummond:2022efc} by using the analytical solution of parallel transport of the spin found by Marck \cite{marck1983solution,vandeMeent:2019cam}. However, one of the issues of the approach of Drummond \& Hughes is the fact that Fourier decomposition leads to a coupled system of algebraic equations, which, in return, fundamentally limits the maximum relative accuracy of the scheme to $\sim 10^{-5}$ \cite{Drummond:2022efc,Skoupy:2023lih}. 

An alternative approach was offered by one of us in Ref. \cite{Witzany:2019nml}, where one first uses the approximate integrability of the MPD equations and their reduction to first-order form. Specifically, in Ref. \cite{Witzany:2019nml}, the Hamiltonian formalism of Ref. \cite{Witzany:2018ahb} was used to formulate a Hamilton-Jacobi equation for the motion of spinning particles in Kerr space-time. The approximate solution of the Hamilton-Jacobi equation revealed two separation constants identical to the constants of motion found by R{\"u}diger \cite{Rudiger:1981,Rudiger:1983} and a reduction of the equations of motion to first order. This approach, while the only fully semi-analytical scheme for general spinning test particle orbits, was not practically implemented to date. We present the first implementation of the solutions of the MPD equations within the Hamilton-Jacobi approach here, surpassing previous approaches in efficiency and accuracy.

GW fluxes sourced by spinning particles were computed in Refs. \cite{Harms:2016ctx,Akcay:2019bvk} for spinning particles on circular orbits in Schwarzschild spacetime, and in Refs. \cite{Tanaka:1996ht,Han:2010tp,Harms:2015ixa,Lukes-Gerakopoulos:2017vkj,Piovano:2020zin} in Kerr spacetime. This was generalized to eccentric orbits in Schwarzschild and Kerr space-time in Refs. \cite{Skoupy:2021asz,Skoupy:2022adh,Skoupy:2024jsi}. Finally, in Kerr spacetime, orbits with different inclinations with respect to the Kerr spin axis will produce different fluxes. To date, there is a single work in which fluxes for spinning particles on eccentric and inclined orbit were computed, and that is the work of Skoup{\'y} et al.~\cite{Skoupy:2023lih}. There, the authors generated the fluxes based on the frequency-domain decomposition of the orbits by Drummond \& Hughes~\cite{Drummond:2022xej,Drummond:2022efc}, which implies that the resulting fluxes are limited by the same $\gtrsim 10^{-5}$ relative error as the orbital solutions. We improve on their computation here by using our new orbital solutions and by including the precession of the secondary spin in the waveform model. 

The computation of such fluxes allows one to drive inspirals in special configurations such as circular orbits or when the spins of the primary and secondary are aligned with orbital angular momentum \cite{Warburton:2017sxk,Piovano:2021iwv,Skoupy:2022adh}.  However, the Kerr field is not spherically symmetric due to the rotation of the central black hole, and this becomes relevant for inspirals with generic initial conditions. This is because then only one component of angular momentum is conserved even at geodesic order, and one has to figure out the adiabatic and post-1-adiabatic evolution of one additional constant not related to explicit symmetries of the space-time. In the case of geodesics, this is the so-called Carter constant \cite{Carter:1968rr}, which can be evolved at the adiabatic order thanks to prescriptions due to Mino and Sago et al. \cite{Mino:1996nk,Sago:2005fn}. In the case of spinning secondaries, there are two issues. First, the spin carries an additional degree of freedom \cite{Witzany:2018ahb}, which has an associated additional constant of motion that can be viewed as a component of the secondary spin aligned with the orbital angular momentum \cite{Rudiger:1981,Witzany:2019nml}. Fortuitously, the evolution of the aligned spin component was recently proven to be zero at 1PA order by Skoup{\'y} \& Witzany \cite{Skoupy:2024jsi}. However, the second issue is that a generalization of the prescription for the evolution of a Carter-like constant \cite{Rudiger:1983,Witzany:2019nml} is required for evolution of EMRIs with spinning secondaries. Unfortunately, such a prescription is currently not known (but see the progress on the topic by Grant \cite{Grant:2024ivt}). We thus present only the GW fluxes of energy and angular momentum and will adjust our scheme to deal with the Carter-like constant once an evolution prescription is found in the future.

\subsection{Summary of results}
\begin{itemize}
    \item We implemented a novel algorithm generating the orbits of spinning test particles in Kerr space-time. The key improvement lies in the first implementation of the analytical perturbative solution of the Hamilton-Jacobi equation found in Ref. \cite{Witzany:2019nml}, which reduce the MPD equations to first order form. We did so through a frequency-domain expansion of deviations of the spinning particle motion from certain referential Kerr geodesics to linear order in the particle spin. Our solutions of the 1st order equations of motion are valid for general orbits, spin configurations, and referential geodesic (see Section~\ref{sec:EoM1storder}-\ref{subsec:parametrization}).
    \item The orbits we obtained agree with previous results to high precision. Our orbit-generation algorithm isthe first \textit{semi-analytical} scheme for orbit generation in the sense that it can provide results to an arbitrary number of significant digits and can generate an arbitrary number of Fourier modes of the deviations without any non-linear scaling in computation costs.
    Moreover, we provide novel frequency corrections in semi-analytic form  (see Appendix~\ref{app:rad_pol_freq_shifts} and~\ref{app:time_azi_freq_shifts}), and analytic expressions for the constants of motion.
    \item Then, we coupled the orbital solutions to a Teukolsky solver and computed gravitational-wave fluxes of energy and azimuthal angular momentum sourced by the motion of the spinning particle. Additionally, we computed waveform snapshots corresponding to generic orbital configurations of the particle, including fully precessing secondary spins.
    \item By judicious implementation choices, we ensured that the efficiency and exponential convergence properties of the novel orbital solver translate also to the Teukolsky computation. In particular, we obtained the fluxes and the snapshots in an explicit decomposition into pieces corresponding to the referential geodesic and a spin correction. This makes the results breakthrough building blocks for 1PA EMRI waveforms for LISA and other detectors. 
\end{itemize}

\subsection{Organization of paper}

This paper is organised as follows. Section~\ref{sec:geodesic_motion} discusses geodesic motion in the Kerr black hole spacetime. This is followed by Section~\ref{sec:spinning_motion} which introduces the motion of spinning test particle in Kerr and describes the calculation of the linear in spin part of the motion. In doing so, we discuss different ways of fixing the referential geodesic. Then, Section~\ref{sec:asymptotic_amplitudes} presents the computation of asymptotic fluxes and amplitudes from the orbits calculated previously. Section~\ref{sec:Numerics} showcases some of our numerical results for the linear in spin corrections to the orbits, partial amplitudes and fluxes. Finally, Section~\ref{sec:conclusions}  summarizes our work and provides an outlook for possible extensions.
\subsection{Notation}
In this work, we use geometrized units $G=c=1$. Moreover, given $M$ the mass of the primary, we set $M=1$.
All spacetime indices are denoted by Greek letters, while tetrad legs are denoted with uppercase Latin letters. Partial and covariant derivatives are indicated with a comma and semicolon, respectively, i.e. $f_{\mu,\nu}= \partial_\nu f_\mu$ and  $f_{\mu;\nu}= \nabla_\nu f_\mu$
The metric signature is $(-,+,+,+)$, with the Riemann tensor defined as 
\begin{equation}
 \tensor{R}{^\delta_\sigma_\mu_\nu} \omega_\delta = 2\nabla_{[\mu} \nabla_{\nu]} \omega_\sigma
\end{equation}
where $\omega_\delta$ an arbitrary 1-form, while the square brackets denote antisymmetrization. We use the Mathematica convention for Legendre elliptic integrals and Jacobi elliptic functions, see Table \ref{tab:elliptic_integrals} in Appendix \ref{app:roots_elliptic}.

\section{Geodesic motion in Kerr spacetime}\label{sec:geodesic_motion}

The Kerr spacetime line element in Boyer-Lindquist coordinates reads
\begin{align}
   \dd s^2 &= -\left(1 - \frac{2 r}{\Sigma}  \right) \dd t^2 + \frac{\Sigma}{\Delta} \dd r^2 + \frac{\Sigma}{1-z^2} \dd z^2 + \nonumber \\ 
   &+\frac{1-z^2}{\Sigma}\left[2 a^2 r (1-z^2)+(a^2+r^2)\Sigma\right] \dd \phi^2 - \nonumber\\
   &- \frac{4 a r (1-z^2)}{\Sigma} \dd t \dd \phi \, ,
\end{align}
where $a$ is the spin of the primary, $z = \cos(\theta)$ and $\theta$ is the usual polar Boyer-Lindquist coordinate, while $\Delta = r^2 -2 r + a^2$ and $\Sigma = r^2 + a^2 z^2$. The spin of the primary black hole is parallel to the z-axis of a Cartesian coordinate frame centered on the primary, and it is (anti-) aligned when $a>0~ (a<0)$. 

The geodesic motion is completely integrable in the Liouville sense~\cite{Schmidt:2002qk} thanks to the existence of 4 constants of motion. One of them is the norm of the $4-$velocity $v^\nu_{\rm g} = \mathrm{d} x^\nu_{\rm g} / \mathrm{d}\tau$. 
The Kerr spacetime symmetries ensure the existence of two Killing vectors, $\Xi^\nu_{(t)}=(-1,0,0,0)$ and $\Xi^\nu_{(\phi)}=(0,0,0,1)$, which imply that $E_{\rm g} = v_{\nu \rm g} \Xi^\nu_{(t)}$ and $L_{z\rm g} = v_{\nu \rm g} \Xi^\nu_{(\phi)}$ are conserved. We have introduced the subscript ``g'' to denote geodesic-order variables.
These quantities can be identified, respectively, as the orbital energy (normalized by $\mu$) and orbital angular momentum parallel to the z-axis (rescaled by $\mu M$), as measured by an observer at infinity.
Finally, a fourth constant of motion, the Carter constant, is given by the Killing tensor $K_{\mu \nu}$ as $K_{\rm g} = v^\mu_{\rm g} K_{\mu \nu} v^\nu_{\rm g}$ \cite{Walker:1970un}, where 
\begin{equation}
K_{\mu \nu} \coloneqq \mathcal F_{\mu \alpha} \mathcal F^\alpha_{~\nu} \, ,
\end{equation}
and $\mathcal F_{\mu \nu}$ is the anti-symmetric Killing-Yano tensor \cite{Penrose:1973um}.

Using these 4 independent conserved quantities, $E_{\rm g}, L_{z \rm g}, K_{\rm g}$ and $v^\nu_{\rm g} v_{\nu \rm g}$, we can recast the second order equations of motion 
into first order form \cite{Carter:1968rr}. Specifically, defining the Carter-Mino time $\lambda$ as  $\dd \tau/\dd \lambda = \Sigma$ allows us to write the first-order equations of motion in separable form as
\begin{subequations}
\label{eq:1st-order-EoMgeo}
  \begin{empheq}[]{align}
  \totder{r_{\rm g}}{\lambda} & = \pm \sqrt{R_{\rm g}(r_{\rm g})} \label{eq:geo_radial_velocity} \, ,\\
  \totder{z_{\rm g}}{\lambda} & =\pm \sqrt{Z_{\rm g} (z_{\rm g})} \label{eq:geo_polar_velocity} \, ,\\
  \totder{t_{\rm g}}{\lambda} &= V^t_{r \rm g}(r_{\rm g}) + V^t_{z\rm g}(z_{\rm g}) \, , \\
  \totder{\phi_{\rm g}}{\lambda} &= V^\phi_{r \rm g}(r_{\rm g}) + V^\phi_{z \rm g}(z_{\rm g}) \, ,
  \end{empheq}
\end{subequations} 
where the functions on the right hand sides are defined as
\begin{align}
& V^t_{r \rm g}(r_{\rm g}) = E_{\rm g} \frac{(r_{\rm g}^2 + a^2)^2}{\Delta} - 2 \frac{a\, r_{\rm g}}{\Delta} L_{z \rm g}  \, ,\\
& V^t_{z\rm g}(z_{\rm g}) =  - a^2 E_{\rm g}  (1-z_{\rm g}^2) \, ,\\
& R_{\rm g}(r_{\rm g}) = \mathcal R(r_{\rm g})^2 - \Delta (K_{\rm g} + r_{\rm g}^2 )  \, , \\
& Z_{\rm g} (z_{\rm g}) = - \mathcal Z(z_{\rm g})^2 + (1-z^2_{\rm g})(K_{\rm g}-a^2 z^2_{\rm g}) \, , \\
& V^\phi_{r \rm g}(r_{\rm g})  = \frac{a}{\Delta} \mathcal R(r_{\rm g}) -a E_{\rm g}  \, , \\
& V^\phi_{z \rm g}(z_{\rm g})  = \frac{L_{z \rm g}}{1-z_{\rm g}^2}  \,,
\end{align}
and we have further introduced 
\begin{align}
    \mathcal R(r_{\rm g}) &= E_{\rm g}(r^2_{\rm g}+a^2)- aL_{z \rm g} \, , \\
    \mathcal Z(z_{\rm g}) &=  L_{z\rm g} - (1-z^2_{\rm g}) a E_{\rm g} \, .
\end{align}
The radial and polar functions $R_{\rm g}(r_{\rm g})$ and $Z_{\rm g}(z_{\rm g})$ are quartic polynomials of their arguments. As such, they are conveniently factorized in terms of their roots as \cite{Schmidt:2002qk}
\begin{align}
&R_{\rm g}(r_{\rm g}) = (r_{\rm g} - r_{1\rm g})(r_{2\rm g} - r_{\rm g})Y_{r\rm g}^2(r_{\rm g})  \, , \\
&Z_{\rm g}(z_{\rm g}) = (z_{1\rm g}^2 - z^2_{\rm g}) Y_{z\rm g}^2(z_{\rm g})\, ,
\end{align}
where 
\begin{align}
&Y_{r\rm g}(r_{\rm g})= \sqrt{(1- E^2_{\rm g})(r_{\rm g} - r_{3\rm g})(r_{\rm g} - r_{4\rm g})} \, , \\
&Y_{z\rm g}(z_{\rm g}) = \sqrt{z^2_{2 \rm g} - a^2(1-E^2_{\rm g})z^2_{\rm g}} \, .
\end{align}
The roots are ordered by magnitude as $r_{4\rm g} < r_{3\rm g}  < r_{2\rm g} \leq r_{1\rm g}$ while $0 \leq z_{1\rm g} \leq 1 < z_{2\rm g}$. Bound geodesic motion occurs for $r_{2 \rm g}\leq r_{\rm g} \leq r_{1 \rm g}$ and $-z_{ 1 \rm g}\leq z_{\rm g} \leq z_{ 1 \rm g}$. It is useful to parametrise the motion in terms of the semi-latus rectum $p_{\rm g}$, eccentricity $e_{\rm g}$ and inclination parameter $x_{\rm g}$, which are related to the turning points as
\begin{align}
r_{1\rm g} = \frac{p_{\rm g}}{1-e_{\rm g}}\,,  \;\,  r_{2\rm g} = \frac{p_{\rm g}}{1+ e_{\rm g}} \,,\; \, z_{1\rm g} = \sqrt{1 - x_{\rm g}^2}\, .
\end{align}
The parameter $x_{\rm g} \in (-1,1)$ provides a double-covering of the $z_{1 \rm g} \in (0,1) $ root. This is resolved by the negative $x_{\rm g}$ corresponding to counter-rotating orbits and the positive orbit corresponding to corotating ones. This can be summarized by the relation
\begin{equation}
x_{\rm g}= \sgn (L_{z \rm g}) \sqrt{1- z_{1\rm g}^2} \, .
\end{equation}
Technically, one picks the parameters $p_{\rm g},\ e_{\rm g},\ x_{\rm g}$, from which the constants of motion and the remaining roots are found by the method outlined in Appendix B of the seminal treatment of Schmidt \cite{Schmidt:2002qk}. Refs.~\cite{Fujita:2009bp,vandeMeent:2019cam} provide analytic solutions for the geodesic bound motion in terms of Legendre elliptic integrals and Jacobi elliptic functions. We present these analytic solutions in Sec.~\ref{subsec:radial_polar_trajectories} in a slightly different form as compared to Refs.~\cite{vandeMeent:2019cam,Fujita:2009bp}.

\subsection{Radial and polar geodesic trajectories} \label{subsec:radial_polar_trajectories}

It is convenient to parametrize the radial and polar trajectories by introducing the anomalies $\chi_{r\rm g} \in [0, 2\pi)$ and $\chi_{z\rm g} \in [0, 2\pi)$. The trajectories are then given by
\begin{align}
r_{\rm g}(\lambda) &=\frac{r_{1\rm g}+r_{2\rm g}}{2} + \frac{r_{1\rm g} - r_{2\rm g}}{2} \sin (\chi_{r\rm g}(\lambda))  \, , \\
z_{\rm g}(\lambda) &=z_{1\rm g} \sin (\chi_{z\rm g}(\lambda)),
\end{align}
where the anomalies are solutions of
\begin{align}
  \totder{\chi_{y\rm g}}{\lambda} & = Y_{y \rm g}(\chi_{y\rm g}(\lambda)) \qquad y = r,z \, . \label{eq:geo_anomaly}
\end{align}
The associated Mino-time frequencies for the radial and polar motion, $\Upsilon_{r\rm g}$ and $\Upsilon_{z\rm g}$,  can be expressed as
\begin{align}
  \Upsilon_{y\rm g} & = 2\pi \Bigg(\int_{0}^{2\pi} \frac{\dd \chi'_{y\rm g}}{Y_{y\rm g}( \chi'_{y\rm g})} \Bigg)^{\!-1}   \qquad y = r,z   \, .
\end{align}
Analytic expressions for $ \Upsilon_{r\rm g}$ and $ \Upsilon_{z\rm g}$ can be found in Refs. \cite{Fujita:2009bp,vandeMeent:2019cam}.
The solutions $\chi_{r\rm g}(\lambda)$ and  $\chi_{z\rm g}(\lambda)$ of the equations~\eqref{eq:geo_anomaly}  can be written in terms of the mean anomalies 
$w_{r\rm g}$ and $w_{z\rm g}$:
\begin{align} \label{eq:wydef}
 w_{y \rm g}(\lambda) & = \Upsilon_{y\rm g} \int_{0}^{\chi_{y\rm g}(\lambda)} \frac{\dd \chi'_{y\rm g}}{Y_{y\rm g}( \chi'_{y\rm g})}   \qquad y = r,z \, .
\end{align}
Analytic formulas for $w_{y \rm g}(\chi_{y \rm g})$ and the inversion $\chi_{y \rm g}(w_{y \rm g})$ are presented in Appendix~\ref{app:geo_freq}.
Using Leibniz's integral rule, it is immediate to see that 
\begin{equation}
\totder{w_y}{\lambda} =\frac{\Upsilon_{y\rm g}}{Y_{y \rm g}(\chi_{y\rm g})} \totder{\chi_{y\rm g}}{\lambda}  = \Upsilon_{y\rm g} \qquad y = r,z \, .
\end{equation}
The analytic solutions of the radial and polar trajectories in terms of the mean anomalies $w_{r\rm g}$ and $w_{z \rm g}$ can be written as
\begin{align}
 r_{\rm g}(w_{r\rm g}) &= \frac{r_{3 \rm g}(r_{1\rm g}-r_{2\rm g})\mathcal W_{r \rm g}(w_{r \rm g}) - r_{2\rm g}}{(r_{1\rm g}-r_{2\rm g})\mathcal W_{r \rm g}(w_{r \rm g})-1}  \, , \\
 z_{\rm g}(w_{z\rm g}) &= z_{1 \rm g} \mathsf{sn} (\alpha_{z \rm g}(w_{z \rm g})|k_{z \rm g}) \, ,
\end{align}
where the expressions for $k_{z \rm g}$ is given in Appendix~\ref{app:roots_elliptic}, while  $\mathcal W_{r \rm g}(w_{r \rm g}), \alpha_{z \rm g}(w_{z \rm g})$ are given in Appendix~\ref{app:geo_freq}.

\subsection{Fourier series expansions in Mino-time}

A generic function $f(r_{\rm g}, z_{\rm g})$ of the geodesic bound trajectories $r_{\rm g}$ and $z_{\rm g}$ can be expanded as a Fourier series:
\begin{align}
f(r_{\rm g}, z_{\rm g}) &= \displaystyle \sum_{n = -\infty}^{\infty} \sum_{k = -\infty}^{\infty} f_{nk} e^{-i \vec b \cdot \wmean_{\rm g}} = \langle f \rangle + \osc f (\wmean_{\rm g}) \, ,
\end{align}
where $\wmean_{\rm g} = (w_{r \rm g}, w_{z \rm g})$, $\vec b = (n,k)$, and  the Fourier coefficients are defined as
\begin{align}
f_{nk} &= \frac{1}{(2\pi)^2} \int_{[0,2\pi)}\dd^2 w_{\rm g}  f(\wmean_r) e^{i \vec b \cdot \wmean_{\rm g}}  \, ,  
\end{align}
while $\langle f \rangle = f_{00}$ is the geodesic average and $\osc  f(\wmean_{\rm g})$ is a purely oscillating function. For instance, the coordinate time and azimuthal rate of change, $\Upsilon_{t\rm g}$  and  $\Upsilon_{\phi \rm g}$ respectively, are given as
\begin{align}
\Upsilon_{t\rm g} & =  \langle V^t_{\rm g} \rangle  \, , \quad  \Upsilon_{\phi \rm g} =  \langle V^\phi_{\rm g} \rangle  \, .
\end{align}
The velocities $ \dd t_{\rm g}(\lambda)/ \dd \lambda$ and $\dd \phi_{\rm g}(\lambda)/ \dd \lambda$ can be expanded in separable Fourier series, which allow to efficiently compute
the trajectories $t_{\rm g}(\lambda)$ and $\phi_{\rm g}(\lambda)$ as 
\begin{align}
&t_{\rm g} = \Upsilon_{t\rm g}\lambda - \displaystyle \sum_{n \neq 0} \frac{(V^t_{r \rm g})_n}{i n \Upsilon_{r\rm g}} e^{-i n w_{r \rm g}} - \displaystyle \sum_{k \neq 0} \frac{(V^t_{z\rm g})_k}{i k\Upsilon_{z\rm g}} e^{-i n w_{z \rm g}}   \, , \label{eq:geotFourierseries} \\
&\phi_{\rm g} = \Upsilon_{\phi \rm g}\lambda - \displaystyle \sum_{n \neq 0}\frac{(V^\phi_{r \rm g})_n}{i n \Upsilon_{r\rm g}} e^{-i n w_{r \rm g}} - \displaystyle \sum_{k \neq 0} \frac{(V^\phi_{z \rm g})_k}{i k\Upsilon_{z\rm g}} e^{-i n w_{z \rm g}} \, . \label{eq:geophiFourierseries}
\end{align}
Closed-form expressions for $t_{\rm g}(\lambda), \phi_{\rm g}(\lambda)$ in terms of elliptic integrals can also be found in Refs. \cite{Fujita:2009bp,vandeMeent:2019cam}.

\section{Spinning particles in Kerr spacetime}\label{sec:spinning_motion}

An extended body can be modeled, under certain assumptions, as a point particle plus smaller corrections called multipoles.
In classical mechanics and electrodynamics, it is customary to expand the potential of a finite-size body in multipoles when it is located far away from the observer. 
In the context of general relativity, the stress-energy tensor of an extended body can be written as a multipolar expansion, which forms the ``gravitational skeleton'' of the body \cite{tulczyjew1959motion}. EMRI models accurate at 1PA order only require the first two multipoles of the secondary, which comprise the mass (monopole) and spin (dipole). 
At the pole-dipole order, the stress-energy tensor can be written as 
\begin{align} \label{eq:stress_energy_tensor}
T^{\mu \nu} &= \int\!\! \dd \tau \bigg( p^{(\mu}v^{\nu)} \frac{\delta^4(x^\alpha-\tilde z^\alpha(\tau))}{\sqrt{-g}} \nonumber \\
                       & -\nabla_\rho \left( S^{\rho (\mu}v^{\nu)} \frac{\delta^4(x^\alpha-\tilde z^\alpha(\tau))}{\sqrt{-g}} \right) \!\bigg) \, .
\end{align}
where $\tau$ is the proper time,  $v^\mu = \dd \tilde z^\mu/\dd \lambda$ is the tangent vector to a representative worldline of the body, $p^\mu$ is the four-momentum, $S^{\mu \nu}$ is the anti-symmetric spin-tensor, and $g$ is the metric determinant. The MPD equations of motion for a spinning body can be obtained from the conservation law ${T^{\mu \nu}}_{;\nu}=0$ \cite{tulczyjew1959motion,Steinhoff:2009tk}
\begin{subequations}
\label{eq:MPDeq}
  \begin{empheq}{align}
     \frac{D p^\mu}{\dd \tau} &= -\frac{1}{2} R^{\mu}_{~\nu \rho \sigma} v^\nu S^{\rho \sigma}  \, , \\
     \frac{D S^{\mu\nu}}{\dd \tau} &= 2 p^{[\mu}v^{\nu]}   \, , 
  \end{empheq}
\end{subequations}
with $D/\dd \tau = v^\mu \nabla_\mu$. The spin parameter $S$ is defined as 
\begin{equation}
S^2 = \frac{1}{2} S^{\mu \nu}S_{\mu \nu} = S^\mu S_\mu \, ,
\end{equation}
where $S^\mu$  is the spin vector, which can be defined in terms of the spin-tensor as
\begin{equation}
S_\mu \equiv -\frac{1}{2\mu} \epsilon_{\mu\nu\alpha \beta}p^\nu S^{\alpha\beta} \, ,
\end{equation}
with $\epsilon_{\alpha\beta\gamma\delta} = \sqrt{-g} e_{\alpha\beta\gamma\delta}$ the antisymmetric Levi-Civita tensor density and $e_{\alpha\beta\gamma \delta}$ the Levi-Civita symbol ($e_{0123} =1$).

Neither the mass $\mu =  - p^\mu p_\mu$ nor $S$ need to be constants of motion (a clear discussion of these facts was given by Semerák \cite{Semerak:1999qc}). Moreover, the MPD system of equations is undetermined due to the freedom in choosing a referential worldline with respect to which the multipoles are computed. A reference worldine can be fixed by choosing a supplementary spin condition (SSC).  We choose the Tulczyjew-Dixon condition \cite{tulczyjew1959motion,Dixon:1970I}:
\begin{equation}
S^{\mu\nu}p_\nu =0 \, ,
\end{equation}
which fixes a unique wordline, and ensures that the dynamical mass $\mu$ and spin $S$ are conserved. In addition, the Tulczyjew-Dixon SSC provides a relation between the 4-velocity $v^\mu$ and the linear momentum $p^\mu$ \cite{Dixon:1970I,Ehlers:1977}.
Under the Tulczyjew-Dixon SSC, the spin-tensor can be expressed only in terms of the spin-vector and momentum as
\begin{equation}
S^{\mu\nu} = \frac{1}{\mu}\epsilon^{\mu\nu\alpha\beta}p_\alpha S_{\beta} \, . \label{eq:spintensordef}
\end{equation}

In geometric units, the spin parameter $S$ has the same dimension of the angular momentum, which implies that 
\begin{equation}
\frac{S}{M\mu} = q s \,,
\end{equation} 
where $s \equiv S/ \mu^2$ is a dimensionless number related to the ratio of the physical size of the object with respect to its gravitational radius and we have restored $M$ for clarity. In the following, we assume $s = \mathcal{O}(1)$. This is because the extreme and intermediate mass ratio binaries observable by LISA are expected to satisfy $s\lesssim 1$ or at least $s \ll 1/q$. For reference, the Earth has $s \simeq 200$, while $s \simeq 0.3$ for a millisecond pulsar and typical neutron star models reach the mass-shedding limit of rotation before $s\sim 0.7$ (e.g. \cite{Chakrabarti:2013tca}). Sub-extremal black holes in General Relativity have $s \leq 1$. 

For extreme mass-ratio binaries, the velocity $v^\mu$ and linear momentum $u^\mu$ are parallel at order $ \mathcal O(s^2q^2)$~\cite{Semerak:1999qc}. 
Moreover, the 4-velocity can be expanded as $v^\mu = v^\mu_{\rm g} + s q \delta v^\mu$ where $v^\mu_{\rm g}$ is the geodesic 4-velocity and $\delta v^\mu$ is the linear correction to the 4-velocity induced by secondary spin. The MPD  equations Eq.~\eqref{eq:MPDeq} to linear order in $s q$ then reduce to
\begin{subequations}
\label{eq:MPDeqlin}
  \begin{empheq}{align}
     \frac{D_{\rm g} v^\mu_{\rm g}}{\dd \tau} &= 0  \, , \label{eq:2ndEoMgeo}\\
     \frac{D_{\rm g} \delta v^\mu}{\dd \tau} &  = -\frac{1}{2} R^{\mu}_{~\nu \rho \sigma} v^\nu_{\rm g} s^{\rho \sigma}  \, , \label{eq:2ndEoMspincorr} \\
    \frac{D_{\rm g} s^{\mu}}{\dd \tau} &= 0 \, ,  \label{eq:partransport}  \\
   s^{\mu\nu} v^{\rm g}_ \nu &= 0 \, ,  \label{eq:linTD-SSC}
 \end{empheq}
\end{subequations}
where $s^{\rho \sigma}  = S^{\rho \sigma} /\mu$, $s^\rho =S^\rho/\mu$ and  $D_g/\dd \tau = v^\mu_{\rm g} \nabla_\mu$. Eq.~\eqref{eq:partransport} is equivalent to $D_{\rm g} s^{\mu\nu}/\dd \tau =0$, while Eq.~\eqref{eq:linTD-SSC} is the linearised Tulczyjew-Dixon SSC at $\mathcal O(sq)$.
 The equations of motion in Eq.~\eqref{eq:MPDeqlin} are a system of second-order linear differential equations, where the spin vector $s^\mu$ (or, equivalently, the spin tensor $s^{\mu \nu}$) is parallel-transported along a geodesic trajectory. The equations are valid to linear order in $s q$ in the sense that the resulting trajectory will be accurate to $\mathcal{O}(s q)$, while the evolution of the spin degree of freedom is only valid to leading ``geodesic'' order. In the rest of the paper, we will solve an equivalent system of equations of motion formed by \textit{first-order} differential equations in a Kerr spacetime background.

\subsection{Constants of motion}

A spinning test body in Kerr background admits two first-integrals of motion
\begin{align} 
    E &= E_{\rm g} - \frac{1}{2} \Xi^{(t)}_{\mu;\nu}s^{\mu\nu} \label{eq:energyspin},
    \\ 
    J_z &= L_{z \rm g} - \frac{1}{2} \Xi^{(\phi)}_{\mu;\nu}s^{\mu \nu} \label{eq:angmomspin},
\end{align} 
which are conserved at any order in $sq$ and do not depend on the choice of the SSC \cite{Dixon:1970I}. In contrast, $ E_{\rm g} $ and $L_{z\rm g}$ are not conserved at $\mathcal O(sq)$. Moreover, the Carter constant is no longer a conserved quantity for the MPD system of equations~\eqref{eq:MPDeq}. For the Tulczyjew-Dixon SSC, it was shown by R{\"u}diger~\cite{Rudiger:1981,Rudiger:1983} that the following quantities
\begin{align}
    K_{\rm R} &= v^\mu K_{\mu \nu} v^\nu - 2 v^\mu s^{\rho \sigma} (\mathcal F_{\mu \rho;\kappa} \mathcal F^\kappa_{~\sigma}+ \mathcal F_{\rho \sigma;\kappa}\mathcal F^\kappa_{~\mu}) \, , \label{eq:Carterspin}  
    \\
   C_{\mathcal F} &= \mathcal F_{\mu \nu} v^\mu s^\nu = l_\nu s^\nu  \, , 
\end{align}
are conserved at $\mathcal O(sq)$.  $K_{\rm R}$ is analogous to the geodesic Carter constant $K_{\rm g} =  v^\mu_{\rm g} K_{\mu \nu}v^\nu_{\rm g}$ while  $C_{\mathcal F}$ can be interpreted as a projection of the spin vector $s^\mu$ onto the geodesic angular-momentum vector $l^\mu = \mathcal F_\alpha^{~\mu} v^\alpha_{\rm g}$. Therefore, we define 
\begin{equation}
s _\parallel =\frac{S_\parallel}{\mu^2} \coloneqq \frac{C_{\mathcal F}}{\sqrt{K_{\rm g}}} \,, \label{eq:spinpara}
 \end{equation}
 while the orthogonal component of $s^\mu$ to $l^\mu$ is $s _\perp = \sqrt{s^2-s^2_\parallel}$. 
In total, a spinning particle in  Kerr spacetime has six conserved quantities at $\mathcal O (sq)$: $\mu, E, J_z, K_{\rm R}, s_\parallel, s$.

\subsection{Equations of motion in first-order form} \label{sec:EoM1storder}

We now consider the linearised MPD system of equations~\eqref{eq:MPDeqlin} in a Kerr spacetime background.  First of all, we notice that Eq.~\eqref{eq:2ndEoMgeo} and Eq.~\eqref{eq:partransport}  do not depend on the 4-velocity spin-correction $\delta v^\mu$, thus they can be solved separately from Eq.~\eqref{eq:2ndEoMspincorr}. Eq.~\eqref{eq:2ndEoMgeo} and Eq.~\eqref{eq:partransport} describe the motion of a free-falling gyroscope (i.e. a spinning point particle where the spin-curvature is zero), which is parallel transported along a geodesic. The solution of Eq.~\eqref{eq:partransport} under the constraint~\eqref{eq:linTD-SSC} is
 \begin{equation}
 s^\mu =s_\perp\big ( \tilde e^\mu_{~(1)}\cos \psi_{\rm p}  + \tilde e^\mu_{~(2)} \sin \psi_{\rm p}  \big) + s_\parallel e^\mu_{~(3)} \,,\label{eq:spinvector}
 \end{equation}
where $\tilde e^\mu_{~(1)}, \tilde e^\mu_{~(2)}$ and $e^\mu_{~(3)} = l^\mu/\sqrt{K_{\rm g}}$ are the tetrad legs of an adapted  orthonormal frame, called the Marck tetrad \cite{marck1983solution}. The zeroth leg is given by $e^\mu_{~(0)} = v^\mu_g$, while $\tilde e^\mu_{~(1)}$ and $\tilde e^\mu_{~(2)}$ are given in~\cite{marck1983solution} (useful reformulations can also be found in~\cite{Witzany:2019nml} and~\cite{vandeMeent:2019cam}). The non-zero projections of the Marck tetrad onto the spin tensor,  $s^{(c)(d)} = s^{\mu \nu}e_\mu^{~(c)} e_\nu^{~(d)}$ are given by
\begin{align}
  s^{(1)(2)} = s_\parallel \ , \ \,  s^{(3)(2)}=s_\perp \cos\psi_{\rm p} \ ,\  \, s^{(3)(1)} = s_\perp \sin\psi_{\rm p} \, , \label{eq:sptMarck}
\end{align}
which can be obtained by using Eq.~\eqref{eq:spinvector} and the definition~\eqref{eq:spintensordef}. The precession phase $\psi_{\rm p}$ satisfies the following equation
\begin{equation}
\totder{\psi_{\rm p}}{\lambda} = \Psi_r(r_{\rm g}) +  \Psi_z(z_{\rm g})  \, , \label{eq:spin-precession-angle}
\end{equation}
where
\begin{align}
\Psi_r(r_{\rm g}) &= \sqrt{K_{\rm g}}\frac{(r^2_{\rm g}+a^2)E_{\rm g}-a L_{z \rm g}}{K+r^2_{\rm g}}  \, , \\
\Psi_z(z_{\rm g}) &= a\sqrt{K_{\rm g}}\frac{L_{z \rm g} - a (1-z^2_{\rm g})E_{\rm g}}{K_{\rm g} - a^2 z^2_{\rm g}} \, .
\end{align}
Thus, the spin vector $s^\mu$ precesses around the geodesic total angular momentum $l^\mu$ with Mino-time frequency $\Upsilon_{\rm p}$. Although Marck \cite{marck1983solution} readily recognized that $\psi_{\rm p}(\lambda)$ has a solution in terms of elliptic integrals, the explicit expression of this solution in Legendre's canonical form was worked out later in~\cite{vandeMeent:2019cam}.
Finally, it should be noted that $e^\nu_{~(2);\mu} e_{\nu(1)}e^\mu_{~(0)} = \dd \psi_{\rm p}/ \dd \lambda$.

At linear order in the mass-ratio and secondary spin, we denote the radial and polar trajectories $r(\lambda)$ and $z(\lambda)$, respectively, as
\begin{align}
    r(\lambda) &= r_{\rm g}(\lambda) + q \delta r(\lambda) \, ,\\
    \delta r(\lambda) &= s_\parallel \delta r_\parallel(\lambda) + s_\perp \delta r_\perp(\lambda) \, , \\
    z(\lambda) &= z_{\rm g}(\lambda) + q \delta z(\lambda) \, , \\
    \delta z(\lambda) &= s_\parallel \delta z_{\parallel}(\lambda) + s_\perp \delta z_\perp(\lambda) \, ,
\end{align}
with $\delta r_\parallel(\lambda)$ and $\delta r_\perp(\lambda)$ the corrections to the trajectories that include the parallel and orthogonal components of the secondary spin, respectively. Equivalent relations also hold for the polar correction to the trajectories.

Ref.~\cite{Witzany:2019nml} showed that equations~\eqref{eq:2ndEoMspincorr} can be recast as a system of first order equations of motion 
\begin{subequations}
\label{eq:1st-order-linMPDeq}
  \begin{empheq}{align}
  \totder{\delta t}{\lambda} &= \delta V^t \, , \\
  \totder{\delta r}{\lambda} & = \pm \frac{\delta R}{2\sqrt{R_{\rm g}(r_{\rm g})}} \, , \label{eq:lin-corr-radial-motion} \\
  \totder{\delta z}{\lambda}  & =\pm  \frac{\delta Z}{2\sqrt{Z_{\rm g}(z_{\rm g})}} \, , \label{eq:lin-corr-polar-motion} \\
  \totder{\delta \phi}{\lambda} &= \delta V^\phi \, ,
  \end{empheq}
\end{subequations} 
where Eq.~\eqref{eq:lin-corr-radial-motion} and Eq.~\eqref{eq:lin-corr-polar-motion} are obtained by linearising Eq.$(46a)$ and Eq.$(46b)$ of~\cite{Witzany:2019nml}. The variations $ \delta R$ and $ \delta Z$ are given by
\begin{align}
  \delta R &= R_s + \frac{\partial R_{\rm g}}{\partial r_{\rm g}} \delta r + \displaystyle \sum^3_{i =1} \frac{\partial R_{\rm g}}{\partial C_{i \rm g}} C_{i\rm s} \, ,  \label{eq:diff_radial_potential} \\
   R_s &= 2\Delta \big(\Psi_r(r_{\rm g}) + \Delta \mathcal S^r + a s_\parallel\sgn(L_{z \rm g} - a E_{\rm g})\big) \, , \\
  \delta Z &= Z_s + \frac{\partial Z_{\rm g}}{\partial z_{\rm g}} \delta z + \displaystyle \sum^3_{i =1} \frac{\partial Z_{\rm g}}{\partial C_{i \rm g}} C_{i\rm s} \, , \label{eq:diff_polar_potential} \\
  Z_s &= 2(1 - z^2_{\rm g}) \big(\Psi_z(z_{\rm g}) + \mathcal S^z - a s_\parallel \sgn(L_{z \rm g} - a E_{\rm g}) \big) \, .
\end{align}
with $(C_{1 \rm g}, C_{2 \rm g}, C_{3 \rm g}) =(E_{\rm g}, L_{z \rm g}, K_{\rm g})$. $C_{i \rm s}$ represents the shifts induced to the constants of motion by the spinning particle, that is, $ (C_{1 \rm s}, C_{2 \rm s}, C_{3 \rm s}) = (E_s, J_{z \rm s}, K_{\rm s})$.
The spin connections $\mathcal S^y = \mathcal S^y(r_{\rm g},z_{\rm g},\psi_{\rm g})$ for $y =r,z$ are the only source of non-separability of the equations of motion. Their expressions are known in closed form but they are too long for the main text, so we relegate them to Appendix~\ref{app:spin_coeffs_Christoffel_sym}. The terms proportional to $2a s_\parallel \sgn(L_{z \rm g} - a E_{\rm g})$ in Eqs.~\eqref{eq:diff_radial_potential} and~\eqref{eq:diff_polar_potential} ensure that $z_{1 \rm s} \to 0$ for $z_{1 \rm g} \to 0$.

The variations $\delta V^t$ and $\delta V^\phi$ are given by
\begin{align}
  \delta V^t &= -\frac{\Sigma}{2}\Upgamma^t + \frac{\partial V^t_{r \rm g}}{\partial r_{\rm g}} \delta r + \frac{\partial V^t_{z\rm g}}{\partial z_{\rm g}} \delta z +\displaystyle \sum^2_{i =1} \frac{\partial V^t_{\rm g}}{\partial C_{i \rm g}} C_{i\rm s} \, ,  \label{eq:diff_time_potential} \\
  \delta V^\phi &= -\frac{\Sigma}{2}\Upgamma^\phi + \frac{\partial V^\phi_{r \rm g}}{\partial r_{\rm g}} \delta r +  \frac{\partial V^\phi_{z \rm g}}{\partial z_{\rm g}} \delta z +\displaystyle \sum^2_{i =1} \frac{\partial V^\phi_{\rm g}}{\partial C_{i \rm g}} C_{i\rm s}  \, . \label{eq:diff_azimuthal_potential}
\end{align}
The expressions for $\Upgamma^t$ and $\Upgamma^\phi$, representing the projections of the Christoffel symbols $\Upgamma^t_{\mu \nu}$ and $\Upgamma^\phi_{\mu \nu}$ , respectively, onto the Marck tetrad, are lengthy, and can be found in Appendix~\ref{app:spin_coeffs_Christoffel_sym}.

As already noted, the first-order corrections to the radial and polar motion, described by Eq.~\eqref{eq:lin-corr-radial-motion} and Eq.~\eqref{eq:lin-corr-polar-motion} are \textit{not separable} for generic orbits due to the spin coefficients $\mathcal{S}^y$ with $y = r,z$, whereas their geodesic counterparts are decoupled. Eq.~\eqref{eq:lin-corr-radial-motion} is independent of the polar motion only in the limit of almost-equatorial orbits (compare with Ref. \cite{Witzany:2023bmq,Drummond:2022xej}).

Furthermore, the right-hand sides of Eq.~\eqref{eq:lin-corr-radial-motion} and Eq.~\eqref{eq:lin-corr-polar-motion} are not defined, respectively, for $r_{\rm g} = r_{1\rm g}$, $r_{\rm g} = r_{2\rm g}$, and $z_{\rm g} = z_{1\rm g}$, i.e. at the geodesic turning points. Nevertheless,  the system~\eqref{eq:1st-order-linMPDeq} has at most removable singularities at the geodesic turning points. In other words, the limits $r_{\rm g} \to r_{1\rm g}$, $r_{\rm g} \to r_{2\rm g}$ and $z_{\rm g} \to z_{1\rm g}$ for Eq.~\eqref{eq:lin-corr-radial-motion} and Eq.~\eqref{eq:lin-corr-polar-motion}, respectively, exist and are finite.\footnote{Note: in Ref.~\cite{Witzany:2019nml} the Marck tetrad is constructed using congruence geodesics, which in general are different from the referential geodesics. Eq.~\eqref{eq:lin-corr-radial-motion} and Eq.~\eqref{eq:lin-corr-polar-motion} have finite limits at the geodesic turning points only if the congruence geodesics and the referential geodesics coincide.}

Finally, the linearised MPD system~\eqref{eq:MPDeqlin}, and its first-order counterparts~\eqref{eq:1st-order-linMPDeq} are not unique, but depend on how a referential geodesic is fixed. Different choices lead to different spin corrections at $\mathcal O (sq)$, which are equivalent and can be mapped onto each other using appropriate transformations. 
Section~\ref{sec:solutionsEoM} presents the solutions of the system~\eqref{eq:1st-order-linMPDeq} valid for any choice of referential geodesics, i.e. the corrections to the constants of motion, turning points and frequencies are not specified. In Section~\ref{subsec:parametrization} we show how to compute the spin corrections to the orbits in two parametrizations: with fixed constants of motion, and fixed turning points (on average). Appendix~\ref{app:DHmapFC} presents the transformations between the orbital dynamics computed in these two parametrizations.
\subsection{General solutions of the equations of motion} \label{sec:solutionsEoM}
We parametrize the corrections to the radial and polar trajectories in terms of the shifts to the turning points and a set of anomalies $\chi_r \in [0,2\pi)$ and $\chi_z \in [0,2\pi)$ together with the precession angle $\psi_{\rm p}$:
\begin{align}
\delta r(\lambda) &= r_{\rm s}(\chi_r(\lambda),\chi_z(\lambda),\psi_{\rm p}(\lambda))   \, , \label{eq:rad_shift_turning_points} \\
\delta z(\lambda) &= z_{\rm s}(\chi_r(\lambda),\chi_z(\lambda),\psi_{\rm p}(\lambda)) \, , \label{eq:pol_shift_turning_points}
\end{align}
where
\begin{align}
    r_{\rm s}(\chi_r ,\chi_z , \psi_{\rm p}) &= \frac{r_{1 \rm s}(z_{\rm g},\psi_{\rm p}) + r_{2 \rm s}(z_{\rm g},\psi_{\rm p})}{2}  \nonumber  \\
    &+\frac{r_{1 \rm s}(z_{\rm g},\psi_{\rm p}) - r_{2 \rm s}(z_{\rm g},\psi_{\rm p})}{2}\sin{\chi_r}  \, , \\
    z_{\rm s}(\chi_r ,\chi_z , \psi_{\rm p})  &= z_{1{\rm s}\perp}(r_{\rm g}) + z_{1{\rm s}\parallel}(r_{\rm g},\psi_{\rm p})\sin{\chi_z}   \, , 
\end{align}
The shifts to the turning points are derived by setting
\begin{align}
  \left.  \delta R_{\rm s} \right \rvert_{r_{\rm g} = r_{1\rm g},r_{2\rm g}}  = 0 \, , \\
  \left. \delta Z_{\rm s} \right \rvert_{z_{\rm g} = z_{1\rm g}, - z_{1\rm g}} = 0 \, .
\end{align}
These shifts are not constant for generic orbits, but depend on the chosen referential geodesics and the precession angle $\psi_{\rm p}$:  $r_{1 \rm s}(z_{\rm g}, \psi_{\rm p}) , r_{2 \rm s}(z_{\rm g}, \psi_{\rm p})$ and $z_{1 \rm s}(r_{\rm g}, \psi_{\rm p}) ,z_{2 \rm s}(r_{\rm g}, \psi_{\rm p})$. Only in the limit of spin-aligned, equatorial orbits, $r_{1 \rm s} , r_{2 \rm s}$ are constant while $z_{1 \rm s} = z_{2 \rm s} = 0 $. The closed form expressions can be found in Appendix~\ref{app:turnp}.

The anomalies $\chi_r = \chi_{r\rm g} + q \delta\chi_r$ and $\chi_z = \chi_{z\rm g} + q\delta\chi_z$ receive $\mathcal O (sq)$ corrections that satisfy the following equations of motion:
\begin{equation}
   \totder{\delta\chi_{y}}{\lambda} =  \delta Y_y(\lambda) \, , \qquad y = r,z \, ,
\end{equation}
where
\begin{align}
    \delta Y_r (\lambda) &= Y_{r \rm s}(\lambda) + r_{\rm s} \totder{Y_{r\rm g}}{r_{\rm g}} - \frac{2 \mathrm h_r(\chi_r, \chi_z, \psi_{\rm p})}{(r_{1\rm g} - r_{2\rm g})\cos(\chi_r)} \, , \\
    \delta Y_z(\lambda) &= Y_{z \rm s}(\lambda) + z_{\rm s} \totder{Y_{z\rm g}}{z_{\rm g}} - \frac{\mathrm h_z(\chi_r, \chi_z, \psi_{\rm p})}{z_{1\rm g}\cos(\chi_z)} \, ,
\end{align}
and
\begin{align}
    Y_{r \rm s}(\lambda) &= \frac{R_s}{2(r_{1 \rm g} - r)(r - r_{2 \rm g})Y_{r\rm g}} \nonumber \\
    &+ \bigg(\frac{r_{1 \rm s}}{r_{\rm g} - r_{1\rm g}} - \frac{r_{2 \rm s}}{r_{2\rm g} - r_{\rm g}} \bigg) \frac{Y_{r\rm g}}{2}  \, , \\
    Y_{z \rm s}(\lambda) &= \frac{Z_s}{z^2_{1 \rm g} \cos^2(\chi_z) Y_{z\rm g}} -\frac{z_{1{\rm s}\parallel} + z_{1{\rm s}\perp}\sin \chi_z}{z_{1\rm g}\cos^2(\chi_z)} Y_{z\rm g}  \, , 
\end{align}
\begin{align}
   & \mathrm h_r(\chi_r, \chi_z, \psi_{\rm p}) = \totder{\psi_{\rm p}}{\lambda}\parder{r_{\rm s}}{\psi_{\rm p}} + \totder{z_{\rm g}}{\lambda} \parder{r_{\rm s}}{z_{\rm g}}  \, , \\ 
   & \mathrm h_z(\chi_r, \chi_z, \psi_{\rm p}) = \totder{\psi_{\rm p}}{\lambda}\parder{z_{\rm s}}{\psi_{\rm p}} + \totder{r_{\rm g}}{\lambda} \parder{z_{\rm s}}{r_{\rm g}} \, .
\end{align}

\subsubsection{Radial and polar trajectories in terms of the mean anomalies} \label{sssec:meananom}
It is convenient to expand the trajectories and velocities of a spinning particle in a Fourier series, especially for the computation of GWs. However, one of the main obstacles of this approach is the non-separability of Eqs.~\eqref{eq:1st-order-linMPDeq}, which in turn implies that the turning points are no-longer separable. In the following, we show a workaround of this problem using a near-identity transformation similar to the one discussed in~\cite{Witzany:2019nml}.
We start by defining  a new set of anomalies $\vec \zeta = ( \zeta_r, \zeta_z, w_{\rm p})$  given by\footnote{The $\zeta_y$ functions could also be defined by expressing  $w_{y \rm g}$ from eq. \eqref{eq:wydef} as a function of $\chi_{y \rm g}$ and replacing it by $\chi_y$. Here we use a different symbol to stress that $\zeta_y$ is a function defined on the phase space of the spinning particle.}
\begin{align}
    \zeta_y & = 
        \Upsilon_{y\rm g} \int_0^{\chi_y}\frac{\dd{\chi'_y}}{Y_{y \rm g}(\chi'_y)} 
    \qquad y = r,z  \, ,  \label{eq:non-hom-mean-anom_int}\\
    w_{\rm p} & = 
        \psi_{\rm p} + \int_0^{\chi_{r\rm g}}\frac{\langle \Psi_r \rangle - \Psi_r(\chi'_{r \rm g})}{Y_{r\rm g}(\chi'_{r \rm g})}\dd{\chi'_{r \rm g}} + \nonumber 
        \\
        & + \int_0^{\chi_{z\rm g}}\frac{\langle \Psi_z \rangle - \Psi_z(\chi'_{z \rm g})}{Y_{z\rm g}(\chi'_{z \rm g})}\dd{\chi'_{z \rm g}}  \, ,
        \\
    \langle \Psi_y \rangle &= 
        \frac{\Upsilon_{y\rm g}}{2\pi} \int_0^{2\pi}\frac{\Psi_y(\chi'_{y \rm g})}{Y_{y \rm g}(\chi'_{y \rm g})}\dd{\chi'_{y \rm g}} 
    \qquad y = r,z \, , 
\end{align} 
where the spin-precession frequency is given by $\Upsilon_{\rm p}  = \langle \Psi_r \rangle  + \langle \Psi_z \rangle $ whose analytic expression is given in Ref.~\cite{vandeMeent:2019cam}.   
After applying  Leibniz's integral rule, we see that
\begin{align}
    \totder{\zeta_y}{\lambda} &= 
        \Upsilon_{y\rm g} \bigg(1 + q \frac{1}{Y_{y \rm g}(\lambda)} \totder{\chi_{y \rm s}}{\lambda}\bigg) \, ,  \label{eq:non-hom-mean-anom} 
    \qquad y=r,z  \, , 
    \\
    \totder{w_{\rm p}}{\lambda} &= \Upsilon_{\rm p} \ .
\end{align}
Clearly, Eq.~\eqref{eq:non-hom-mean-anom} are not in homogeneous form, unlike their geodesic counterparts.
These equations can be recast in a homogeneous form by means of a near-identity transformation $\vec \zeta = \wmean - q\vec \xi(\wmean)$, where $\wmean = (w_r, w_z, w_{\rm p})$ while $w_r, w_z$ are a new set of mean anomalies that satisfy homogeneous ODEs. The formal solution for $\vec \xi = (\xi_r, \xi_z, \xi_{\rm p})$ is
\begin{align} \label{eq:xi_nit}
   & \xi_y = \displaystyle \sum_{n = -\infty}^{\infty} \sum_{k = -\infty}^{\infty} \sum_{j = -1}^{1} \frac{i e^{-i \vec \kappa \cdot \wmean}}{\vec \kappa \cdot \vec \Upsilon_{\rm g}}\left(\frac{\delta Y_y}{Y_{y \rm g}}\right)_{\!nkj}  \qquad y=r,z \, ,   \\
   & \xi_{\rm p} = 0 \, , \\
   & \left(\frac{\delta Y_y}{Y_{y \rm g}}\right)_{\!nkj} = \frac{1}{(2\pi)^3} \int_{(0,2\pi]^3} \dd^3 \mathsf{w} \frac{\delta Y_y(\wmean)}{Y_{y \rm g}(w_y)} e^{i \vec \kappa \cdot \wmean} 
\end{align}
where $\vec\Upsilon_{\rm g} = (\Upsilon_{r\rm g},\Upsilon_{z\rm g}, \Upsilon_{\rm p})$ and $\vec \kappa = (n,k, j)$ with $j \in (-1,0,1)$. The vector $\vec \xi$ is well-defined as long as the geodesic trajectory does not cross a resonance $\vec \kappa \cdot \vec \Upsilon_{\rm g} = 0$. 

The new equations of motion for $\wmean = (w_r, w_z, w_{\rm p})$ are
\begin{align}
    \totder{w_y}{\lambda} &= \Upsilon_{y\rm g} \bigg(1+ q \left \langle\frac{\delta Y_y}{Y_{y \rm g}} \right\rangle \! \! \bigg) \qquad y = r,z  \, ,  \\
    \totder{w_{\rm p}}{\lambda} &= \Upsilon_{\rm p} \, ,
\end{align}
where
\begin{equation}
    \left \langle\frac{\delta Y_y}{Y_{y \rm g}} \right\rangle \equiv \frac{1}{(2\pi)^3} \int_{(0,2\pi]^3}\dd^3 \mathsf{w} \frac{\delta Y_y(\wmean)}{Y_{y \rm g}(w_y)}  \qquad y=r,z \, .
\end{equation}
The corrections to the radial and polar frequencies,  $\Upsilon_{r \rm s}$ and $\Upsilon_{z \rm s}$ respectively, are given by
\begin{equation}
   \Upsilon_{y \rm s} \equiv \Upsilon_{y\rm g}  \left \langle\frac{\delta Y_y}{Y_{y \rm g}} \right\rangle  \qquad y=r,z \, .
\end{equation}
It is easy to show that $\Upsilon_{r \rm s}$ and $\Upsilon_{z \rm s}$ do not depend on  $s_\perp$. The semi-analytical expressions for the frequency shifts can be found in Appendix \ref{app:rad_pol_freq_shifts}.

We can now conveniently expand any function $f(r,z)$ of the spin-perturbed trajectories $r,z$ in Fourier series using the new mean anomalies $\wmean$, in a fashion similar to Refs.~\cite{Drummond:2022xej,Drummond:2022efc}. To do so, we first need to write the trajectories $r,z$ in terms of $\wmean$. We first notice that the anomalies $\chi_r (\zeta_r)$ and $\chi_z (\zeta_z)$ can be obtained analytically by simply inverting Eqs.~\eqref{eq:non-hom-mean-anom_int} (see  Appendix~\ref{app:geo_freq} for their explicit expressions). By substituting $\vec{\zeta} = \wmean - q\vec \xi(\wmean)$ into $\chi_r(\zeta_r)$ and $\chi_z(\zeta_z)$, and expanding in $q$, we can write $r(\wmean) $ and  $z(\wmean) $ as 
\begin{align}
r(\wmean) &= r_{\rm g}(w_r) + q \delta r(\wmean) \, , \\
z(\wmean) &= z_{\rm g}(w_z)  + q\delta z(\wmean) \, ,
\end{align}
where
\begin{align}
\delta r(\wmean) &= r_{\rm s}(\wmean)  - \totder{r_{\rm g}}{\chi_r} \totder{\chi_r}{w_r} \xi_r (\wmean) \, , \label{eq:spin_radial_trajectory} \\
\delta z(\wmean) &= z_{\rm s}(\wmean)  - \totder{z_{\rm g}}{\chi_z} \totder{\chi_z}{w_z} \xi_z(\wmean) \, . \label{eq:spin_polar_trajectory} 
\end{align}

\subsubsection{Fourier series expansions in Mino-time}
A generic function $f(r, z, \psi_{\rm p})$ of the spin trajectories $r(\wmean)$ and $z(\wmean)$, and spin vector $s^\mu(r_{\rm g},z_{\rm g}, \psi_{\rm p})$ can be expanded as 
\begin{align}
f(r, z, \psi_{\rm p}) &= f_{\rm g}(r_{\rm g}(w_r),z_{\rm g}(w_z)) + q \delta f(\wmean) \, , \\
\delta f(\wmean) &= f_{\rm s}(\wmean) + \frac{\partial f_{\rm g}}{\partial r_{\rm g}} \delta r(\wmean) + \frac{\partial f_{\rm g}}{\partial z_{\rm g}} \delta z(\wmean) \nonumber \\ 
&+ \displaystyle \sum^3_{i =1} \frac{\partial f_{\rm g}}{\partial C_{i \rm g}} C_{i\rm s} \, \label{eq:spin_differential}
\end{align}
where $f_{\rm g}(r_{\rm g}, z_{\rm g})$ is the geodesic contribution to the function $f(r, z) $ whereas $f_{\rm s}(\wmean)$ is the explicit correction due to the secondary spin. We can then expand the function $f(r, z, \psi_{\rm p})$ in Fourier series as
\begin{align}
f_{\rm g}(r_{\rm g}(w_r),z_{\rm g}(w_z)) &= \displaystyle \sum_{n = -\infty}^{\infty} \sum_{k = -\infty}^{\infty} (f_{\rm g})_{nk} e^{-i \vec b \cdot \vec w} \nonumber\\
&= \langle f_{\rm g} \rangle + \osc  f_{\rm g}(\vec w)  \, , \\
\delta f(\wmean) &= \displaystyle \sum_{n = -\infty}^{\infty} \sum_{k = -\infty}^{\infty} \sum_{j = -1}^{1}(\delta f)_{nkj} e^{-i \vec \kappa \cdot \wmean} \nonumber\\
&= \langle \delta f \rangle + \osc (\delta f)(\wmean)  \, ,
\end{align}
where $\vec w = (w_r,w_z)$ while the Fourier coefficients are defined as
\begin{align}
(f_{\rm g})_{nk} &= \frac{1}{(2\pi)^2} \int_{(0,2\pi]^2}\dd^2 w  f_{\rm g}(\vec w) e^{i \vec b \cdot \vec w}  \, , \\
(\delta f)_{nkj} &= \frac{1}{(2\pi)^3} \int_{(0,2\pi]^3}\dd^3 \mathrm{w} \delta f(\wmean) e^{i \vec \kappa \cdot \wmean}  \, , \\
& \langle f_{\rm g} \rangle   =  (f_{\rm g})_{00}  \, ,\qquad  \langle \delta f \rangle   = (\delta f)_{000} \, ,
\end{align}
To reduce clutter, we introduce the short-hand notation
\begin{equation}
\displaystyle \sum_{\vec \kappa} \equiv \displaystyle \sum_{n = -\infty}^{\infty} \sum_{k = -\infty}^{\infty} \sum_{j = -1}^{1}   
\end{equation}
For instance, the spin corrections to the coordinate time and azimuthal rate of change, $\Upsilon_{t \rm s}$  and  $\Upsilon_{\phi \rm s}$ respectively, are given as
\begin{align}
\Upsilon_{t \rm s} & = \left\langle \totder{\delta t}{\lambda} \right \rangle  \, , \quad  \Upsilon_{\phi \rm s} = \left \langle \totder{\delta \phi}{\lambda} \right\rangle \, .
\end{align}
As shown in \cite{Witzany:2019nml}, these averages can be computed without the explicit knowledge of the shift vectors $\xi_y$. We repeat this argument and derive the explicit expressions for $\Upsilon_{t \rm s}$ and $\Upsilon_{\phi \rm s}$ in Appendix \ref{app:upstphi}.

The corrections to the velocities $ \dd \delta t/ \dd \lambda$ and $\dd \delta \phi/ \dd \lambda$ can then be expanded in Fourier series as
\begin{align}
\totder{\delta t}{\lambda} &= \displaystyle \sum_{\vec \kappa}(\delta V^t)_{nkj} e^{-i\vec \kappa \cdot \wmean}  \, , \label{eq:dtsdlFour} \\
\totder{\delta \phi}{\lambda} &= \displaystyle \sum_{\vec \kappa}(\delta V^\phi)_{nkj} e^{-i\vec \kappa \cdot \wmean} \, . \label{eq:dphisdlFour}
\end{align}
After integrating
\begin{align}
\totder{t}{\lambda} &= \totder{t_{\rm g}}{\lambda} + q \totder{\delta t}{\lambda} \, , \\
\totder{\phi}{\lambda} &= \totder{\phi_{\rm g}}{\lambda} + q \totder{\delta \phi}{\lambda} \, ,
\end{align}
with respect to $\lambda$ and considering Eqs.~\eqref{eq:dtsdlFour} and \eqref{eq:dphisdlFour}, the purely oscillating spin corrections to the trajectories, $\osc(\delta t)(\wmean)$ and $\osc(\delta\phi)(\wmean)$, respectively, can be efficiently computed as
\begin{align}
\osc(\delta t) &= \sum_{n \neq 0} \frac{i(V^t_{r \rm g})_n\Upsilon_{r \rm s}}{n (\Upsilon_{r\rm g})^2} e^{-i n w_r} + \sum_{k \neq 0} \frac{i(V^t_{z\rm g})_k \Upsilon_{z \rm s}}{k (\Upsilon_{z\rm g})^2} e^{-i k w_z} \nonumber \\
                       & + \displaystyle \sum_{\vec \kappa \neq 0} \frac{i(\delta V^t)_{nkj}}{ \vec \kappa \cdot \vec \Upsilon_{\rm g}} e^{-i \vec \kappa \cdot \wmean} \, , \label{eq:spintFourierseries} \\
\osc(\delta\phi) &= \sum_{n \neq 0} \frac{i(V^\phi_{r \rm g})_n \Upsilon_{r \rm s}}{n (\Upsilon_{r\rm g})^2} e^{-i n w_r} + \sum_{k \neq 0} \frac{i(V^\phi_{z \rm g})_k \Upsilon_{z \rm s}}{k (\Upsilon_{z\rm g})^2} e^{-i k w_z} \nonumber \\
                        & + \displaystyle \sum_{\vec \kappa \neq 0} \frac{i(\delta V^\phi)_{nkj}}{\vec \kappa \cdot \vec \Upsilon_{\rm g}} e^{-i \vec \kappa \cdot \wmean}  \, . \label{eq:spinphiFourierseries}
\end{align} 
The previous spectral solutions are valid for any $\vec \kappa $ such that $\vec \kappa \cdot \vec \Upsilon_{\rm g} \neq 0$ , i.e. for non-resonant geodesics orbits. 
The trajectories $t(\wmean)$ and $\phi(\wmean)$ up to first order in the spin are then described by the following expressions:
\begin{align}
 t(\lambda) &= \Upsilon_t\lambda + \osc t_{\rm g}(\vec w) + q \osc(\delta t)(\wmean) \, , \label{eq:t_trajectory}\\
 \phi(\lambda) &= \Upsilon_\phi\lambda + \osc \phi_{\rm g}(\vec w) + q \osc(\delta \phi)(\wmean) \, ,  \label{eq:phi_trajectory}
\end{align}
with $\Upsilon_t = (\Upsilon_{t\rm g} + q s_\parallel \Upsilon_{t \rm s})$ and $\Upsilon_\phi = (\Upsilon_{\phi \rm g} + q s_\parallel \Upsilon_{\phi \rm s})$.

It is worth comparing our expressions for the trajectories with the expressions in Section IV.3 of Ref. \cite{Drummond:2022efc}. In both cases, geodesics and spin corrections are parameterised in terms of mean anomalies $\wmean$, which include the frequency shifts at $\mathcal O(sq)$.

\subsection{Fixing the referential geodesic} \label{subsec:parametrization}
Practical computations of the spin orbital corrections require choosing a fiducial referential geodesic. In our formalism, this is equivalent to specifying the shifts to constants of motion $E_{\rm s}$, $J_{z\rm s}$ and $K_{\rm s}$. The corrections to the orbital motion can then be computed using the expressions shown in Section~\ref{sec:solutionsEoM}, Appendix~\ref{app:turnp} (for the shifts to the turning points), and Appendices~\ref{app:rad_pol_freq_shifts} and~\ref{app:time_azi_freq_shifts} for the shifts to the frequencies. 
Before presenting some of the most common choices of fiducial geodesics, we can make a general statement. The shifts to the constants of motion $E_{\rm s}$, $J_{z\rm s}$ and $K_{\rm s}$ can be written in terms of averaged quantities, which only depend on $s_\parallel$ since terms proportional to $s_\perp$ average to zero at order $\mathcal O(q s)$. Hence, the spin corrections to the orbits for $s_\perp$ are the same for any parametrization. The choice of the referential geodesics only affect the spin corrections proportional to $s_\parallel$.
Furthermore,  we noticed that, the correction to the polar motion $\delta z$ for almost equatorial orbits (i.e. when $z_{1\rm g} \sim 0$) reduces to
\begin{equation}
\delta z \sim - s_\perp \frac{\cos \psi_{\rm p} \sqrt{r^2_{\rm g} + (L_{z\rm g} - a E)^2}}{r_{\rm g} (L_{z\rm g} - a E_{\rm g})} \qquad z_{1\rm g} \sim 0
\end{equation}
which for $a=0$ agrees with Eq. (S41) of \cite{Witzany:2023bmq}. It is easy to see that this solution satisfies Eq.(5.15) of Ref.~\cite{Drummond:2022xej}, which can be simplified as\footnote{The coefficient $(\hat L_z - a \hat E)$ in front of the second line of Eq.(5.15) of Ref.~\cite{Drummond:2022xej} is incorrect.}
\begin{equation}
    \frac{\dd^2 \delta z}{\dd \lambda^2} + \Upsilon^2_{z\rm g}\delta z = - 3 s_\perp\frac{\cos\psi_{\rm p}}{r^2}L_{z\text{rd}}\sqrt{L^2_{z\text{rd}} + r^2_{\rm g}} \, ,
\end{equation}
where $L_{z\text{rd}} \equiv L_{z\rm g} - a E_{\rm g}$, and by using the equation of motion in the limit of equatorial orbits, we have that 
\begin{equation} 
    \frac{\dd^2 \delta z}{\dd \lambda^2} = -s_\perp\frac{\cos(\psi_{\rm p})\sqrt{r^2_{\rm g} + L^2_{z\text{rd}} }}{ L^2_{z\text{rd}}r^2_{\rm g}}\big(3L^2_{z\text{rd}} - r_{\rm g}\Upsilon^2_{z\rm g}\big) \, .
\end{equation}

\subsubsection{Fixed constants of motion}
The first parametrization we consider fixes the fiducial geodesic with constants of motion $E_{\rm g}, L_{z \rm g}, K_{\rm g}$ in such a way that $E_{\rm g} =E$ , $L_{z \rm g} = J_z$ , $K_{\rm g} = K$ (or $E_{\rm s} = J_{z\rm s} = K_{\rm s} = 0$).
This choice of the fiducial geodesic was first proposed in Ref.~\cite{Witzany:2019nml}. However, it also naturally arises when considering the alternative definition of the Carter constant $Q_{\rm g} \equiv K_{\rm g} - (L_{z \rm g} - a E_{\rm g})^2$ and its spin-perturbed counterpart~\cite{Tanaka:1996ht}. We label with the superscript ``FC'' all quantities computed in the fixed constants of motion parametrization. 
Due to its simplicity, this scheme is well suited to compute the shifts of constants of motion in other parametrizations.

\subsubsection{Fixed turning points on average}
The second parametrization we consider was used in the already discussed paper by Drummond \& Hughes~\cite{Drummond:2022efc}, which solved the second-order set of MPD equations using a Fourier series ansatz. There, the referential geodesic was fixed by imposing that the averaged spin corrections to the turning points are instead zero. We refer to this choice of the referential geodesic as the fixed turning points parametrization, and label all related quantities with the superscript ``DH". 

The expressions for $E^\text{DH}_{\rm s}$, $J^\text{DH}_{z\rm s}$ and $K^\text{DH}_{\rm s}$ can be obtained as follows.  Given a function of the constants of motion $f(E,J,K)$, the differential $\delta f^{\text{DH}}_s$ due to the constants of motion shifts  $E^\text{DH}_{\rm s}$, $J_{z \rm s}^\text{DH}$, $K^\text{DH}_{\rm s}$ as computed in~\cite{Drummond:2022xej,Drummond:2022efc} is
\begin{equation}
\delta f^\text{DH} = \delta f^\text{FC} + \displaystyle \sum^3_{i=1} \frac{\partial f_{\rm g}}{\partial C_{i\rm g}} C^\text{DH}_{i\rm s}
\end{equation}
where $ \big(C^\text{DH}_{1\rm s}, C^\text{DH}_{2\rm s}, C^\text{DH}_{3\rm s}\big) = \big(E^\text{DH}_{\rm s}, J^\text{DH}_{z\rm s}, K^\text{DH}_{\rm s} \big)$. Applying the previous expression to the turning points $r_1$, $r_2$ and $z_1$ and taking the average gives
\begin{equation}
\begin{pmatrix}
  E^\text{DH}_{\rm s} \\
  J^\text{DH}_{z\rm s} \\
  K^\text{DH}_{\rm s} \\
 \end{pmatrix} = -
 \begin{pmatrix}
  \frac{\partial E_{\rm g}}{\partial r_{1\rm g}} & \frac{\partial E_{\rm g}}{\partial r_{2\rm g}} & \frac{\partial E_{\rm g}}{\partial z_{1\rm g}}\\
  \frac{\partial L_{z\rm g}}{\partial r_{1\rm g}} & \frac{\partial L_{z\rm g}}{\partial r_{2\rm g}} & \frac{\partial L_{z\rm g}}{\partial z_{1\rm g}} \\
  \frac{\partial K_{\rm g}}{\partial r_{1\rm g}} & \frac{\partial K_{\rm g}}{\partial r_{2\rm g}} & \frac{\partial K_{\rm g}}{\partial z_{1\rm g}}
 \end{pmatrix}
 \!\cdot \!
 \begin{pmatrix}
  \langle r^\text{FC}_{1\rm s} \rangle \\
  \langle r^\text{FC}_{2\rm s} \rangle \\
  \langle z^\text{FC}_{1\rm s} \rangle \\
 \end{pmatrix}
 \label{eq:DH_constant_of_motion}
\end{equation}
The shifts~\eqref{eq:DH_constant_of_motion} are expressed in closed form since the the averages $\langle r^\text{FC}_{1 \rm s} \rangle$, $\langle r^\text{FC}_{2 \rm s} \rangle$ and $\langle z^\text{FC}_{1 \rm s} \rangle$ can be written in closed form in terms of Legendre elliptic integrals (see Appendix~\ref{app:turnp}). Since the map between the roots $r_{1\rm g}$, $r_{2\rm g}$ and $z_{1 \rm g}$ and the geodesics constants of motion is unique, the Jacobian determinant of the transformation~\eqref{eq:DH_constant_of_motion} is always non-zero, except for spherical and circular orbits.

Finally, Appendix~\ref{app:DHmapFC} presents the transformations of the orbital corrections between the ``FC'' and ``DH'' parametrizations, which we used to cross-check our calculations.

\subsubsection{Fixed frequencies}
The referential geodesics obtained by the parametrizations above have fundamental orbital frequencies that are generally not equal to the frequencies of the spinning particle motion (cf.~\cite{Drummond:2022efc,Drummond:2022xej}). This leads to secularly growing differences in the phases between the trajectories. To avoid such issues, a third alternative parametrization is to choose the referential geodesics with the same frequencies as the spin-trajectories. We will call this the fixed frequency parametrization, or ``FF" parametrization in short (for more details, see also Refs.~\cite{Mathews:2021rod,Skoupy:2022adh}). 
Specifically, the radial, polar and azimuthal frequencies shifts, respectively, $\Omega_{r\rm s}$ $\Omega_{z\rm s}$ and $\Omega_{\phi \rm s}$ defined in Eqs.~\eqref{eq:freq_shifts_coordinate_time} are set to zero. These are the frequency shifts measured in a reference frame located at infinity with respect to the primary BH. The shifts to constants of motion that fix $\Omega_{r\rm s} =\Omega_{z\rm s} =\Omega_{\phi \rm s} =0$ are given by

\begin{equation}
\begin{pmatrix}
  E^\text{FF}_{\rm s} \\
  J^\text{FF}_{z\rm s} \\
  K^\text{FF}_{\rm s} \\
 \end{pmatrix} = -
 \begin{pmatrix}
  \frac{\partial \Omega_{r\rm g}}{\partial E_{\rm g}} & \frac{\partial \Omega_{r\rm g}}{\partial L_{z\rm g}} & \frac{\partial \Omega_{r\rm g}}{\partial K_{\rm g}}\\
  \frac{\partial \Omega_{z\rm g}}{\partial E_{\rm g}} & \frac{\partial \Omega_{z\rm g}}{\partial L_{z\rm g}} & \frac{\partial \Omega_{z\rm g}}{\partial K_{\rm g}} \\
  \frac{\partial \Omega_{\phi\rm g}}{\partial E_{\rm g}} & \frac{\Omega_{\phi\rm g}}{\partial L_{z\rm g}} & \frac{\Omega_{\phi\rm g}}{\partial K_{\rm g}}
 \end{pmatrix}^{\!\!-1}
 \!\cdot \!
 \begin{pmatrix}
  \Omega^\text{FC}_{r\rm s} \\
  \Omega^\text{FC}_{z\rm s} \\
  \Omega^\text{FC}_{\phi\rm s} \\
 \end{pmatrix}
 \label{eq:FF_constant_of_motion}
\end{equation}
The determinant of the Jacobian matrix in the previous transformation may be zero in certain regions of the parameter space due to the iso-frequency pairing phenomenon~\cite{Warburton:2013yj}. In such a situation, the matrix inverse in~\eqref{eq:FF_constant_of_motion} does not exist.

We have not implemented the solutions of the EoM in the fixed frequency parametrization, but it is easy to do so using our general formalism and the shifts~\eqref{eq:FF_constant_of_motion}.

\subsubsection{Alternative parametrizations}
For completeness, let us also mention yet another possible parametrization, in which the initial conditions of the reference geodesic match those of the spinning-body orbit \textemdash{ } for more details, see Refs.~\cite{Bini:2011nhv,Bini:2014soa}. We will refer to this parametrization as “IC”. For example, Ref.~\cite{Bini:2011nhv} obtained analytic expressions for the orbit of a spinning body with the same initial position and 4-velocity as a circular equatorial reference geodesic. 

Given a set of initial conditions $x^\mu_{(0)}, u^\mu_{(0)}, s^{\mu\nu}_{(0)}$, one can compute the corresponding constants of motion $E,\, J_z,\, K, s_\parallel$ from equations \eqref{eq:energyspin}, \eqref{eq:angmomspin}, \eqref{eq:Carterspin} and \eqref{eq:spinpara}. Similarly, one can compute the $E_{\rm g}, L_{\rm g}, K_{\rm g}$ corresponding to $x^\mu_{(0)}, u^\mu_{(0)}$ by setting $s = 0$ in the formulas.  As a result, we see that the shifts of the constants of motion are
\begin{align}
    & E^{\rm IC}_{\rm s} = -\frac{1}{2} \Xi_{\mu;\nu}(x^\mu_{(0)}) s^{\mu\nu}_{(0)}\,,
    \\
    & J^{\rm IC}_{z \rm s} =  -\frac{1}{2} \Xi_{\mu;\nu} (x^\mu_{(0)}) s^{\mu\nu}_{(0)}
    \\
    & K^{\rm IC}_{\rm s} = - 2 u^\mu_{(0)} s^{\rho \sigma}_{(0)} (\mathcal F_{\mu \rho;\kappa} \mathcal F^\kappa_{~\sigma}+ \mathcal F_{\rho \sigma;\kappa}\mathcal F^\kappa_{~\mu})\Big|_{x^\mu = x^\mu_{(0)}} \,.
\end{align}
 For the computation of average observables such as fundamental frequencies and fluxes, or even the Fourier modes of the trajectories, the shifts of the constants of motion are sufficient. However, one can also find the initial phase shift of the mean anomalies in order to plot the geodesic and the spinning particle as truly starting from the same point. For this one first finds the mean anomalies $w_{r \rm g}, w_{z \rm g}$ of the geodesic by using equations \eqref{eq:wrg} and \eqref{eq:wzg} with initial data $x^\mu_{(0)}, u^\mu_{(0)}, s^{\mu\nu}_{(0)}$. For $w_{\rm p}$ one then needs to find the spin-precession angle from projecting into the Marck tetrad as in \eqref{eq:sptMarck}, and then using the relations (56)-(62) in Ref. \cite{vandeMeent:2019cam}.\footnote{Note that the quantities denoted as $q_\psi,q_r,q_z$ in Ref. \cite{vandeMeent:2019cam} must be replaced by $w_{\rm p}, w_{r \rm g}, w_{z \rm g}$ for this computation.}   Then the spin-corrected anomalies corresponding to the same initial conditions will be $w_y = w_{y \rm g}  + q \xi_y(w_{r \rm g},w_{z \rm g},w_{\rm p})$.

\subsubsection{The case of quasi-spherical orbits}
When the referential geodesic is a spherical orbit (e.g. $e_{\rm g} =0$), the shifts to the turning points diverge because of terms $\sim 1/e_{\rm g}$. This problem arises in all the aforementioned parametrizations, including the fixed-turning-point parametrization. Indeed, in such a scheme the turning points are fixed only on average, so the divergence $1/e_{\rm g}$ is still present in the fluctuating terms. However, the limit $e_{\rm g} \to 0$ is finite and well defined. It is possible to remove such singularities from all parametrizations by a suitable expansion in $e_{\rm g}$. 
We stress that, when the fiducial geodesic has $e_{\rm g} =0$, the spin perturbed trajectory is not a spherical orbit, that is, the Boyer-Lindquist radius is non-constant. We leave the case of quasi-spherical orbits for future work. 

\section{Asymptotic fluxes and amplitudes for a spinning particle} \label{sec:asymptotic_amplitudes}
\subsection{Teukolsky formalism}
The metric perturbations induced by the secondary on the background spacetime can be computed by solving a master equation known as the Teukolsky equation \cite{Teukolsky:1972my,Teukolsky:1973ha}. This equation was derived using the Newman-Penrose formalism, which allows us to single out the non-trivial degree of freedom of the Riemann tensor. The perturbation to the Weyl scalar $\Psi_4$ at spatial infinity reads
\begin{equation}
    \Psi_4 (r \to \infty) = \frac{1}{2} \frac{\partial^2}{\partial t^2}(h_+ - i h_\times) \, . \label{eq:psi_4_inf}
\end{equation}
Using the following Fourier ansatz
\begin{equation}
    \Psi_4 = \rho^4 \displaystyle \sum_{\ell =2}^{\infty}\sum_{m=-\ell}^{\ell} \int_{-\infty}^{\infty} \dd \omega R_{\ell m \omega}(r) \nswsh (\theta) e^{i (m \phi - \omega t)}\,,
\end{equation}
the Teukolsky equation reduces from a partial differential equation to a tower of decoupled ordinary differential equations, known as the mode-decomposed Teukolsky equation in the Fourier domain. The functions $\nswsh (\theta)$ are the spin-weighted spheroidal harmonics, which are eigenfunctions of the following (irregular) Sturm-Liouville problem
\begin{align}
&\left.\Bigg[\frac{1}{\sin\theta}\totder{}{\theta}\left(\sin\theta\totder{}{\theta} \right) - c^2\sin^2\theta - \left( \frac{m - 2 \cos\theta}{\sin\theta}\right)^{\!\!2} \right.\nn\\
&+\left. 4 c \cos\theta- 2 
+2 mc\right.\!\Bigg]\rswsh = -\lambda_{\ell m\omega} \rswsh  \, ,\label{eq:swsweq}
\end{align}
with eigenvalue $\lambda_{\ell m \omega}$. To reduce clutter in the expressions, we introduced the notation $c = a \omega$ . Moreover, from now on, the spin-weight -2, which refers to metric perturbations of the background spacetime, is understood.
The eigenvalues and eigenfunctions have the following symmetries: $\lambda_{\ell m -\omega} = \lambda_{\ell -m \omega}$ and 
\begin{equation}
 S^{- c}_{\ell -m}(\theta) = (-1)^l \!\!\, S^c_{\ell m}(\pi-\theta) \, .
\end{equation}
For $c =0$, $\rswsh$ reduce to the spin-weighted spherical harmonics. 
The radial function $R_{\ell m \omega}(r)$ is a solution of the following differential equation
\begin{equation}
\Delta^2 \totder{}{r}\left(\!\frac{1}{\Delta}\totder{R_{\ell m\omega}}{r}\right) - V(r) R_{\ell m \omega}(r) = \mathcal T_{\ell m \omega}\, , \label{eq:radialTeueq}
\end{equation}
where the potential $V(r)$ reads
\begin{align}
V(r) &= -\frac{\mathcal K^2(r) + 4i(r-1)\mathcal K(r)}{\Delta} + 8i\omega r + \lambda_{\ell m \omega} \ 
,\\
\mathcal K(r) &\equiv (r^2 + a^2)\omega -a m \, ,
\end{align}
while $\mathcal T_{\ell m \omega} = \mathcal T_{\ell m \omega}(T_{\mu \nu})$ is the source Teukolsky term, which depends on the stress-energy tensor $T_{\mu \nu}$. The inhomogeneous solutions of~\eqref{eq:radialTeueq} satisfy the following boundary conditions\footnote{Our convention for the boundary conditions is the opposite of Ref~\cite{Drasco:2005kz}}
\begin{align}
 R_{\ell m\omega}(r \to r_+) &= Z^H_{\ell m \omega} \Delta^2 e^{- i \kappa_{\mathcal H} r^*(r)} \, , \\
 R_{\ell m\omega}(r \to \infty) &= Z^\infty_{\ell m \omega} r^3 e^{i \omega r^*(r)} \, ,
\end{align}
with $\kappa_{\mathcal{H}}= \omega - m a/(2 r_+)$, $r_\pm = 1 \pm \sqrt{1- a^2}$ and 
\begin{equation}
    r^*(r) = r + \frac{2 r_+}{r_+ - r_-} \ln\!\Big(\frac{r- r_+}{2}\Big) - \frac{2 r_-}{r_+ - r_-} \ln\!\Big(\frac{r- r_-}{2}\Big) \, .
\end{equation}

The asymptotic partial amplitudes $Z^\infty_{\ell m \omega}$ and $ Z^H_{\ell m \omega}$ are given as
\begin{align}
    Z^{H,\infty}_{\ell m \omega} &= \frac{1}{W_r} \int_{r_+}^\infty \dd r' \frac{R^{\text{up,in}}_{\ell m \omega}(r')}{\Delta^2} \mathcal T_{\ell m \omega}(r') \, ,
\end{align}
where $\Rin$ and $\Rup$ are the two homogeneous solutions of Eq.~\eqref{eq:radialTeueq}, which satisfy the the following asymptotic values at horizon $ r_+$ and at infinity:
\begin{align}
R^\text{in}_{\ell m\omega} \sim
\begin{cases}
B^{\textup{tran}}_{\ell m \hat\omega} \Delta^2 e^{-i \kappa_{\mathcal H} r^\ast} \quad  &r \to 
r_+ \ ,\\[0.5mm]
B^{\textup{out}}_{\ell m \omega} r^3 e^{i \omega r^{\ast}} + B^{\textup{in}}_{\ell m \omega} r^{-1}  e^{-i \hat\omega\hat r^\ast} \, \quad & r \to \infty  \ ,\label{eq:inBCTeu}
\end{cases}
\end{align}
\begin{align}
R^{\text{up}}_{\ell m\omega} \sim 
\begin{cases}
D^{\textup{out}}_{\ell m \omega} e^{i\kappa_{\mathcal H} r^\ast} + D^{\textup{in}}_{\ell m \omega}\Delta^2 e^{- i\kappa_{\mathcal H} r^\ast} \quad   & r\to r_+ \ , \\[0.5mm]
D^{\textup{tran}}_{\ell m \omega} r^3  e^{i \omega \hat r^\ast}  \quad & r \to \infty  \ 
, \label{eq:upBCTeu}
\end{cases}
\end{align}
while 
\begin{align}
W_r &\equiv\frac{1}{\Delta}\left(\! {\Rin}\totder{\Rup}{r} - {\Rup}{}\totder{\Rin}{r} \!\right) \\
    &= 2 i\omega D^{\textup{tran}}_{\ell m \omega} B^{\textup{in}}_{\ell m \omega} \nonumber
\end{align}
is the corresponding Wronskian. 

When the dust settles, we can write the asymptotic partial amplitudes $Z^{H,\infty}_{\ell m \omega}$ as integrals over particular source terms \cite{Drasco:2003ky,Drasco:2005kz}
\begin{equation}
Z^{H,\infty}_{\ell m \omega} = \int_{-\infty}^{\infty} \dd t \, e^{i (\omega t - m \phi(t))}I^{H,\infty}_{\ell m}(t) \, ,
\end{equation}
where 
\begin{equation}
    I^{H,\infty}_{\ell m}(t) =  I^{H,\infty}_{\ell m}(r(t),z(t),\psi_{\rm p}(t), R^{\text{up,in}}_{\ell m \omega}, \lambda_{\ell m \omega}, S^c_{\ell m}) \, ,
\end{equation}
is a complicated function that includes the source term $\mathcal T_{\ell m \omega}(T^{\mu \nu})$. See the following Section \ref{subsec:asymampspin}, Appendix \ref{app:source_term} and Refs. \cite{Drasco:2003ky,Drasco:2005kz,Fujita:2009us} for more details. After a change of variable from $t$ to Carter-Mino time $\lambda$, we can write
\begin{equation}
Z^{H,\infty}_{\ell m \omega} = \int_{-\infty}^{\infty} \dd \lambda \, \totder{t}{\lambda} e^{i (\omega t(\lambda) - m \phi(\lambda))}I^{H,\infty}_{\ell m}(t) \, .
\end{equation}
By plugging the expansions~\eqref{eq:t_trajectory} and \eqref{eq:phi_trajectory} into the previous expressions, we get
\begin{equation}
Z^{H,\infty}_{\ell m \omega} = \int_{-\infty}^{\infty} d \lambda \,e^{i (\omega \Upsilon_t - m \Upsilon_\phi)\lambda}J^{H,\infty}_{\ell m}(\lambda) \, ,\label{eq:Teuamp}
\end{equation}
where 
\begin{equation}
J^{H,\infty}_{\ell m}(\lambda) = \totder{t}{\lambda} e^{i (\omega \osc t(\lambda) - m \osc \phi(\lambda))}I^{H,\infty}_{\ell m}(\lambda) \, ,
\end{equation}
Since $J^{H,\infty}_{\ell m}(\lambda)$ is triperiodic, it can be expanded in Fourier series as
\begin{align}
J^{H,\infty}_{\ell m \omega}(\lambda) &= \sum_{\vec \kappa}  J^{H,\infty}_{\ell m \vec \kappa} e^{-i \vec \kappa \cdot \wmean} \, , \label{eq:Jfourierseries} \\ 
J^{H,\infty}_{\ell m \vec \kappa} &=  \frac{1}{(2\pi)^3 \Upsilon_t} \int_{(0,2\pi]} \dd^3 \mathrm{w} J^{H,\infty}_{\ell m}(\wmean) e^{i \vec \kappa \cdot \wmean} \, ,
\end{align} 
By inserting Eq~\eqref{eq:Jfourierseries} in \eqref{eq:Teuamp}, the integral becomes a sum of delta functions:
\begin{align}
Z^{H,\infty}_{\ell m \omega} &= \sum_{\vec \kappa}Z^{H,\infty}_{\ell m \vec \kappa} \, \delta(\omega - \omega_{m\vec \kappa}) \, , \\
 Z^{H,\infty}_{\ell m \vec \kappa} &\equiv  J^{H,\infty}_{\ell m \vec \kappa} \, ,
\end{align}
where 
\begin{align}
    \omega_{m\vec \kappa} &= \omega^{\rm g}_ {m\vec\kappa} + q  s_\parallel \omega^{\rm s}_ {m\vec\kappa} \, , \label{eq:gw_spectrum}\\
    \omega^{\rm g}_{m\vec\kappa} & = m \Omega_{\phi \rm g} +  n \Omega_{r \rm g} +  k \Omega_{z\rm g} + j\Omega_{\rm pg} \, , \\
    \omega^{\rm s}_ {m \vec\kappa} & = m \Omega_{\phi \rm s} +  n \Omega_{r \rm s} +  k \Omega_{z\rm s} + j\Omega_{\rm ps}  \, , 
\end{align}
and
\begin{align}  \label{eq:freq_shifts_coordinate_time}
&\Omega_{y \rm g} = \frac{\Upsilon_{y\rm g}}{\Upsilon_{t\rm g}} \, , \quad \Omega_{y\rm s} = \frac{\Upsilon_{y\rm s}}{\Upsilon_{t\rm g}} - \frac{\Upsilon_{y\rm g}\Upsilon_{t\rm s}}{\Upsilon^2_{t\rm g}} \, , \qquad y = r,z  \, , \\
&\Omega_{\phi\rm g} = \frac{\Upsilon_{\phi\rm g}}{\Upsilon_{t\rm g}} \, , \quad \Omega_{\phi\rm s} = \frac{\Upsilon_{\phi\rm s}}{\Upsilon_{t\rm g}} - \frac{\Upsilon_{\phi\rm g}\Upsilon_{t\rm s}}{\Upsilon^2_{t\rm g}} \, , \\
&\Omega_{\rm pg} = \frac{\Upsilon_{\rm p}}{\Upsilon_{t\rm g}} \, , \quad \Omega_{\rm ps} = -\frac{\Upsilon_{\rm p}\Upsilon_{t \rm s}}{\Upsilon^2_{t \rm g}}
\end{align}
are the coordinate-time frequencies as measured by an observer at infinity. Using~\eqref{eq:gw_spectrum}, the asymptotic partial amplitudes $Z^{H,\infty}_{\ell m \omega}$ can be expanded as follows
\begin{align}
Z^{H, \infty}_{\ell m \vec \kappa} &= \hat Z^{H, \infty}_{\ell m \vec\kappa} + q \delta Z^{H, \infty}_{\ell m \vec \kappa} \, , \label{eq:partial_amp_expand}
\end{align}
with $\hat Z^{H, \infty}_{\ell m \vec\kappa}$ denoting the adiabatic leading order term, and $\delta Z^{H, \infty}_{\ell m \vec \kappa}$ the secondary spin-induced corrections. By combining Eqs.~\eqref{eq:partial_amp_expand},~\eqref{eq:gw_spectrum} and~\eqref{eq:psi_4_inf}, we can write the waveform $h = h_+- i h_\times$ as
\begin{equation}\label{eq:strain}
h  =  -\frac{2\mu}{d_{\mathrm{L}}} \sum_{\ell m\vec\kappa} \mathcal A_{\ell m \vec \kappa} e^{i m\varphi} e^{-i(\omega^{\rm g}_ {m \vec\kappa} + q  s_\parallel \omega^{\rm s}_ {m \vec\kappa})u} \, , 
\end{equation}
where $u = t- r^*(d_{\rm L}),$ and $(d_{\mathrm{L}},\vartheta,\varphi)$ are the polar coordinates of the source's sky location, and
\begin{align}
 \mathcal A_{\ell m \vec k} &= \mathcal A^{\rm g}_{\ell m \vec \kappa} + q \delta\mathcal A_{\ell m \vec \kappa} \, , \\
\mathcal A^{\rm g}_{\ell m \vec \kappa} &= \frac{\hat Z^\infty_{\ell m \vec\kappa}}{(\omega^{\rm g}_{m \vec\kappa})^2} S^{\rm g}_{\ell m \vec\kappa}(\vartheta) \, , \\
\delta\mathcal A_{\ell m \vec\kappa} &= \frac{\hat Z^\infty_{\ell m \vec\kappa}}{(\omega^{\rm g}_ {m \vec\kappa})^2} \delta S_{\ell m \vec\kappa}(\vartheta) + \frac{\delta Z^\infty_{\ell m \vec \kappa}}{(\omega^{\rm g}_ {m \vec\kappa})^2} S^{\rm g}_{\ell m \vec\kappa}(\vartheta)  \nonumber \\
& -2\frac{\omega^{\rm s}_ {m\vec\kappa}}{\omega^{\rm g}_ {m \vec\kappa}}\mathcal A^{\rm g}_{\ell m \vec\kappa}  \, , 
\end{align}
where we have introduced the short-hand notation 
\begin{align}
\displaystyle \sum_{\ell m\vec \kappa} & \equiv \displaystyle \sum_{\ell = 2}^{\infty} \sum_{m = -\ell}^{\ell} \sum_{n = -\infty}^{\infty} \sum_{k = -\infty}^{\infty} \sum_{j = -1}^{1}    \, .
\end{align}
Thanks to Eq.~\eqref{eq:gw_spectrum}, we can expand the eigenvalues and eigenfunctions of Eq.~\eqref{eq:swsweq} as
\begin{align}
\lambda_{\ell m \omega} &= \lambda^{\rm g}_ {\ell m \vec \kappa} + q \delta\lambda_ {\ell m \vec \kappa} \, \label{eq:Teukolsky_polar_eigenvalue_lin} \, ,\\
S^c_{\ell m}(\theta) &= S^{\rm g}_ {\ell m \vec\kappa}(\theta) + q \delta S_{\ell m \vec \kappa}(\theta) \, . \label{eq:Teukolsky_polar_eigenfunction_lin}
\end{align}
Likewise, the homogeneous solutions of the radial Teukolsky equation can be expanded as
\begin{align}
R^{\text{up,in}}_{\ell m \omega}(r) &= \hat R^{\text{up,in}}_{\ell m \vec\kappa}(r)  + q \delta R^{\text{up,in}}_{\ell m \vec\kappa}(r) \, . \label{eq:Teukolsky_radial_sol_lin}
\end{align}
The functions $\delta R^{\text{up,in}}_{\ell m \vec\kappa}(r)$ were obtained by first rewriting the homogeneous version of Eq.~\eqref{eq:radialTeueq} with the ansatz of Ref.~\cite{Zenginoglu:2011jz}, and then expanding the resulting equations and boundary conditions  in the small mass-ratio $q$ using regular perturbation theory. We use the algorithms described in Sec. C and Appendix B of Ref.~\cite{Piovano:2021iwv} to compute the corrections $\delta R^{\text{up,in}}_{\ell m \vec\kappa}(r)$, $\delta \lambda_{\ell m \vec \kappa}$ and $\delta S_{\ell m \vec \kappa}(\theta)$. These corrections vanish for any harmonic mode $n,m,k,j$ in the case of the fixed frequency parametrization, i.e when $\omega^{\rm s}_{m\vec\kappa} = 0$.

\subsubsection{Adiabatic and post-adiabatic waveforms} \label{sec:waveform}

The waveform~\eqref{eq:strain} is not suitable for modeling the GW signal emitted by a binary system since its amplitudes and frequencies are constant in time. Rather, Eq.~\eqref{eq:strain} can be interpreted as a ``snapshot'' for a given time $t$ and a geodesic with orbital elements $a(t), p_{\rm g}(t), e_{\rm g}(t), x_{\rm g}(t)$ and frequencies $\omega_{m \vec \kappa}(t)$, which are no longer constants due to radiation reaction. 

Given that the evolution of the EMRI configuration is  slow compared to the timescale of the orbital motion of the secondary, one can show that, at leading order in the mass ratio, the so-called adiabatic approximation can be used~\cite{Pound:2021qin,Miller:2020bft}. In this approximation, the small body smoothly decays from one geodesic with orbital elements $a(t), p_{\rm g}(t), e_{\rm g}(t), x_{\rm g}(t)$ to another geodesic at  time  $t + \epsilon$ with $a(t + \epsilon), p(t + \epsilon)_{\rm g}, e_{\rm g}(t + \epsilon), x_{\rm g}(t + \epsilon)$ with $\epsilon \ll 1$. Thus, the GW signal emitted by the secondary can be modelled as a sequence of  ``snapshots''. In the case of a spinning body,  ~\eqref{eq:strain} becomes
\begin{equation}\label{eq:strain_time_evol}
h(t)  =  -\frac{2\mu}{d_{\mathrm{L}}} \sum_{\ell m\vec\kappa} \mathcal A_{\ell m \vec \kappa} e^{i m\varphi} e^{-i\Phi_{\ell m \vec k}(t)} \, , 
\end{equation}
where
\begin{align}
 \Phi_{\ell m \vec k}(t) &= \Phi^{\rm g}_ {\ell m \vec k}(t) + q \Phi^{\rm s}_ {\ell m \vec k}(t) \, , \\
 \Phi^{\rm g}_ {\ell m \vec \kappa}(t) &=  \int_{t_0}^t \dd t\omega^{\rm g}_ {\ell m \vec \kappa}(t)  \, , \\
 \Phi^{\rm s}_ {\ell m \vec \kappa}(t) &=  \int_{t_0}^t \dd t\omega^{\rm s}_ {\ell m \vec \kappa}(t)    \, .
\end{align}
The rate of change of the would-be constants of motion $E_{\rm g}$ and $L_{z \rm g}$ is approximated by the asymptotic GW fluxes $\mathcal F$ with flux-balance laws~\cite{Sago:2005fn}, which also hold in the case of a spinning secondary~\cite{Akcay:2019bvk}. For a spinless test-body, one can write the rate of change of the Carter constant using similar flux balance laws; for a spinning particle, such a balance law is currently unknown and one currently cannot accurately drive the evolution.

\subsection{Spin corrections to the asymptotic amplitudes } \label{subsec:asymampspin}
By leveraging Eqs.~\eqref{eq:Teukolsky_polar_eigenvalue_lin}, \eqref{eq:Teukolsky_polar_eigenfunction_lin} and~\eqref{eq:Teukolsky_radial_sol_lin}, it is possible to fully linearize the function $I^{H,\infty}_{\ell m}(\wmean)$ in the secondary spin as 
\begin{align}
I^{H,\infty}_{\ell m}(\wmean) &= \hat I^{H,\infty}_{\ell m}(\vec w) + q \delta I^{H,\infty}_{\ell m}(\wmean)  \, , \\
\delta I^{H,\infty}_{\ell m}(\wmean) &= s_\parallel \delta I^{H,\infty}_{\ell m \parallel}(\vec w) + s_\perp \delta I^{H,\infty}_{\ell m \perp}(\wmean) \, ,
\end{align}
In the same way, the function $J^{H,\infty}_{\ell m}(\wmean)$ can be expanded
\begin{align}
&J^{H,\infty}_{\ell m}(\wmean) = \hat J^{H,\infty}_{\ell m}(\vec w) + q \delta J^{H,\infty}_{\ell m}(\wmean)  \, , \\
& \hat J^{H,\infty}_{\ell m}(\vec w) = \totder{t_{\rm g}}{\lambda} e^{i\beta_{\rm g}(\vec w)} \hat I^{H,\infty}_{\ell m}(\vec w) \, , \\
&\delta J^{H,\infty}_{\ell m}(\wmean) = s_\parallel \delta J^{H,\infty}_{\ell m \parallel}(\vec w) + s_\perp \delta J^{H,\infty}_{\ell m \perp}(\wmean)  \, ,
\end{align}
where
\begin{equation}
\beta_{\rm g}(\vec w) =  \omega^{\rm g}_ {m \vec\kappa}\osc t_{\rm g}(\vec w)  - m \osc \phi_{\rm g}(\vec w) \, . 
\end{equation}
Finally, we separate the corrections to the asymptotic partial amplitudes as
\begin{equation}
\delta Z^{H,\infty}_{\ell m \vec \kappa} = s_\parallel \delta Z^{H,\infty}_{\ell m \vec b \parallel} + s_\perp \delta Z^{H,\infty}_{\ell m \vec b \perp}  \, ,
\end{equation}
with $\delta Z^{H,\infty}_{\ell m \vec b \parallel} \equiv \delta Z^{H,\infty}_{\ell mnk0 \parallel}$ while $\delta Z^{H,\infty}_{\ell m \vec b \perp}$ includes all terms with index $j \neq 0$.

\subsubsection{Leading order asymptotic amplitudes}
The leading order amplitudes are given by
\begin{equation}
\hat Z^{H,\infty}_{\ell m \vec\kappa} =  \frac{1}{(2\pi)^3 \Upsilon_{t\rm g}} \int_{(0,2\pi]^3} \dd^3 \mathrm{w} \hat J^{H,\infty}_{\ell m}(\vec w) e^{i\vec\kappa \cdot \wmean} \, ,
\end{equation}
The function $I^{o,\rm g}_{\ell m}(\vec w)$ is given by 
\begin{equation}
    \hat I^{H,\infty}_{\ell m}(\vec w) = \left. \frac{\Sigma}{W_{r\rm g}} \bigg(A_{0\rm g} - A_{1\rm g} \totder{}{r} + A_{2\rm g} \totderhigh{}{r}{2} \bigg) \hat R^{\text{up,in}}_{\ell m \vec \kappa} \right \rvert_{r = r_{\rm g}} 
\end{equation}
with 
\begin{equation}
W_{r \rm g} \equiv\frac{1}{\Delta}\bigg(\hat R^{\text{in}}_{\ell m \vec\kappa}\totder{\hat R^{\text{up}}_{\ell m \vec \kappa}}{r} - \hat R^{\text{up}}_{\ell m \vec\kappa}\totder{\hat R^{\text{in}}_{\ell m \vec\kappa}}{r} \bigg) \, ,
\end{equation}
The coefficients $A_{0\rm g}$, $A_{1\rm g}$, $A_{2\rm g}$ are cumbersome functions of $r_{\rm g}$ and $z_{\rm g}$ that are provided in the Appendix~\ref{app:source_term}. For more details, see also Refs~\cite{Fujita:2009us}.
Since the function $\hat J^{H,\infty}_{\ell m}(\vec w)$ does not depend on the mean anomaly $w_{\rm p}$, the leading order partial amplitudes $\hat Z^{H,\infty}_{\ell m \vec\kappa}$ satisfy the following relation
\begin{equation}
\hat Z^{H,\infty}_{\ell mnk1} = \hat Z^{H,\infty}_{\ell mnk-1}=0  \, ,
\end{equation}
hence the adiabatic waveform amplitudes $\mathcal A^{\rm g}_{\ell m \vec k}$ are zero when the indices $j = +1$ or $j = -1$. Thus, an adiabatic waveform does not include any spin-precession terms due to the secondary spin in either its amplitudes or phase.

\subsubsection{Asymptotic amplitudes proportional to $s_\parallel$}
Here we consider the corrections to the asymptotic amplitudes arising from the parallel component of spin $s_\parallel$, i.e. for all the corrections with $j =0$, which are given as
\begin{align} \label{eq:partial_amplitude_corr_parallel}
\delta Z^{H,\infty}_{\ell m\vec b\parallel} &=  \frac{1}{(2\pi)^2 \Upsilon_{t\rm g}} \int_{(0,2\pi]^2} \delta J^{H,\infty}_{\ell m \parallel}(\vec w) e^{i\vec b \cdot \vec w} \dd^2 w  \nonumber \\
 &- \frac{\Upsilon_{t \rm s}}{\Upsilon_{t\rm g}} \hat Z^{H,\infty}_{\ell m \vec b} \, ,
\end{align}
where $\hat Z^{H,\infty}_{\ell m \vec b} = \hat Z^{H,\infty}_{\ell mnk0}$ and
\begin{align}
\delta J^{H,\infty}_{\ell m \parallel}(\vec w) &=  e^{i \beta_{\rm g}(\vec w)} \bigg( \delta I^{H,\infty}_{\ell m \parallel}(\vec w)\totder{t_{\rm g}}{\lambda} + \hat I^{H,\infty}_{\ell m}(\vec w)\totder{\delta t_\parallel}{\lambda} \bigg)    \nonumber \\
& + i \beta_{{\rm s}\parallel}(\vec w) \hat J^{H,\infty}_{\ell m}(\vec w) \, ,
\end{align}
and
\begin{equation}
\beta_{{\rm s}\parallel}(\vec w) = \omega^{\rm s}_{m \vec b}\osc t_{\rm g}(\vec w) +  \omega^{\rm g}_{m \vec b}\osc (\delta t_\parallel)(\vec w) - m \osc (\delta\phi_\parallel)(\vec w) \, ,
\end{equation}
with $ \omega^{\rm g}_ {m \vec b} \equiv \omega^{\rm g}_ {mnk0}$, and $ \omega^{\rm s}_ {m \vec b} \equiv  \omega^{\rm s}_ {mnk0}$. The differential $\delta I^{H,\infty}_{\ell m \parallel}(\vec w)$ is given by
\begin{align}
\delta I^{H,\infty}_{\ell m \parallel}(\vec w) &= I^{H,\infty}_{\ell m \rm s\parallel}(\vec w) + \frac{\partial \hat I^{H,\infty}_{\ell m}}{\partial r_{\rm g}} \delta r_\parallel + \frac{\partial \hat I^{H,\infty}_{\ell m}}{\partial z_{\rm g}} \delta z_\parallel \nonumber \\
& +  \omega^{\rm s}_{m \vec b}\left. \frac{\partial \hat I^{H,\infty}_{\ell m}}{\partial z_{\rm g}} \right \rvert_{\omega =\omega^{\rm g}_{m \vec b}}  \, , \label{eq:deltaI_source_term_parallel}
\end{align}
while $\delta r_\parallel$ and $\delta z_\parallel$ are the radial and polar shifts in the trajectories~\eqref{eq:spin_radial_trajectory} and~\eqref{eq:spin_polar_trajectory}, respectively, which are proportional to $s_\parallel$. The derivative with respect to $\omega$ in Eq.~\eqref{eq:deltaI_source_term_parallel} accounts for the spin corrections due to the explicit dependence to $\omega$ in the function $I^{H,\infty}_{\ell m}$. The function $I^{H,\infty}_{\ell m \rm s\parallel}(\vec w)$ is defined as 
\begin{widetext}
\begin{align}
    I^{H,\infty}_{\ell m \rm s\parallel}(\vec w) &= \left. \frac{\Sigma}{W_{r \rm g}}\bigg ( A_{0{\rm s}\parallel} - \big( A_{1{\rm s}\parallel} + B_{1{\rm s}\parallel} \big)\totder{}{r} + \big( A_{2{\rm s}\parallel} + B_{2{\rm s}\parallel} \big) \totderhigh{}{r}{2} - B_{3{\rm s}\parallel} \totderhigh{}{r}{3} \bigg) \hat R^{\text{up,in}}_{\ell m \vec b} \right \rvert_{r = r_{\rm g}} \nonumber \\
   &+ \left. \frac{\Sigma}{W_{r \rm g}}\bigg ( A_{0\rm g} - A_{1\rm g} \totder{}{r} + A_{2\rm g} \totderhigh{}{r}{2} \bigg) \delta R^{\text{up,in}}_{\ell m \vec b\parallel} \right \rvert_{r = r_{\rm g}} - \frac{W_{r \rm s\parallel}}{W_{r \rm g}} \hat I^{H,\infty}_{\ell m}(\vec w) \, ,
\end{align}
\end{widetext}
where we defined $\hat R^{\text{up,in}}_{\ell m \vec b}(r) \equiv \hat R^{\text{up,in}}_{\ell mnk0}(r)$, $\delta R^{\text{up,in}}_{\ell m \vec b\parallel}(r) \equiv \delta R^{\text{up,in}}_{\ell mnk0\parallel}(r)$ and
\begin{align}
    W_{r \rm s\parallel} &\equiv\frac{1}{\Delta}\bigg(\delta R^{\text{in}}_{\ell m \vec b\parallel}\totder{\hat R^{\text{up}}_{\ell m \vec b}}{r} - \delta R^{\text{up}}_{\ell m \vec b\parallel}\totder{\hat R^{\text{in}}_{\ell m \vec b}}{r} \bigg)  \nonumber \\
    & + \frac{1}{\Delta}\bigg(\hat R^{\text{in}}_{\ell m \vec b}\totder{\delta R^{\text{up}}_{\ell m \vec b \parallel}}{r} - \hat R^{\text{up}}_{\ell m \vec b}\totder{\delta R^{\text{in}}_{\ell m \vec b\parallel}}{r} \bigg) 
\end{align}
The coefficients $A_{0{\rm s}\parallel}$, $A_{1{\rm s}\parallel}$, $A_{2{\rm s}\parallel}$ and $B_{1{\rm s}\parallel}$, $B_{2{\rm s}\parallel}$, $B_{3{\rm s}\parallel}$ are quite lengthy functions of $r_{\rm g}$ and $z_{\rm g}$, which can found in Appendix~\ref{app:source_term}. These terms were derived in Ref.~\cite{Piovano:2020zin} in the Carter tetrad frame and Ref.~\cite{Skoupy:2021asz} in a Boyer-Linquist base (see also \cite{Skoupy:2022adh}). Finally, we stress that the spin-corrections to the partial amplitudes~\eqref{eq:partial_amplitude_corr_parallel} do not specify a fiducial geodesic and are thus valid for any parametrization. After fixing a fiducial geodesics and computing the shifts to the constants of motion, it is then straightforward to compute the orbital corrections using the general expressions of Sec.\ref{sec:solutionsEoM}, which in turns can be directly applied to compute~\eqref{eq:partial_amplitude_corr_parallel}. 

\subsubsection{Asymptotic amplitudes proportional to $s_\perp$}
We now consider the corrections to the asymptotic amplitudes due to the orthogonal component of spin $s_\perp$, i.e. for all the corrections with either $j = +1$ or $j = -1$, which are defined as
\begin{equation}
\delta Z^{H,\infty}_{\ell m \vec \kappa \perp} =  \frac{1}{(2\pi)^3 \Upsilon_{t\rm g}} \int_{(0,2\pi]^3} \dd^3 \mathrm{w} \delta J^{H,\infty}_{\ell m \perp}(\wmean) e^{i\vec\kappa \cdot \wmean} \, . \label{eq:Zsperp}
\end{equation}
The function $\delta J^{H,\infty}_{\ell m \perp}(\wmean)$ is given by
\begin{align}
\delta J^{H,\infty}_{\ell m \perp}(\wmean) &=  e^{i\beta_{\rm g}(\vec w)} \bigg( \delta I^{H,\infty}_{\ell m \perp}(\wmean)\totder{t_{\rm g}}{\lambda} + \hat I^{H,\infty}_{\ell m}(\vec w)\totder{\delta t_\perp}{\lambda} \bigg)    \nonumber \\
& + i \beta_{{\rm s} \perp}(\wmean) \hat J^{H,\infty}_{\ell m}(\vec w) \, ,
\end{align}
with $\beta_{{\rm s} \perp}(\wmean) = \omega^{\rm g}_{m \vec\kappa}\osc (\delta t_\perp)(\wmean) - m \osc (\delta\phi_\perp)(\wmean)$.
The differential $\delta I^{H,\infty}_{\ell m \perp}(\wmean)$ is given by
\begin{align}
\delta I^{H,\infty}_{\ell m \perp}(\wmean) &= I^{H,\infty}_{\ell m s \perp}(\wmean) + \frac{\partial \hat I^{H,\infty}_{\ell m}}{\partial r_{\rm g}} \delta r_\perp + \frac{\partial \hat I^{H,\infty}_{\ell m}}{\partial z_{\rm g}} \delta z_\perp\, , 
\end{align}
while $\delta r_\perp$ and $\delta z_\perp$ are the radial and polar shifts in the trajectories~\eqref{eq:spin_radial_trajectory} and~\eqref{eq:spin_polar_trajectory}, respectively, which are proportional to $s_\perp$. The function $I^{H,\infty}_{\ell m \rm s \perp}(\wmean)$ is defined as 
\begin{widetext}
\begin{align}
    I^{H,\infty}_{\ell m \rm s \perp}(\wmean) &= \left. \frac{\Sigma}{W_{r \rm g}}\bigg ( A_{0{\rm s}\perp} - \big( A_{1{\rm s}\perp} + B_{1{\rm s}\perp} \big)\totder{}{r} + \big( A_{2{\rm s}\perp} + B_{2{\rm s}\perp} \big) \totderhigh{}{r}{2} - B_{3{\rm s}\perp} \totderhigh{}{r}{3} \bigg) \hat R^{\text{up,in}}_{\ell m \vec b} \right \rvert_{r = r_{\rm g}}  \, ,
\end{align}
\end{widetext}
The coefficients $A_{0{\rm s}\perp}$, $A_{1{\rm s}\perp}$, $A_{2{\rm s} \perp}$ and $B_{1{\rm s} \perp}$, $B_{2{\rm s} \perp}$, $B_{3{\rm s} \perp}$ are quite lengthy functions of $r_{\rm g}$, $z_{\rm g}$ and $\psi_{\rm p}$. Their explicit expressions as functions are presented in Appendix~\ref{app:source_term}.

All terms in the function $\delta J^{H,\infty}_{\ell m \perp}(\wmean)$ are proportional to either $\cos \psi_{\rm p}$ or $\sin \psi_{\rm p}$. Moreover, since $\psi_{\rm p}(\wmean) = w_{\rm p} + \psi_{{\rm p}r}(w_r)+ \psi_{{\rm p}z}(w_z)$
, we expand $\sin(\psi_{\rm p}(\wmean))$ and $\cos(\psi_{\rm p}(\wmean))$ 
and collect all terms in $\delta J^{H,\infty}_{\ell m \perp}(\wmean)$ proportional to either $\cos w_{\rm p}$ or $\sin w_{\rm p}$, obtaining
\begin{equation}
 \delta J^{H,\infty}_{\ell m \perp}(\wmean) = \cos(w_{\rm p}) \delta J^{H,\infty}_{\ell m \perp,\mathsf{cn}}(\vec w) + \sin(w_{\rm p}) \delta J^{H,\infty}_{\ell m \perp,\mathsf{sn}}(\vec w) \, .
\end{equation}
Finally, by plugging the previous expression in Eq.~\eqref{eq:Zsperp}, we get 
\begin{align}
\delta Z^{H,\infty}_{\ell m \vec b1 \perp} &=  \frac{1}{2(2\pi)^2 \Upsilon_{t\rm g}} \int_{(0,2\pi]^2} \dd^2 w \delta J^{H,\infty}_{\ell m \perp,\mathsf{cn}}(\vec w) e^{-i \vec b \cdot \vec w}  \nonumber \\
     &- \frac{i}{2(2\pi)^2 \Upsilon_{t\rm g}} \int_{(0,2\pi]^2} \dd^2 w \delta J^{H,\infty}_{\ell m \perp,\mathsf{sn}}(\vec w) e^{-i \vec b \cdot \vec w} \, .
\end{align}
\begin{align}
\delta Z^{H,\infty}_{\ell m \vec b-1 \perp} &=  \frac{1}{2(2\pi)^2 \Upsilon_{t\rm g}} \int_{(0,2\pi]^2} \dd^2 w \delta J^{H,\infty}_{\ell m \perp,\mathsf{cn}}(\vec w) e^{-i \vec b \cdot \vec w}  \nonumber \\
     &+ \frac{i}{2(2\pi)^2 \Upsilon_{t\rm g}} \int_{(0,2\pi]^2} \dd^2 w \delta J^{H,\infty}_{\ell m \perp,\mathsf{sn}}(\vec w) e^{-i \vec b \cdot \vec w} \, .
\end{align}

We stress that $\delta Z^{H,\infty}_{\ell m \vec \kappa \perp}$ is the same in any parametrization, since it depends only on the spin corrections to the trajectories and velocities proportional to $s_\perp$. Moreover, notice that

\begin{align}
\delta\mathcal A_{\ell m \vec k \perp} &=  \frac{\delta Z^\infty_{\ell m \vec \kappa}}{(\omega^{\rm g}_ {m \vec\kappa})^2} S^{\rm g}_ {\ell m \vec\kappa \perp}(\vartheta)   \, , 
\end{align}

\subsection{GW fluxes}

 The averaged energy and angular momentum fluxes can be derived as
 \begin{align}
\langle\mathcal F^E \rangle&=\langle\mathcal F_{\rm g}^E\rangle +q\, \langle \delta \mathcal F^E\rangle\,,\\
\langle\mathcal F^{J_z}\rangle &=\langle\mathcal F_{\rm g}^{J_z}\rangle +q\,\langle \delta \mathcal F^{J_z}\rangle\,,
\end{align}
where
\begin{align}
\langle \mathcal F_{\rm g}^E\rangle &= \displaystyle \sum_{\ell m \vec\kappa} \hat{\mathcal F}^{E}_{\ell m \vec\kappa}\,, \quad 
\langle\mathcal F_{\rm g}^{J_z}\rangle= \displaystyle \sum_{\ell m \vec\kappa} \hat{\mathcal F}^{J_z}_{\ell m \vec\kappa}  \,, \\
\langle \delta \mathcal F^E \rangle &= \displaystyle \sum_{\ell m \vec\kappa} \delta \mathcal F^{E}_{\ell m \vec\kappa} \,, \quad 
\langle \delta \mathcal F^{J_z}\rangle= \displaystyle \sum_{\ell m \vec\kappa} \delta \mathcal F^{J_z}_{\ell m \vec\kappa}  \,,
\end{align}
with 
\begin{align}
\hat{\mathcal F}^{E}_{\ell m \vec\kappa}&= \frac{|\hat Z^\infty_{\ell m \vec\kappa}|^2+\alpha_{\ell m \vec\kappa}\,|\hat Z^H_{\ell m \vec\kappa} |^2}{4 \pi (\omega^{\rm g}_ {m \vec\kappa})^2}\,, \\
\hat{\mathcal F}^{J_z}_{\ell m \vec\kappa}&=\frac{m(|\hat Z^\infty_{\ell m \vec\kappa}|^2+\alpha_{\ell m \vec\kappa}\,|\hat Z^H_{\ell m \vec\kappa} |^2)}{4 \pi (\omega^{\rm g}_ {m \vec\kappa})^3}\,,\\
\delta \mathcal F^{E}_{\ell m \vec\kappa}&= \frac{|\hat Z^\infty_{\ell m \vec\kappa}||\delta Z^\infty_{\ell m \vec\kappa}|+\alpha_{\ell m \vec\kappa}\,|\hat Z^H_{\ell m \vec\kappa} ||\delta Z^H_{\ell m \vec\kappa} |}{2 \pi (\omega^{\rm g}_ {m \vec\kappa})^2} \nonumber\\
&\phantom{=}+\frac{\delta\alpha_{\ell m \vec\kappa}\,|\hat Z^H_{\ell m \vec\kappa} |^2}{4 \pi (\omega^{\rm g}_ {m \vec\kappa})^2} -2\frac{\omega^{\rm s}_ {m \vec\kappa}}{\omega^{\rm g}_ {m \vec\kappa}}\hat{\mathcal F}^{E}_{\ell m \vec\kappa} \,, \\
\delta{\mathcal F}^{J_z}_{\ell m \vec\kappa}&= \frac{m(|\hat Z^\infty_{\ell m \vec\kappa}||\delta Z^\infty_{\ell m \vec\kappa}|+\alpha_{\ell m \vec\kappa}\,|\hat Z^H_{\ell m \vec\kappa} ||\delta Z^H_{\ell m \vec\kappa} |)}{2 \pi (\omega^{\rm g}_ {m \vec\kappa})^3}\nonumber\\
&\phantom{=}+\frac{m\,\delta\alpha_{\ell m \vec\kappa}\,|\hat Z^H_{\ell m \vec\kappa} |^2}{4 \pi (\omega^{\rm g}_ {m \vec\kappa})^3}-3\frac{\omega^{\rm s}_ {m \vec\kappa}}{\omega^{\rm g}_ {m \vec\kappa}}\hat{\mathcal F}^{J_z}_{\ell m \vec\kappa}\,.
\end{align}
and
\begin{align} 
\alpha_{\ell m \vec\kappa}&=\frac{256(2 r_+)^5 \kappa_{\mathcal H}(\kappa_{\mathcal H}^2+4 \epsilon^2)(\kappa_{\mathcal H}^2+16 \epsilon^2)(\omega^{\rm g}_ {m \vec\kappa})^3}{|\mathcal{C}_{\ell m\omega}|^2}\,,\\
\delta\alpha_{\ell m \vec\kappa}&=3\,\frac{\omega^{\rm s}_ {m \vec\kappa}}{\omega^{\rm g}_ {m \vec\kappa}} \,\alpha_{\ell m \vec\kappa}-\frac{\delta(|\mathcal{C}_{\ell m\omega}|^2)}{ |\mathcal{C}_{\ell m\omega}|^2}\alpha_{\ell m \vec\kappa}\,,\\
\epsilon&=\sqrt{1-a^2}/(4 r_+)\,.
\end{align}
The Teukolsky-Starobinsky constant is given by
\begin{align}\label{eq:TSconstant}
    |\mathcal{C}_{\ell m\omega}|^2&= \big((\lambda_{\ell m\omega}+2)^2+4 a \omega_{m \vec\kappa} (m-a \omega_{m \vec\kappa})\big)\nonumber\\
    &\times\big(\lambda_{\ell m\omega}^2+36 a \omega_{m \vec\kappa}(m-a \omega_{m \vec\kappa})\big)\nonumber\\
    &-(2 \lambda_{\ell m\omega}+3)\big(48a \omega_{m \vec\kappa}(m-2a\omega_{m \vec\kappa})\big)\nonumber\\
    &+144(\omega_{m \vec\kappa})^2(1-a^2)\,,
\end{align}
and its shift $\delta(|\mathcal{C}_{\ell m\omega}|^2)$ is obtained by linearising Eq.~\eqref{eq:TSconstant}.

Three comments are now in order. First, the geodesic partial amplitudes, $\hat Z^\infty_{\ell m \vec\kappa}$, vanish for $j=\pm1$.
Second, all terms in the shifts of the partial amplitudes proportional to the perpendicular part of the spin are oscillatory. Consequently, the only contribution from $s_\perp$  comes from the modes $j=\pm1$. At the level of the fluxes, these components lead to quadratic pieces in $s_\perp$, and we can thus ignore them at leading order in $s$.

The third comment is a remark on the linear momentum flux. In principle, the spin of the primary black hole defines a characteristic direction (the z-direction) into which an overall linear momentum flux $\mathcal{F}^{P_z}$ could flow. However, as shown e.g. in Ref. \cite{Ruiz:2007yx}, such a flux can only arise if the energy flux $\mathcal{F}^E$ through infinity and the horizon violates equatorial-reflection symmetry or, in other words, if it has non-zero content in the $l+m$-odd harmonics. However, formulas for bound Kerr geodesic manifestly obey the equatorial symmetry, so the $\hat{\mathcal{F}}_{lm}^{R}$ for odd $l+m$ are zero. Is this also true for spinning test particles? As discussed in Refs \cite{Witzany:2019nml,Drummond:2022efc,Drummond:2023wqc}, the violation of manifest equatorial symmetry of the spinning-particle orbits appears only in terms proportional to the fully oscillating orthogonal secondary spin $s_\perp$, which averages out at leading order. As such, we also obtain zero contributions for $\delta\mathcal{F}^E_{lm\vec{\kappa}}$ when $l+m$ is odd, and thus also zero linear momentum flux. The only caveat of this discussion are orbital resonances, for which linear momentum fluxes appear even for geodesics \cite{Hirata:2010xn,vandeMeent:2014raa}. We leave the investigation of resonant orbits for future work.   

\section{Numerical results}\label{sec:Numerics}
In this section show some of the numerical results obtained using the above implementation. To ease comparisons with the literature, we introduce a secondary spin alignment angle $\varphi_{\mathrm{s}}$ such that
\begin{equation}
    s_{\perp}=\cos(\varphi_{\mathrm{s}}),\quad s_{||}=\sin(\varphi_{\mathrm{s}}) \ ,
\end{equation}
meaning that $\varphi_{\mathrm{s}}=90^{\circ}$ corresponds to a secondary spin fully aligned to the primary spin and $\varphi_{\mathrm{s}}=0^{\circ}$ to a purely perpendicular secondary spin. Unless otherwise stated, all the figures that follow have orbital parameters $\{a,p_{\rm g},e_{\rm g},x_{\rm g}\}=\{0.9,4,0.3,1/\sqrt{2}\}$, mass ratio $q=1/20$\footnote{This choice of mass ratio is possibly outside of the bounds of the applicability of the 1PA approximation. We choose such a large value here to make the effects of spin clearly visible on the plots. Nevertheless, it is important to notice that perturbation theory results has shown remarkable agreement with numerical relativity simulations even for large $q$~\cite{Wardell:2021fyy}). } , and $\bar n_{\text{max}}=\bar k_{\text{max}} =12$ where $\bar n_{\text{max}}$ ($\bar k_{\text{max}}$) is the maximum number of radial (polar) Fourier harmonics in the orbits. We checked that 12 modes in the orbital Fourier series are sufficient to ensure numerical convergence in all the results presented here. In general, the optimal $\bar n_{\text{max}}$ and $\bar k_{\text{max}}$ needed to achieve a given accuracy depends on orbital elements. In other words, high inclined and/or high eccentric orbits require a larger number of orbital Fourier series to achieve numerical convergence compare to orbits with smaller eccentricity or inclination. 

\begin{figure*}[!htb]
	\centering
     \includegraphics[width=0.88\textwidth]{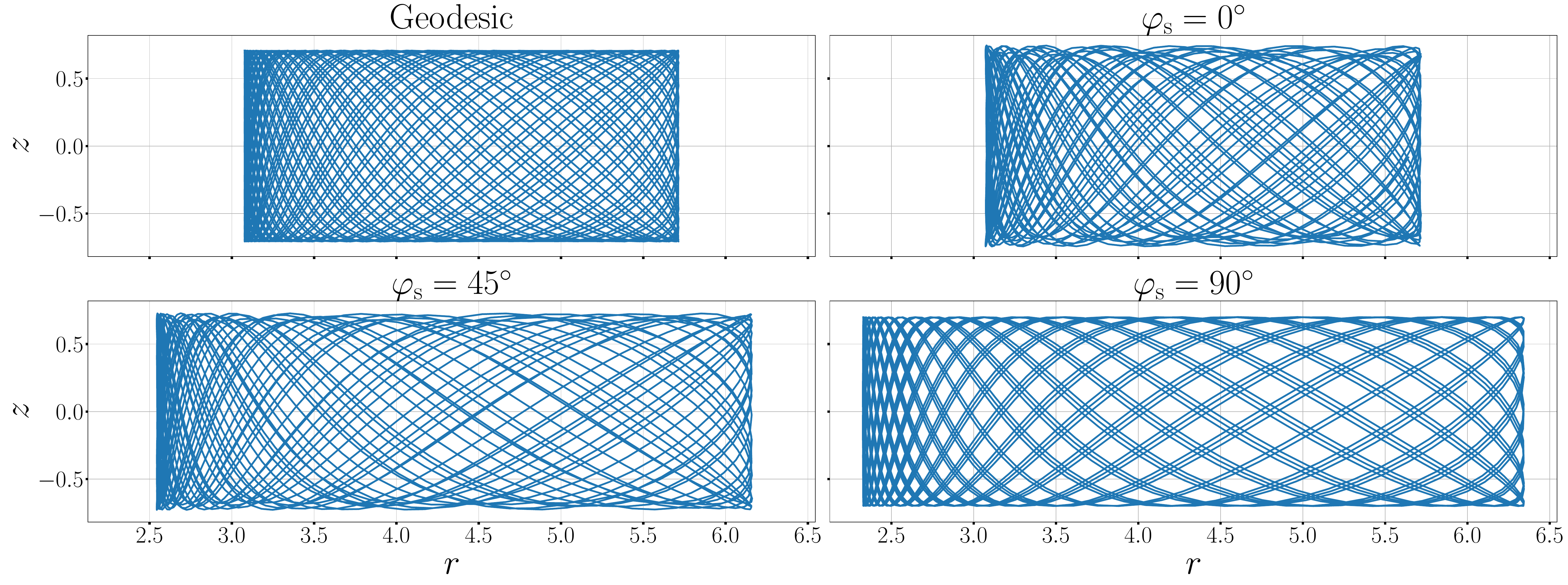}
    	\caption{Plots of trajectories in $z$-$r$ space for increasing spin alignment angle $\varphi_{\mathrm{s}}$ in the ``FC'' parametrization. The top left plot shows geodesic trajectories for comparison. At $\varphi_{\rm s} =0$ the spin is fully misaligned and oscillating with an independent frequency that externally drives the $r,z$ librations. For $\varphi_{\rm s} = 90^\circ$, the  spin is fully aligned and the dynamics are fully determined by the $r,z$ frequencies. Additionally, the radial libration increases while the polar libration decreases as the spin becomes more aligned.}
	\label{fig:z_r_phis_FC}
\end{figure*}

\begin{figure*}[!htb]
	\centering
     \includegraphics[width=0.88\textwidth]{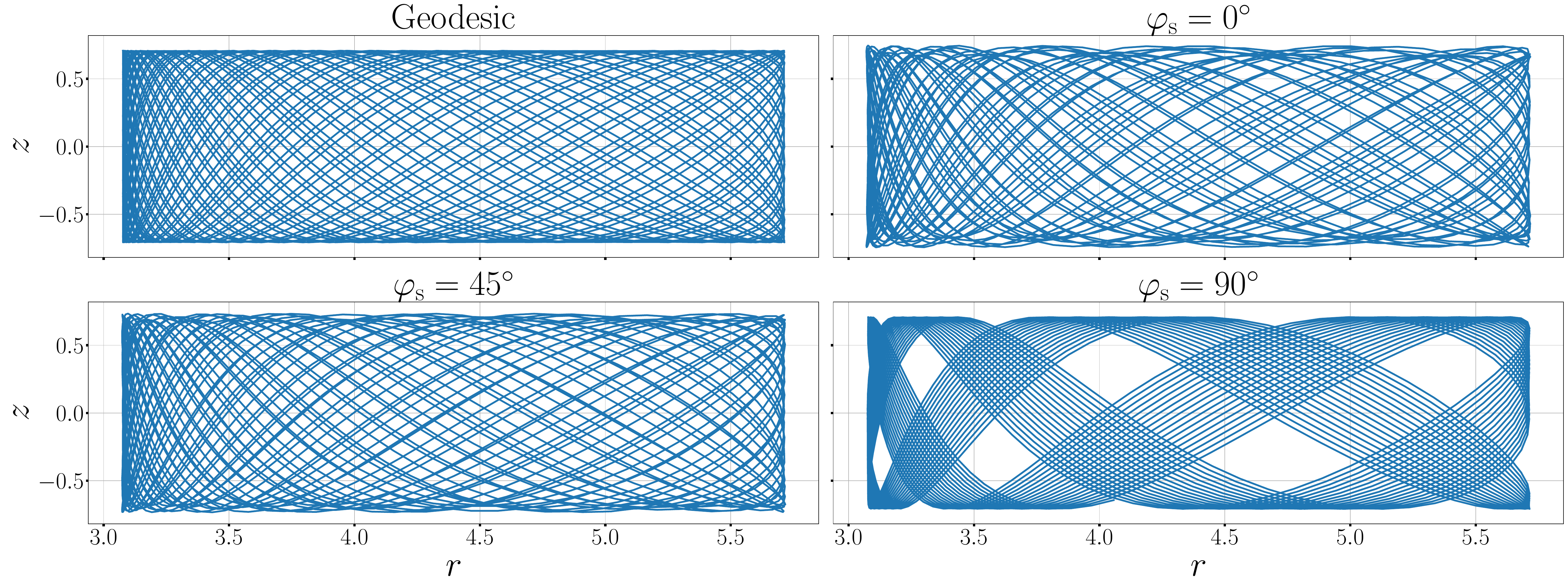}
    	\caption{Same as Fig.\ref{fig:z_r_phis_FC}, but for the ``DH'' parametrization. In this case, both radial and polar libration have oscillations that average to zero by definition.}
	\label{fig:z_r_phis_DH}
\end{figure*}

In Fig.~\ref{fig:z_r_phis_FC} and  Fig.~\ref{fig:z_r_phis_DH} we plot the orbit on the $z$-$r$ plane for 3 values of $\varphi_{\mathrm{s}}=0^{\circ}, 45^\circ, 90^\circ$ in the ``FC" and ``DH" parametrizations, respectively, and compare them to the geodesic case. For $\varphi_{\mathrm{s}}=0^{\circ}$  the spin is fully misaligned and oscillating with an independent frequency that externally drives the $r,z$ librations. On the other hand, for $\varphi_{\rm s} = 90^\circ$, the spin is fully aligned, the independent oscillation of the spin is absent, and the dynamics are fully determined by the $r,z$ frequencies. Note that in the ``FC" parametrization the radial libration increases with $\varphi_{\mathrm{s}}$, while the polar libration decreases. In the ``DH" parametrization, both radial and polar libration have oscillations that average to zero by definition.

\begin{figure}[!htb]
	\centering
     \includegraphics[width=1\linewidth]{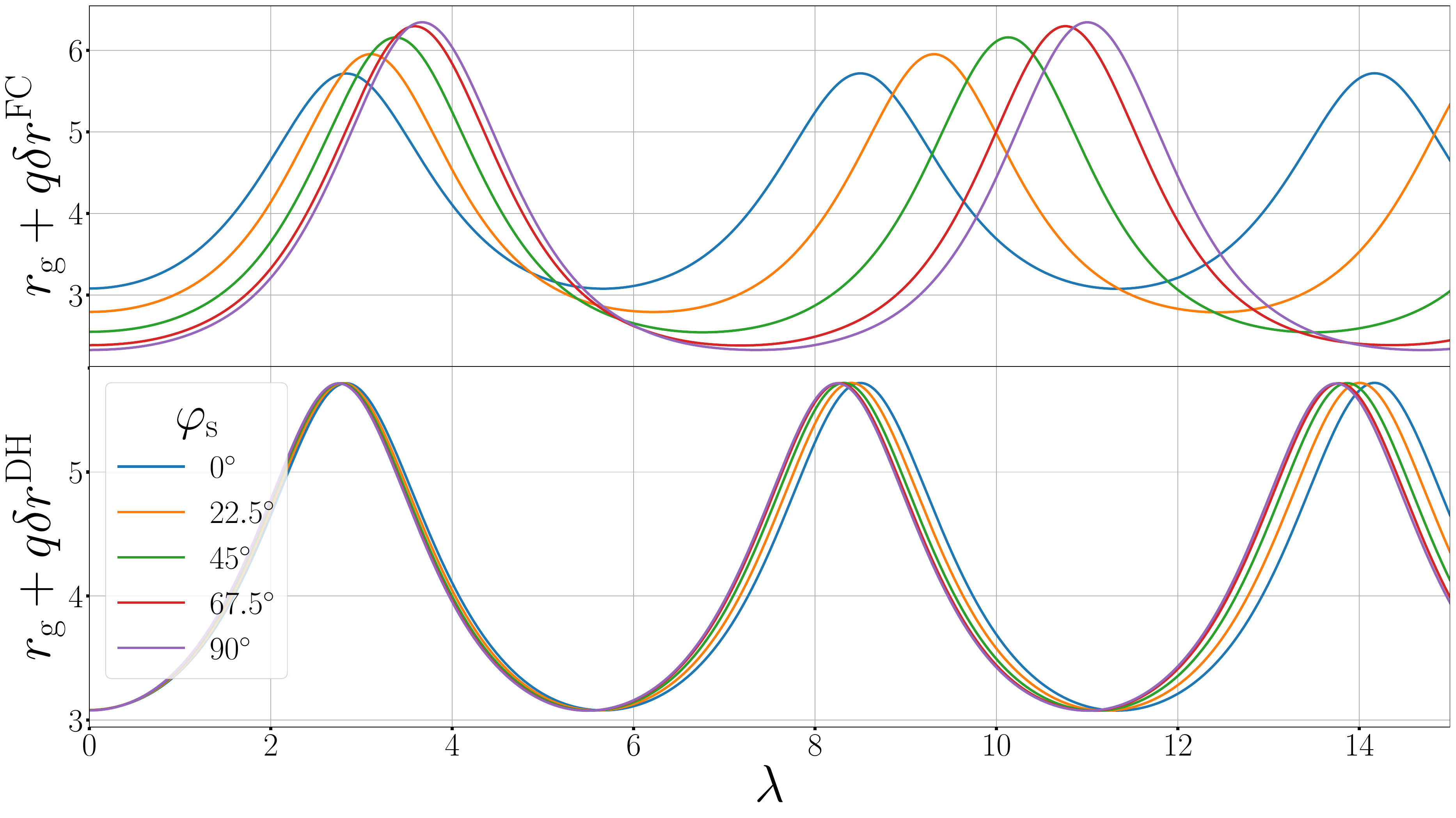}
    	\caption{Top panel: the evolution of $r_{\mathrm{g}}+q \delta r^{\text{FC}}$ for different values of $\varphi_{\mathrm{s}}$ and fixed secondary spin in the ``FC'' scheme. Bottom panel: same plot as above but in the ``DH'' scheme.}
	\label{fig:r_phis_FC}
\end{figure}

In Fig.~\ref{fig:r_phis_FC} we present the evolution of $r_{\rm g}+q\delta r$ in terms of Mino-time for increasing $\varphi_{\rm s}$  in the two parametrizations. We see that the dephasing increases as the spin become more aligned as only the aligned component contributes to the long-term evolution of the anomalies; this effect is significantly more pronounced in the ``FC" scheme. 

\begin{figure}[!htb]
	\centering
     \includegraphics[scale=0.20]{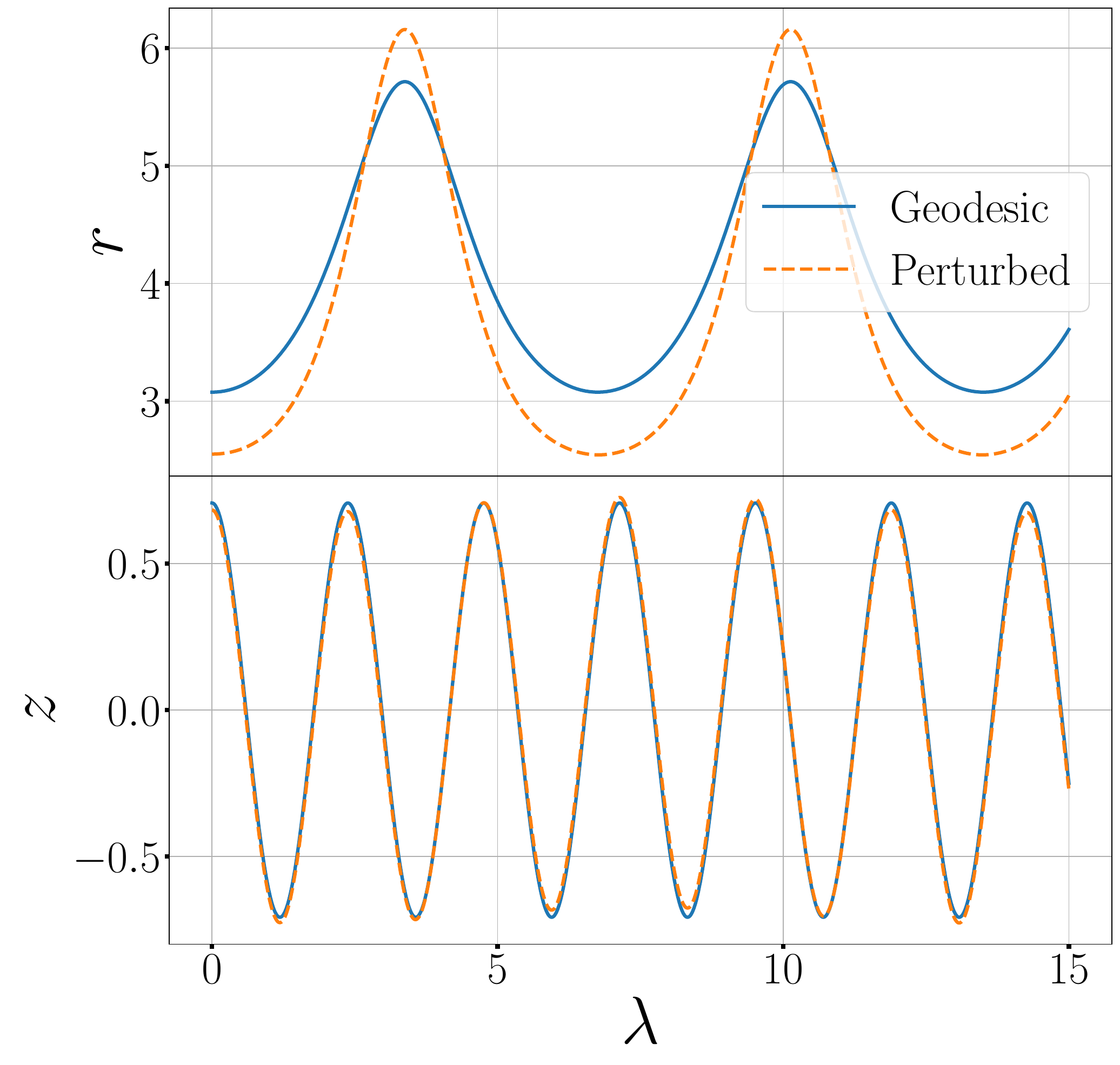}
    	\caption{Comparison of the radial and polar trajectories for a perturbed and geodesic orbit with $\varphi_{\mathrm{s}}=45^{\circ}$ and in the ``FC" scheme. We evaluate the geodesic quantities using the perturbed mean-anomalies to remove dephasing.}
	\label{fig:r_z_plots_FC}
\end{figure}

\begin{figure}[!htb]
	\centering
     \includegraphics[scale=0.20]{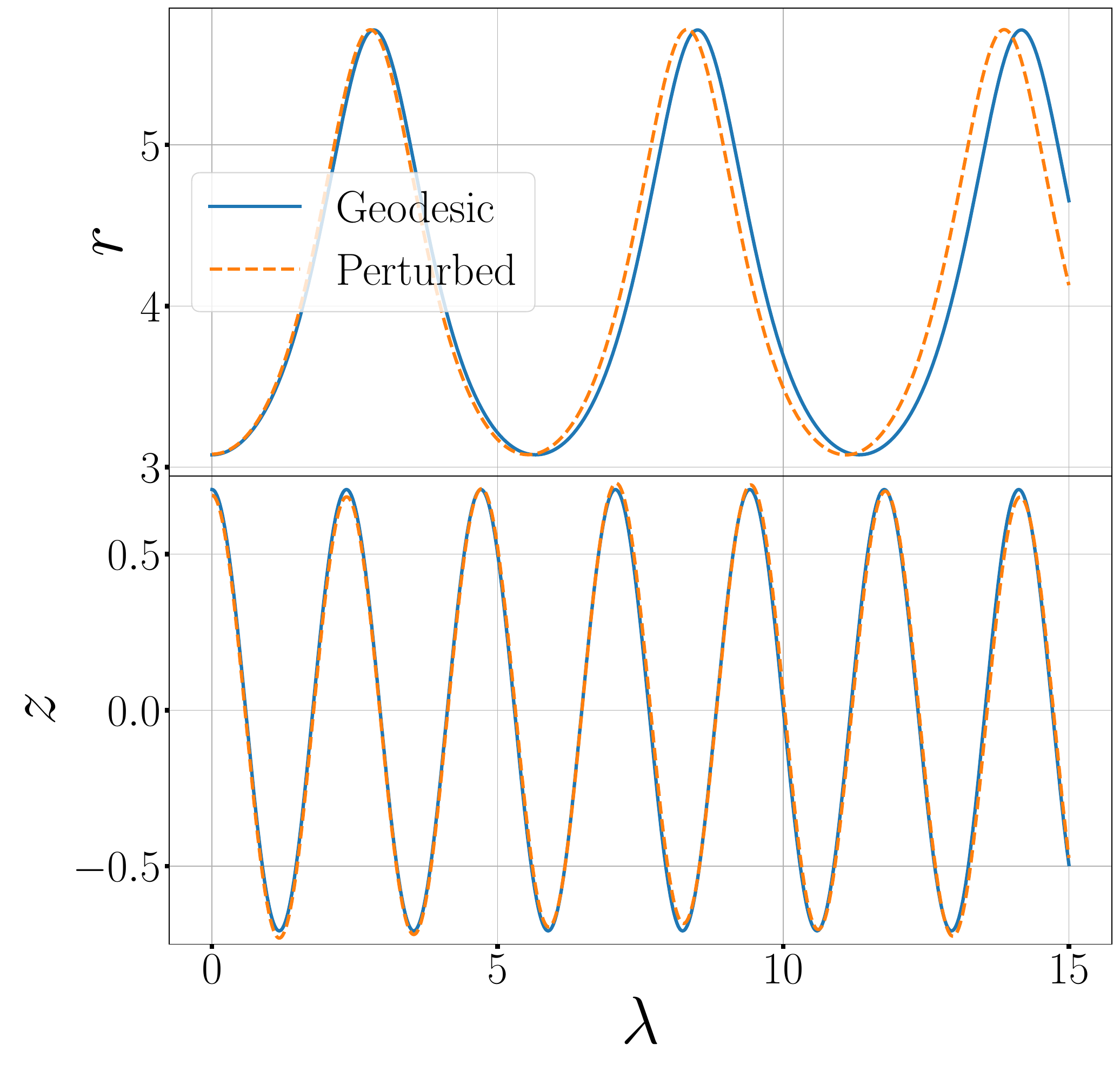}
    	\caption{Comparison of the radial and polar trajectories for a perturbed and geodesic orbit with $\varphi_{\mathrm{s}}=45^{\circ}$ using the fixed  ``DH" parametrization. We evaluate the geodesic quantities using the geodesic mean-anomalies as is done in Fig.~3 in Ref.~\cite{Drummond:2022xej}.}
	\label{fig:r_z_plots_DH}
\end{figure}

Finally, in Fig.~\ref{fig:r_z_plots_FC} and Fig.~\ref{fig:r_z_plots_DH}, we plot the radial and polar trajectories for a perturbed and geodesic orbit in the ``DH'' and ``FC'' parametrizations, respectively. Note that in Fig.~\ref{fig:r_z_plots_DH} we evaluated the geodesic quantities using the perturbed mean-anomalies in order to remove dephasing with the perturbed trajectories, while in the ``FC'' case we used the geodesic mean-anomalies to facilitate comparison with Fig.~3 in Ref.~\cite{Drummond:2022xej}. 

\begin{figure*}[!htb]
	\centering
    \includegraphics[width=0.86\linewidth]{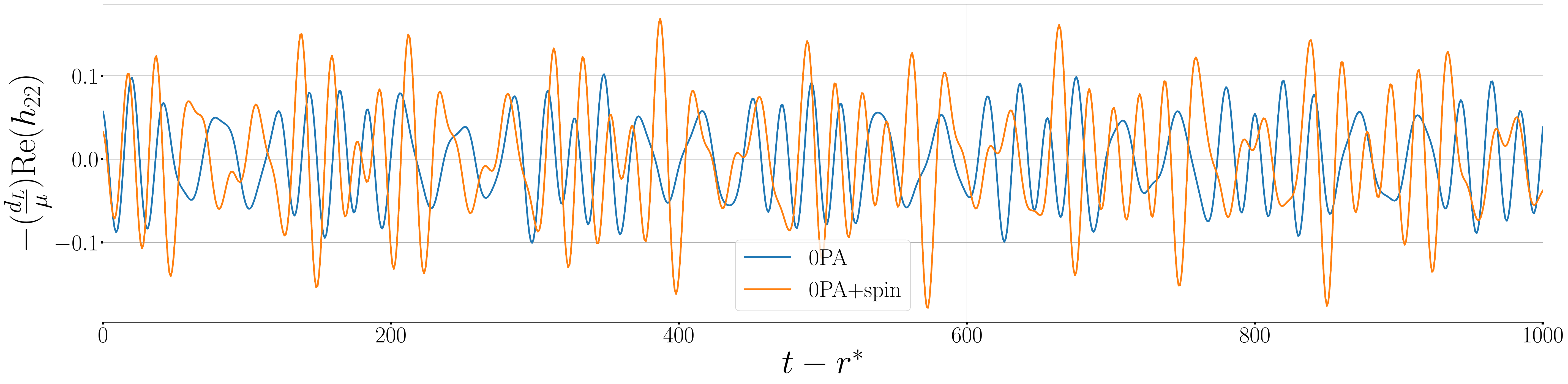}
   	\caption{Comparison between waveform snapshots at 0PA (blue curve) and 0PA with spin-corrections (orange curve). The spin-perturbed orbit were computed in the ``FC'' parametrization, with orbital elements $\{a,p_g, e_g,x_g\}=\{0.9,4,0.3,1/\sqrt{2}\}$. We also set $q=1/20$, $\varphi_s = 45^{\circ}$, and $\{\vartheta,\varphi\}=\{ \pi/2,0 \}$.} 
	\label{fig:waveform_FC}
\end{figure*}

\begin{figure*}[!htb]
	\centering
    \includegraphics[width=0.86\linewidth]{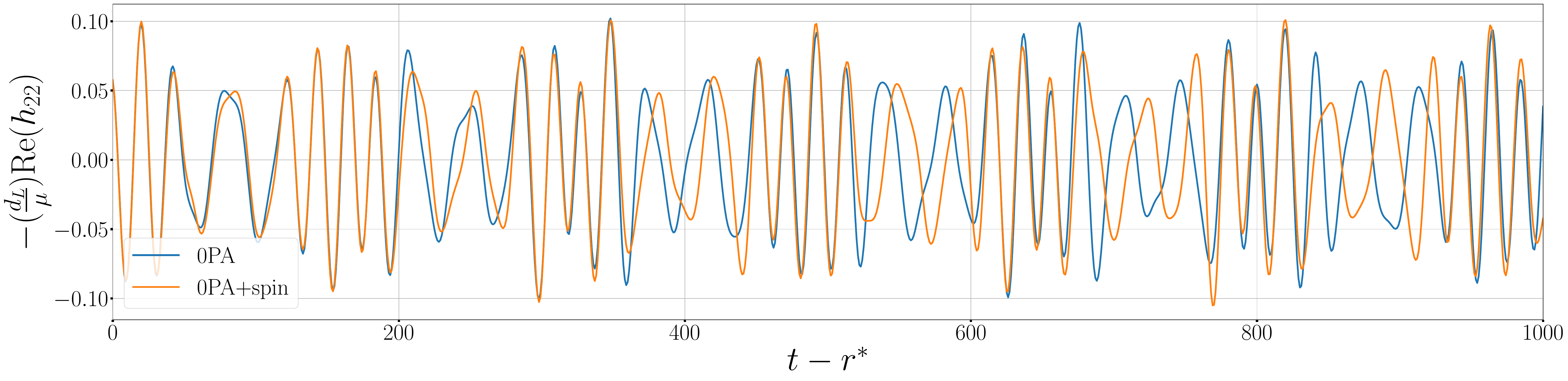}
   	\caption{Same as Fig.~\ref{fig:waveform_DH} but in the ``DH'' parametrization.} 
	\label{fig:waveform_DH}
\end{figure*}

In Fig.~\ref{fig:waveform_FC} and Fig.~\ref{fig:waveform_DH} we present a waveform snapshot given by Eq.~\eqref{eq:partial_amp_expand} for a spin-perturbed orbit in the fixed turning points and fixed constants of motion schemes, respectively. We set $\varphi_s = 45^{\circ}$ and source direction to $\{\vartheta,\varphi\}=\{ \pi/2,0 \}$, while the orbital elements are the same of the previous plots. Here we included contributions from all modes with $l=2, m = 2, n \in [-3,3], k \in [-3,3] $. We emphasize that we have not carried out a detailed analysis of how many modes to include in the snapshot waveforms \textemdash this range was chosen only for illustrative purpose to include radial and polar harmonics. 

Tables with numerical values for the correction to the frequencies, for the partial amplitudes shifts, $\delta Z^{H,\infty}_{\ell m \vec \kappa}$, and for the shifts of the energy and angular momentum fluxes, $\delta \mathcal{F}^{E}_m,\delta\mathcal{F}^{J_z}_m$, are presented in Appendix~\ref{app:data} for future reference.

\subsection{Comparison with the work of Skoupý et al.}

The shifts in the real and imaginary part of the partial amplitudes due to the parallel component of the spin  were studied by Skoupý et al.~\cite{Skoupy:2023lih}. Specifically, in Sec. IV of~\cite{Skoupy:2023lih} one can find numerical results for the partial amplitudes for given $\{l, m, n, k\}$ of a generic orbit with $a = 0.9, p_{\rm g} = 12, e_{\rm g} = 1/5, x_{\rm g} = \sqrt{3}/2$ up to fifth significant digits. Our results in the ``DH" parametrization, shown in Table~\ref{tbl:amplParDH} in Appendix~\ref{app:data},  are in excellent quantitative agreement.

Our implementation improves on the previous works in several ways. In terms of the code architecture, it is flexible and modular. For example, it can tackle any parametrization (both for orbits and fluxes) with the most straightforward implementations for the ``FC" and ``DH" parametrizations. 

Additionally, unlike~\cite{Skoupy:2023lih} that relies on solving the coupled over-constrained system presented in Refs.~\cite{Drummond:2022xej,Drummond:2022efc} to obtain orbital quantities, our code can compute the shifts to the constants of motion, frequencies, trajectories and velocities independently of each other. As such, our computational costs scale linearly with the number of computed modes, and we obtain high precision results already when working at machine precision.  However, it is still necessary to use arbitrary precision in our formalism for the computation of the radial and polar solutions of Eqs.~\eqref{eq:radialTeueq} and~\eqref{eq:swsweq} and their spin corrections.

\begin{figure}[!htb]
	\centering
     \includegraphics[scale=0.126]{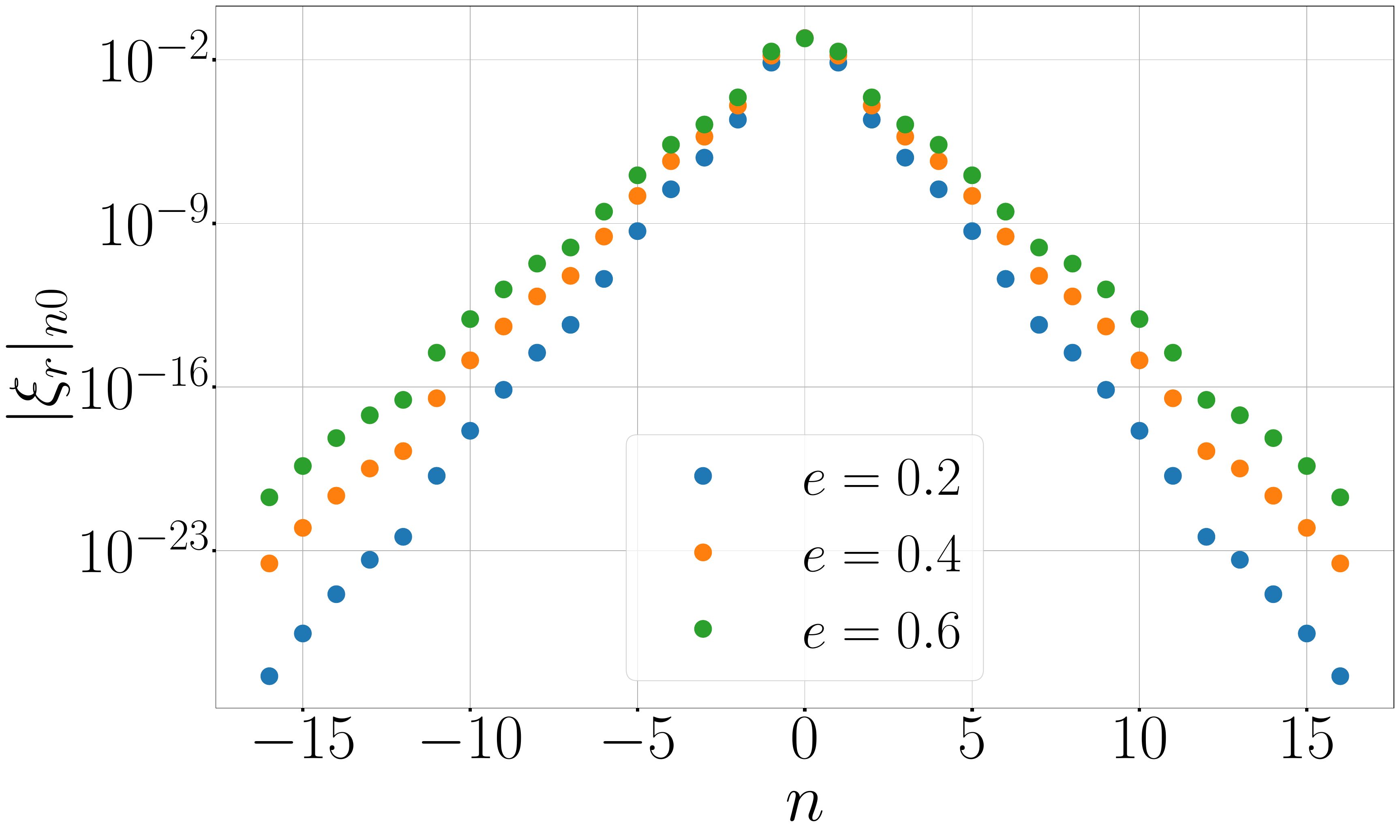}
    	\caption{Convergence of the radial Fourier modes of the anomaly perturbation $\xi_{r}$ for spinning particle orbits in the ``DH" scheme with $\{a,p_{\rm g},x_{\rm g}\}=\{0.9,9,1/2\}$ and different choices of eccentricity. Here we use 45 digits of precision, and $\bar n_{max}=\bar k_{max}=16$.}
	\label{fig:xir_DH_conv_n}
\end{figure}

\begin{figure}[!htb]
	\centering
     \includegraphics[scale=0.126]{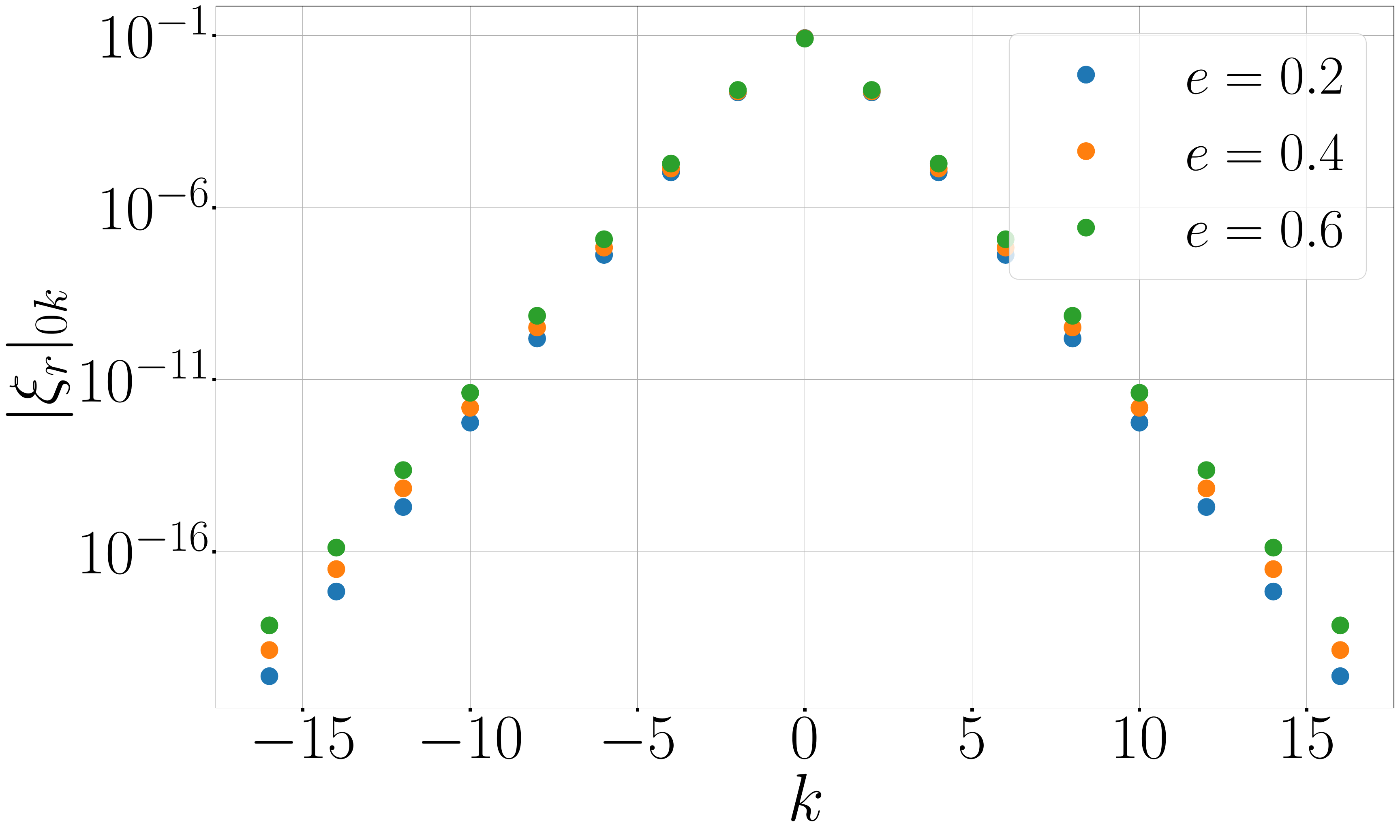}
    	\caption{Same plot as in Fig.~\ref{fig:xir_DH_conv_n} with identical orbital parameters, but for the polar Fourier modes. We note here that for odd $k$ modes the coefficient $(\xi_{r})_{nk}$ vanishes, and similar result hold for $\xi_z$.} 
	\label{fig:xir_DH_conv_k}
\end{figure}

Furthermore, for the computation of the trajectories, our algorithm ensures a convergent result even in situations with high inclination and high eccentricity. Fig.~\ref{fig:xir_DH_conv_n} shows the absolute value of radial Fourier modes of the anomaly perturbation, $(\xi_r)_{n0}$, as a function of $n$ for different values of the eccentricity, $e_{\rm g}$, and $\{a,p_{\rm g}, x_{\rm g}\}=\{0.9, 9, 1/2 \}$. These coefficients were computed in the ``DH" parametrization. We see that $(\xi_r)_{n0}$ monotonically decreases with $|n|$, which indicates convergence in all cases. Similarly, Fig.~\ref{fig:xir_DH_conv_k} shows the polar Fourier modes of the anomaly perturbation, $(\xi_r)_{0k}$, computed for the same orbital elements of~Fig.~\ref{fig:xir_DH_conv_n} and once again we see convergence. $\xi_z$ has a similar fall-off for the Fourier coefficients. This convergent behaviour also holds for the $\xi_r$ and $\xi_z$ coefficients in the ``FC" parametrization. This is to be contrasted with Fig. 1 of~\cite{Skoupy:2023lih} where the convergence stops at certain value of $\pm n$ that strongly depends on the eccentricity. Such a problem in their approach is likely due to numerical errors in the high-order terms of the Fourier series, which are slowly convergent for high-eccentricity.


For the computation of the partial amplitudes and the fluxes, our code uses the same algorithm as~\cite{Skoupy:2023lih}, with the key difference that it directly computes the linear spin corrections. This is to be contrasted with~\cite{Skoupy:2023lih} where the linear in spin part of the amplitudes and fluxes has to be extracted numerically by using fourth-order finite differencing of the value of the total fluxes and amplitudes at finite values of spin (see Eq. 66 of Ref.~\cite{Skoupy:2023lih}). As a result, our method of computing the spin correction to fluxes is inherently faster. 

The time needed to compute the spin-corrections to the orbits in our formalism strongly depends on the maximum number of radial $\bar n_{max}$ and polar $\bar k_{max}$ modes kept in the Fourier series, but it is largely independent of the precision of the input parameters or the chosen parametrization. Specifically, for $\bar n_{max} = \bar k_{max} = 12$, the computation of all orbital parameters (constants of motion and their shifts, Mino-frequencies and their shifts, and the orbits and their parallel component shifts) takes roughly 11 seconds, regardless of the value of eccentricity and/or inclination for the parallel sector. The computation of the corrections due to $s_\perp$ takes longer, around 28 seconds for 12 Fourier modes.
The time cost of the computation of the partial amplitudes and their correction depend on several factors: the precision used in input, the number of orbital Fourier modes $\bar n_{max}$ and $\bar k_{max}$, and, most importantly, the harmonic set $\{\ell, m, n , k\}$.  For instance, given an orbit with 12 Fourier modes, input precision of 45 digits and harmonics $\{\ell, m, n , k\} =\{2,1,2,1\}$, it takes 5 seconds to compute the leading order partial amplitudes and fluxes, and the relative corrections due to the parallel component of the spin. For the same parameters and input precision, the computation of $\delta Z^{H,\infty}_{\ell m \vec \kappa \perp}$ has a much longer runtime: around 80 seconds for $j=+1$ and 65 seconds $j=-1$. In general, the computation of orthogonal corrections $\delta Z^{H,\infty}_{\ell m \vec \kappa \perp}$ is much more expensive than the computation of its counterpart $\delta Z^{H,\infty}_{\ell m \vec b \parallel}$. We stress that our code has not been yet fully optimized, and there is plenty of room for improvements. All computations are performed on a single core of a 12th gen Intel Core i5-1235U $@ 1.30$GHz.

\begin{figure}[!htb]
	\centering
     \includegraphics[width=0.86\linewidth]{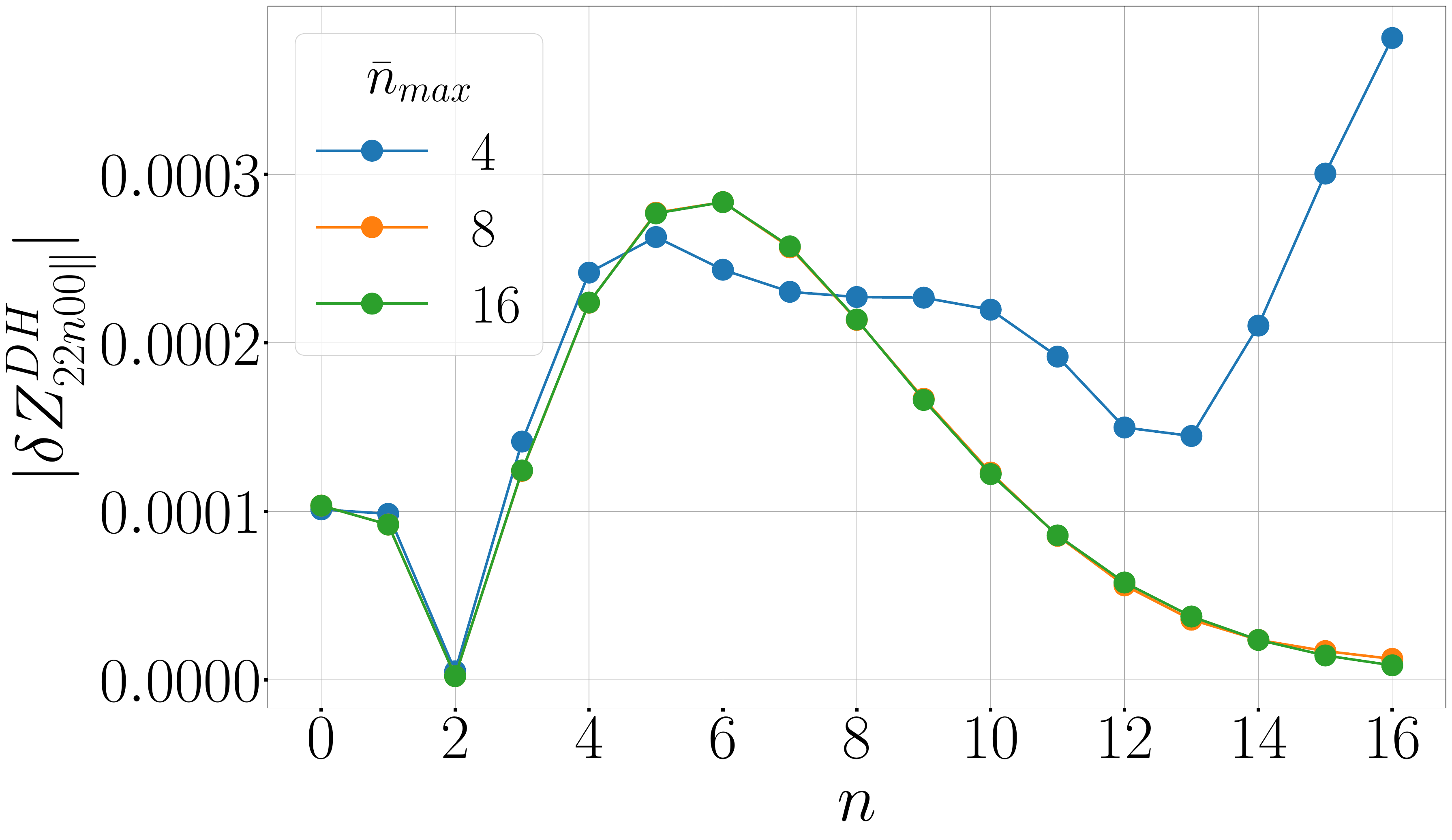}
    	\caption{Absolute value of the corrections to the partial amplitudes $|\delta Z^{\text{DH}}_{22n00\parallel}|$ for increasing value of the waveform harmonic $n$ and different orbital Fourier modes $\bar n_{\text{max}}$. As $n$ increases, more Fourier modes are needed in the computation of the orbits to achieve convergence. Orbital elements: $\{a,p_{\rm g},e_{\rm g},x_{\rm g}\}=\{0.9,9,1/2,1/2\}$. }
	\label{fig:conv_amp_n}
\end{figure}

\begin{figure}[!htb]
	\centering
     \includegraphics[width=0.86\linewidth]{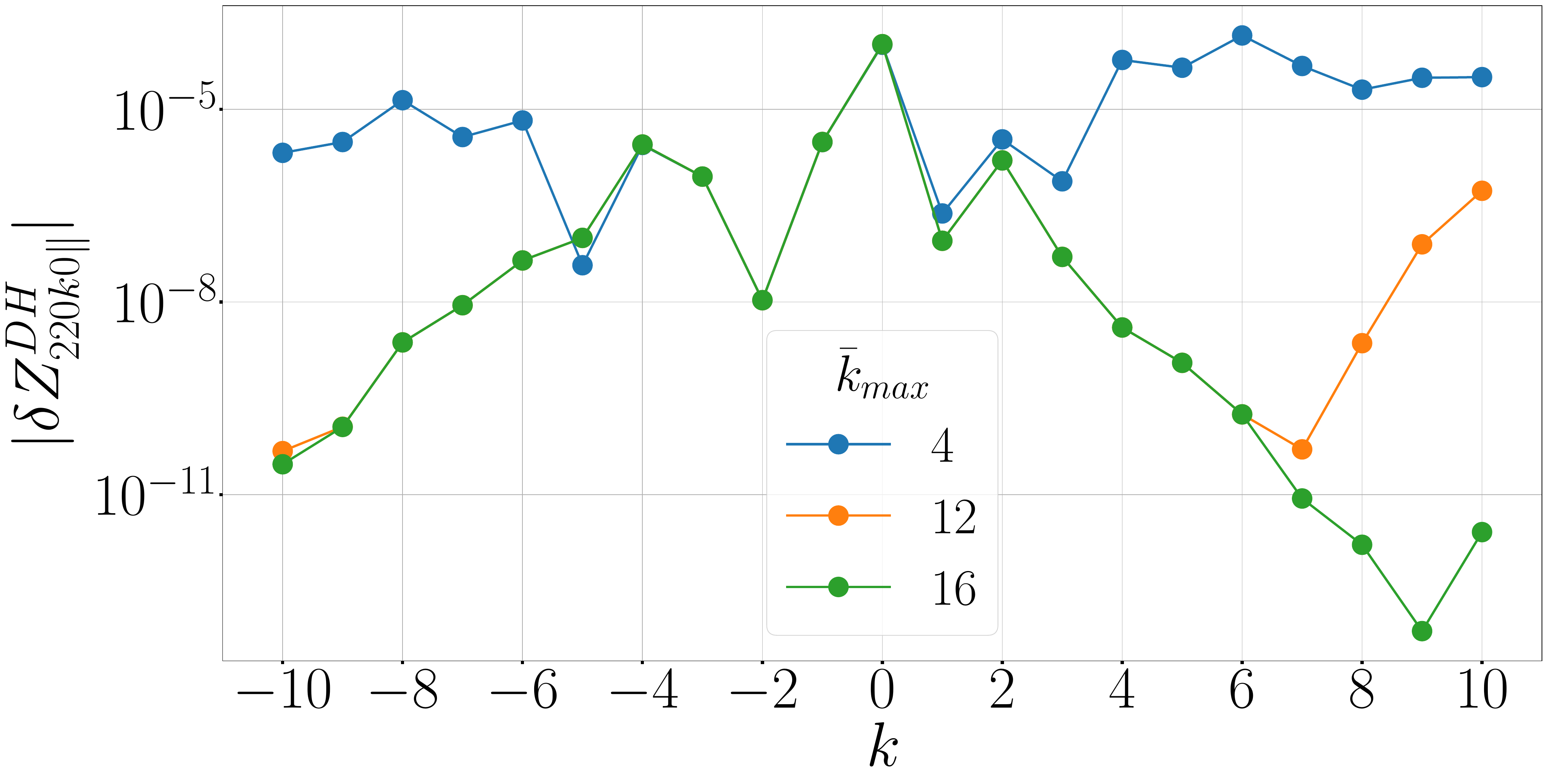}
    	\caption{Absolute value of the corrections to the partial amplitudes $|\delta Z^{\text{DH}}_{220k0\parallel}|$ for increasing value of the waveform harmonic $k$ and different orbital Fourier modes $\bar k_{\text{max}}$. Similar to Fig.\ref{fig:conv_amp_n}, higher values of the waveform harmonic $k$ require an higher number of polar Fourier modes $\bar k_{\text{max}}$. Orbital elements: $\{a,p_{\rm g},e_{\rm g},x_{\rm g}\}=\{0.9,9,1/2,1/2\}$.}
	\label{fig:conv_amp_k}
\end{figure}

As mentioned previously, the precision on the partial amplitudes and their corrections is affected by the number of Fourier modes kept in the orbits. In general, higher $\{\ell, m, n , k\}$ harmonics require a larger number of $\bar n_{\text{max}}$ and $\bar k_{\text{max}}$ Fourier modes for the partial amplitudes to reach numerical convergence. This behavior was already observed in~\cite{Skoupy:2023lih}, and it is presented here in Fig.~\ref{fig:conv_amp_n} and Fig.~\ref{fig:conv_amp_k}. Moreover, in Table~\ref{tbl:FourierAnalysis} we exemplify the impact of the number of modes used on the spin corrections to the fluxes for an orbit with $\{a,p_{\rm g},e_{\rm g},x_{\rm g}\}=\{0.9,9,1/2,1/2\}$, and in the fixed turning point scheme. We set $\ell = m =2$ and summed $\delta \mathcal{F}^{E, DH}_{22nk}$ over $ n \in [-3,3], k \in [-3,3]$\textemdash here we keep 45 digits in the computation. In this case, retaining 12 Fourier modes achieves good compromise between accuracy and speed for the chosen harmonic set.
It is worth mentioning that the eccentricity and inclination of the orbits impact differently the convergence of the fluxes. Indeed, it was observed in Ref.~\cite{Drasco:2005kz} that for high-inclined orbits the fluxes numerically converge by summing over at most 20 $k$ polar harmonics.
By contrast, the fluxes for high eccentric orbits may require a summation over $100$ radial $n$ harmonics. 

\begin{table}[!htb]
    \begin{tabular}{| c|c|c|c|c|}
        \hline
        $\bar n_{\textup{max}}=\bar k_{\textup{max}}$ &  $\delta \mathcal{F}^{E, DH}_{22}\times 10^{7}$\\
         \hline
        4 modes &-7.98830\\
        8 modes &-6.14956\\
        12 modes &-6.16626\\
        16 modes &-6.16638\\
        \hline
    \end{tabular}
    \caption{Impact of the number of Fourier modes kept in the orbits on the calculation of the fluxes for  $\{a,p_{\rm g},e_{\rm g},x_{\rm g}\}=\{0.9,9,1/2,1/2\}$.  We consider $\bar n_{\textup{max}}=\bar k_{\textup{max}}$, set $\ell =m = 2$ and sum the fluxes over $ n \in [-3,3], k \in [-3,3]$. The results are truncated at 6 significant digits.} \label{tbl:FourierAnalysis}
\end{table}

Finally, in the supplementary material available in the repository~\cite{repoHJproject}, we performed detailed tests between the corrections to the orbits computed in our formalism and the results of the code used in~\cite{Skoupy:2023lih}. In these tests, we considered orbits with $\{a,p_{\rm g},e_{\rm g},x_{\rm g}\}=\{0.9,4,5/10,\sqrt{3}/2\}$. The fractional difference between the results of the two codes is at worst $10^{-3} \%$ (in the case of the corrections $\delta t^{\text{DH}}$), and typically much smaller, of order  $10^{-5} \%$.

\section{Discussion and Outlooks}\label{sec:conclusions}

In this paper, we constructed the orbital trajectories of spinning test particles near Kerr black holes within a novel semi-analytical scheme, and derived new analytic expressions for the constants of motion shifts. We then coupled these trajectories to a Teukolsky solver in order to obtain gravitational-wave fluxes of energy and azimuthal angular momentum, and waveform snapshots. Given our explicit decomposition of the fluxes into a referential geodesic (0PA) piece and a spin (1PA) correction, we provide a breakthrough ingredient for 1PA waveforms for EMRIs. Our formalism is flexible, and it can be readily implemented to compute spin-corrections at 1PA in the fixed frequency parametrization (which is convenient for self-force calculations~\cite{Wardell:2021fyy,Mathews:2021rod} and data analysis studies~\cite{Burke:2023lno}) or in the fixed initial conditions scheme (which is useful for computation involving NITs~\cite{Drummond:2023wqc}). The next step should be to incorporate our results into existing infrastructures such as the Fast EMRI Waveforms framework \cite{Katz:2021yft,Chua:2020stf}. 

Phrased differently, our results represent a significant leap in the modeling of eccentric, fully precessing spinning compact binaries at large mass ratios. In the future, it would be interesting to compare our results with those of numerical relativity simulations of fully precessing binaries \cite{Scheel:2014ina,Boyle:2019kee,Ossokine:2020kjp}, and to use our results to inform or validate surrogate, phenomenological or EOB models as was done with numerical relativity simulations e.g. in Refs. \cite{Varma:2019csw,Pratten:2020ceb,Blackman:2017pcm,Ossokine:2020kjp,Ramos-Buades:2023ehm}. At the moment, however, we can only provide information on the fluxes, or the immediate multipolar content of the waveforms and not full inspiral waveforms due to the lack of a balance law for the spin-corrected Carter constant. A more involved set-up would also be needed to calibrate only the spin corrections of the fluxes, since the full 1PA flux includes non-spinning $\mathcal{O}(q)$ corrections, which are currently unknown for generic orbits \cite{Warburton:2021kwk}. 

There is one case where the inspiral can be driven without the knowledge of the Carter constant evolution, and those are quasi-spherical inspirals of the spinning test particle in Kerr spacetime (which corresponds to inclined, non-eccentric precessing binaries). In that case, a flux-balance argument allows to drive the inspiral self-consistently at 1PA using only the energy and angular-momentum fluxes. We are currently preparing a manuscript that provides the details of this argument and adiabatic inspirals of spinning test particles on quasi-spherical orbits \cite{SkoupySpherInprep}.

\begin{acknowledgments}
We thank Josh Mathews for collaboration at the initial stages of this work.
This work makes use of the ``xAct" \textit{Mathematica} package~\cite{xAct} and the following \textit{Mathematica} packages of the Black Hole Perturbation Toolkit~\cite{BHPToolkit}: ``KerrGeodesic"\cite{BHPToolkitKerrGeodesics},``Teukolsky"\cite{BHPToolkitTeukolsky}, and a custom made version of the ``SpinWeightedSpheroidalHarmonics", version 0.3.0~\cite{BHPToolkitSpinWeightedSpheroidalHarmonicsv0.3.0}.
All supplementary material are available in the GitHub repository~\cite{repoHJproject}.

We are grateful to Lisa Drummond, Philip Lynch, Viktor Skoup\'y and Scott Hughes for reading through a draft of this paper and providing helpful feedback.
GAP is deeply grateful to Lisa Drummond and Viktor Skoup\'y for many insightful discussions, for sharing the code developed in~\cite{Skoupy:2023lih} and some of the data used in~\cite{Drummond:2022xej,Drummond:2022efc}.
GAP acknowledges support from an Irish Research Council Fellowship under grant number GOIPD/2022/496. C.P. acknowledges support from a Royal Society – Science Foundation Ireland University Research Fellowship via grant URF/R1/211027. J.MU. acknowledge support from the University College Dublin Ad Astra Fellowship. 
\end{acknowledgments}

\newpage


\appendix

\section{Parametrization by roots and elliptic integrals convention}
\label{app:roots_elliptic}

Table~\ref{tab:elliptic_integrals} list or convention for the Legendre elliptic integral, which follow the Mathematica convention.
\begin{table}[h!]
  \begin{center}
    \caption{Elliptic Integral Conventions}
    \label{tab:elliptic_integrals}
    \begin{tabular}{l|c}
      \toprule 
      \textbf{Function} & \textbf{Definition} \\
      \midrule 
       $\mathsf{F}(\phi|m)$& $\int_{0}^{\phi}\frac{1}{\sqrt{1-m\sin^{2}(\theta)}}\mathrm{d}\theta$\\
       $\mathsf{K}(m)$& $\mathsf{F}(\pi/2|m)$\\
       $\mathsf{E}(\phi|m)$& $\int_{0}^{\phi}\sqrt{1-m\sin^{2}(\theta)}\mathrm{d}\theta$\\
       $\mathsf{E}(m)$& $\mathsf{E}(\pi/2|m)$\\
        $\mathsf{\Pi}(n,\phi|m)$&$\int_{0}^{\phi}\frac{1}{(1-n\sin^{2}(\theta))\sqrt{1-m\sin^{2}(\theta)}}\mathrm{d}\theta$\\
        $\mathsf{\Pi}(n|m)$&$\mathsf{\Pi}(n,\pi/2|m)$\\
       $\mathsf{am}(u|m)$& $u=\mathsf{F}(\mathsf{am}(u|m)|m)$\\
       $\mathsf{sn}(u|m)$& $u=\sin(\mathsf{am}(u|m))$\\

      \bottomrule 
    \end{tabular}
  \end{center}
\end{table}
The $m$ argument in all the elliptic integrals and functions is either $k_{r\rm g}$ or $k_{z\rm g}$
\begin{align}
    k_{r \rm g} &= \frac{(r_{1\rm g} - r_{2\rm g})(r_{3\rm g} - r_{4\rm g})}{(r_{1\rm g} - r_{3\rm g})(r_{2\rm g} - r_{4\rm g})} \, , \\
    k_{z \rm g} &= \frac{a^2 (1 - E^2_{\rm g})z^2_{1 \rm g}}{z^2_{2 \rm g}}  \, , 
\end{align}
while the following terms are common $n$ arguments in $\mathsf{\Pi}(n|m)$ 
\begin{align}
   \gamma(r_{\rm g}) &= - \frac{a^2 z^2_{1\rm g}}{r^2_{\rm g}} \, , \\
    h_r &=\frac{r_{1\rm g}-r_{2\rm g}}{r_{1\rm g}-r_{3\rm g}}\,,\\
    h_\pm&=h_r \frac{r_{3\rm g}-r_\pm}{r_{2\rm g}-r_\pm}\,.
\end{align}

For completeness we present explicit expressions for all radial and polar roots \cite{Fujita:2009bp}. In what follows $Q_{\rm g}=K_{\rm g}-(L_{z\rm g}-aE_{\rm g})^2$.
\begin{align}
    r_{3\mathrm{g}}&=\frac{1}{1-E_{\mathrm{g}}^2}-\frac{r_{1\mathrm{g}}+r_{2\mathrm{g}}}{2} \nonumber \\
    &+\sqrt{\left(\frac{r_{1\mathrm{g}}+r_{2\mathrm{g}}}{2}-\frac{1}{1-E_{\mathrm{g}}^2}\right)^2-\frac{a^2 Q_{\mathrm{g}}}{r_{1\mathrm{g}} r_{2\mathrm{g}} (1-E^2_{\mathrm{g}})}},\\
    r_{4\mathrm{g}}&=\frac{a^2 Q_{\mathrm{g}}}{r_{1\mathrm{g}}r_{2\mathrm{g}}r_{3\mathrm{g}}(1-E_{\rm g}^2)},\\
    z_{2\mathrm{g}}&=\sqrt{a^2(1-E_{\mathrm{g}}^2)+\frac{L_{\mathrm{g}}}{1-z_{1\mathrm{g}}}}.
\end{align}

\section{Analytic expressions for the mean anomalies and geodesics frequencies} \label{app:geo_freq}
\subsection{Mean anomalies}
The integrals of Eqs.~\ref{eq:wydef} can be written in terms of Jacobi elliptic integrals and then inverted to obtain $\chi_{r\rm g}(w_{r \rm g})$ and $\chi_{z\rm g}(w_{z \rm g})$. The result is 
\begin{align}
 w_{r\rm g} & = \frac{\pi}{\mathsf K(k_{r\mathrm{g}})}\mathsf F\Bigg(\!\! \left. \arcsin\! \!\Bigg(\sqrt{\frac{(r_{\rm g}-r_{2\rm g})(r_{1\rm g}-r_{3\rm g})}{(r_{\rm g}-r_{3\rm g})(r_{1\rm g}-r_{2\rm g})}}\Bigg) \right\rvert k_{r \rm g}\! \Bigg) \, , \label{eq:wrg}
 \\
 w_{z\rm g} & =  \frac{\pi}{2\mathsf K(k_{z\rm g})}\mathsf F(\chi_{z\rm g} |k_{z\rm g})  \, , \label{eq:wzg}
\end{align}
and
\begin{align}
\chi_{r\rm g}(w_{r \rm g}) &= \arcsin\! \bigg( \frac{(r_{1\rm g} + r_{2\rm g} - 2 r_{3\rm g} ) \mathcal W_{r \rm g}(w_{r \rm g}) - 1}{1 - (r_{1\rm g}-r_{2\rm g})\mathcal W_{r \rm g}(w_{r \rm g})} \bigg)   \, , \\
\chi_{z\rm g}(w_{z \rm g}) &= \mathsf{am}(\alpha_{z \rm g}(w_{z \rm g})|k_{z \rm g})\, ,
\end{align}
with
\begin{align}
\mathcal W_{r \rm g}(w_{r \rm g}) &= \frac{1}{r_{1\rm g}-r_{3\rm g}}\!\!\left.\mathsf{sn^2}\bigg(\frac{1}{\pi}\mathsf{K}(k_{r \rm g})w_{r \rm g}\right\rvert k_{r \rm g} \bigg)  \, , \\
\alpha_{r \rm g}(w_{z \rm g}) &= \frac{1}{\pi} \mathsf K (k_{r \rm g})w_{r \rm g} \, , \\
\alpha_{z \rm g}(w_{z \rm g}) &= \frac{2}{\pi} \mathsf K (k_{z \rm g})\Big(w_{z \rm g}+\frac{\pi}{2}\Big) \, .
\end{align}
The $\pi/2$ above is introduced to match the convention of the KerrGeodesics package (\cite{KerrGeodesics090}).
The explicit solutions of the integrals~\ref{eq:non-hom-mean-anom_int} and their inversions $\chi_r(\zeta_r), \chi_z(\zeta_z)$ can be obtained by simply replacing $w_{r\rm g}$ with $\zeta_r$ and $w_{z\rm g}$ with $\zeta_z$.

\subsection{Geodesics frequencies}
The geodesic Mino-time frequencies are given by~\cite{Fujita:2009bp,vandeMeent:2019cam}
\begin{align}
 \Upsilon_{r\rm g}&=\frac{\pi}{2\, \mathsf{K}(k_{r\rm g})}\sqrt{(1-E_{\rm g}^2)(r_{1\rm g}-r_{3\rm g})(r_{2\rm g}-r_{4\rm g})}\,,\\
 \Upsilon_{z\rm g}&=\frac{\pi\, z_{2\rm g}}{2 \,\mathsf{K}(k_{z\rm g})}\,,\\
  \Upsilon_{t\rm g}&= \Upsilon^t_{r\rm g} + \Upsilon^t_{z\rm g}\,,\\
  \Upsilon_{\phi \rm g}&= \Upsilon^\phi_{r\rm g} + \Upsilon^\phi_{z\rm g} \,,   
\end{align}
where
\begin{align}
    \Upsilon^t_{z\rm g}&=-a^2 E_{\rm g}+\frac{E_{\rm g} \,Q_{\rm g}}{(1-E_{\rm g}^2)\, z_{1 \rm g}^2}\left(1-\frac{\mathsf{E}(k_{z\rm g})}{\mathsf{K}(k_{z\rm g})}\right)\,,\\
    \Upsilon^\phi_{z\rm g} &=\frac{L_{z\rm g}}{\mathsf{K}(k_{z\rm g})} \mathsf{\Pi}(z_{1\rm g}^2|k_{z\rm g})\,, 
\end{align}
\begin{widetext}
\begin{align}
 \Upsilon^t_{r\rm g}&=(4+a^2)E_{\rm g}+E_{\rm g}\Bigg[(4+r_{1\rm g}+r_{2\rm g}+r_{3\rm g})\frac{r_{3\rm g}}{2}-\frac{r_{1 \rm g} r_{2 \rm g}}{2}+(r_{1 \rm g}-r_{3 \rm g})(r_{2 \rm g}-r_{4 \rm g})\frac{\mathsf{E}(k_{r\rm g})}{2 \mathsf{K}(k_{r\rm g})} \nonumber\\
&+(4+r_{1\rm g}+r_{2\rm g}+r_{3\rm g}+r_{4\rm g})(r_{2\rm g}-r_{3\rm g})\frac{\mathsf{\Pi}(h_r|k_{r\rm g})}{2 \mathsf{K}(k_{r\rm g})}\nonumber\\
&+\frac{2}{(r_+-r_-)}\left(\frac{(4-a L_{z\rm g}/E_{\rm g})r_+-2 a^2}{r_{3 \rm g}-r_+}\left(1-\frac{r_{2\rm g}-r_{3\rm g}}{r_{2\rm g}-r_+}\frac{\mathsf{\Pi}(h_+|k_{r \rm g})}{\mathsf{K}(k_{r \rm g})}\right)-(+\leftrightarrow-)\right)\Bigg]\,, \\
 \Upsilon^\phi_{r\rm g}&= \frac{a}{(r_+-r_-)}\left[\frac{2 E_{\rm g}r_+- a L_{z \rm g}}{r_{3 \rm g}-r_+}\left(1-\frac{r_{2\rm g}-r_{3\rm g}}{r_{2\rm g}-r_+}\frac{\mathsf{\Pi}(h_+|k_{r \rm g})}{\mathsf{K}(k_{r \rm g})}\right)-(+\leftrightarrow-)\right]
\end{align}
\end{widetext}

\newpage
\section{Spin coefficients and projections of the Christoffel symbols onto the Marck tetrad}
\label{app:spin_coeffs_Christoffel_sym}

The spin coefficients $\mathcal S^y$ for $y = r,z$ are defined as
\begin{align}
 \mathcal S^y &= \mathcal{S}^y_\parallel +\mathcal  S^y_\perp \qquad y = r,z \, , \\
  \mathcal S^y_\parallel &\coloneqq  v^y_{\rm g} \tilde e^\mu_{~(1);y} \tilde e_{\mu(2)} s^{(1)(2)} = s_\parallel \mathcal S^y_{(1)(2)}(r_{\rm g}, z_{\rm g})  \, , \\
 \mathcal S^y_\perp  &\coloneqq v^y_{\rm g} \Big( \tilde e^\mu_{~(2);y} e_{\mu(3)} s^{(2)(3)} + e^\mu_{~(3);y} \tilde e_{\mu(1)} s^{(3)(1)} \Big)  \nonumber \\
  & =  s_\perp \Big(\mathcal S^y_{(2)(3)}\cos{\psi_{\rm p}} + \mathcal S^y_{(3)(1)} \sin{\psi_{\rm p}} \Big)  \, , 
\end{align}
with $v^r_{\rm g}$ and $v^z_{\rm g}$ the geodesics velocities defined in Eqs.~\eqref{eq:geo_radial_velocity} and~\eqref{eq:geo_polar_velocity}, respectively. Moreover, notice that the terms
\begin{align}
    &v^r_{\rm g}\frac{\dd v^r_{\rm g}}{\dd r_{\rm g}} = 2 r_{\rm g} E_{\rm g} \mathcal R(r_{\rm g}) - r_{\rm g} \Delta - (r_{\rm g} - 1)(K_{\rm g} + r^2_{\rm g}) \, , \\
    &v^z_{\rm g}\frac{\dd v^z_{\rm g}}{\dd z_{\rm g}} = (1 - E^2_{\rm g})a^2 z_{\rm g}(2 z^2_{\rm g} - z^2_{1 \rm g}) -z_{\rm g} z^2_{2 \rm g} \, ,
\end{align}
are finite at the turning points. The components $\mathcal S^y_{(1)(2)}$ are given by
\begin{widetext}
\begin{align}
 \mathcal S^r_{(1)(2)} &= v^r_{\rm g}v^z_{\rm g}\frac{2 a^2 r_{\rm g} z_{\rm g} \mathcal R(r_{\rm g})}{\sqrt{K_{\rm g}}\Delta^2 \Sigma^2} + v^r_{\rm g}\frac{\dd v^r_{\rm g}}{\dd r_{\rm g}}\frac{r_{\rm g}\mathcal R(r_{\rm g})(K_{\rm g} - a^2 z^2_{\rm g})}{\sqrt{K_{\rm g}}(K_{\rm g} + r^2_{\rm g})\Delta^2\Sigma} +\frac{(v^r_{\rm g})^2}{\sqrt{K_{\rm g}}\Delta^2}\bigg(\frac{\mathcal R(r_{\rm g})}{\Sigma} - \frac{2r^2_{\rm g}\mathcal R(r_{\rm g})}{\Sigma^2}  -\frac{E_{\rm g}(K_{\rm g}-r^2_{\rm g})}{K_{\rm g}+r^2_{\rm g}}  \bigg)\, ,
\end{align}
\begin{align}
 \mathcal S^z_{(1)(2)} &= -(v^z_{\rm g})^2\bigg(\frac{\mathcal R(r_{\rm g}) (r^2_{\rm g}-a^2 z^2_{\rm g})}{\sqrt{K_{\rm g}}(1-z^2_{\rm g})\Sigma^2} + \frac{2 a^2 E_{\rm g} (K_{\rm g} + r^2_{\rm g})z^2_{\rm g} }{\sqrt{K_{\rm g}}(1 - z^2_{\rm g})\Sigma(K_{\rm g} - a^2 z^2_{\rm g})}\bigg) + v^z_{\rm g} v^r_{\rm g}\frac{2 a r_{\rm g} z_{\rm g} \mathcal Z(z_{\rm g})}{\sqrt{K_{\rm g}}(1 - z^2_{\rm g})\Sigma^2} \nonumber \\
 &+ v^z_{\rm g}\frac{\dd v^z_{\rm g}}{\dd z_{\rm g}}\frac{a z_{\rm g} (K_{\rm g} + r^2_{\rm g})\mathcal Z(z_{\rm g})}{\sqrt{K_{\rm g}}(1 - z^2_{\rm g})\Sigma(K_{\rm g} - a^2 z^2_{\rm g})} \, ,
\end{align}
\end{widetext}
The components $ \mathcal S^y_{(2)(3)}$ are given by
\begin{widetext}
\begin{align}
 \mathcal S^r_{(2)(3)} &= v^r_{\rm g}v^z_{\rm g}\frac{a \mathcal R(r_{\rm g})\big(r^2_{\rm g}(K_{\rm g} - a^2 z^2_{\rm g}) - a^2z^2_{\rm g}(K_{\rm g}+r^2_{\rm g})\big)}{\sqrt{K_{\rm g}}\Delta^2\Sigma^2 \sqrt{K_{\rm g} + r^2_{\rm g}}\sqrt{K_{\rm g} - a^2 z^2_{\rm g}}} - v^r_{\rm g}\frac{\dd v^r_{\rm g}}{\dd r_{\rm g}}\frac{a z_{\rm g}\mathcal R(r_{\rm g}) (K_{\rm g} - a^2 z^2_{\rm g})}{\sqrt{K_{\rm g}}\Delta^2\Sigma \sqrt{K_{\rm g} + r^2_{\rm g}}\sqrt{K_{\rm g} - a^2 z^2_{\rm g}}} \nonumber \\
 &+ v^2_{r\rm g}\frac{a r_{\rm g} z_{\rm g} \big(2 K_{\rm g} \mathcal R(r_{\rm g}) - a r^2_{\rm g} \Lzrd - a^4 z^4_{\rm g} E_{\rm g} - (3 a^2 r^2_{\rm g} E_{\rm g} -a^3 \Lzrd)z^2_{\rm g} \big) }{\sqrt{K_{\rm g}}\Delta^2\Sigma^2 \sqrt{K_{\rm g} + r^2_{\rm g}}\sqrt{K_{\rm g} - a^2 z^2_{\rm g}}} \, ,
\end{align}

\begin{align}
 \mathcal S^z_{(2)(3)} &= v^2_{z\rm g} \frac{a r_{\rm g} z_{\rm g}\big(a^2 E_{\rm g}(1-z^2_{\rm g})(r^2_{\rm g} - a^2 z^2_{\rm g}) - 2 a K_{\rm g} \mathcal Z(z_{\rm g}) -E_{\rm g} \Sigma^2 - a L_{z \rm g} (r^2_{\rm g} - a^2 z^2_{\rm g}) \big)}{\sqrt{K_{\rm g}}(1 - z^2_{\rm g})\Sigma^2\sqrt{K_{\rm g} - a^2 z^2_{\rm g}}\sqrt{K_{\rm g} + r^2_{\rm g}}}  \nonumber \\
 &+ v^z_{\rm g}v^r_{\rm g}\frac{\big(r^2_{\rm g}(K_{\rm g} - a^2 z^2_{\rm g}) - a^2 z^2_{\rm g}(K_{\rm g} + r^2_{\rm g})\big)\mathcal Z(z_{\rm g})}{\sqrt{K_{\rm g}}(1-z^2_{\rm g})\Sigma^2\sqrt{K_{\rm g} + r^2_{\rm g}}\sqrt{K_{\rm g} - a^2 z^2_{\rm g}}} + v^z_{\rm g}\frac{\dd v^z_{\rm g}}{\dd z_{\rm g}}\frac{r_{\rm g}\sqrt{K_{\rm g} + r^2_{\rm g}}\mathcal Z(z_{\rm g})}{\sqrt{K_{\rm g}}(1-z^2_{\rm g})\Sigma\sqrt{K_{\rm g}-a^2 z^2_{\rm g}}}  \, ,
\end{align}
\end{widetext}
The components $ \mathcal S^y_{(3)(1)}$ are given by
\begin{widetext}
\begin{align}
 \mathcal S^r_{(3)(1)} &= v^r_{\rm g} \Bigg(\frac{a z_{\rm g} r_{\rm g} \mathcal R(r_{\rm g}) \big( r_{\rm g} ( a L_{z \rm g} - r_{\rm g} E_{\rm g}) - a \Lzrd \big)\sqrt{K_{\rm g} - a^2 z^2_{\rm g}}}{K_{\rm g}\Delta^3 \Sigma\sqrt{K_{\rm g} + r^2_{\rm g}}} - (v^z_{\rm g})^2\frac{a^3 z^3_{\rm g} \sqrt{K_{\rm g} + r^2_{\rm g}}}{K_{\rm g} (1 - z^2_{\rm g})\Delta\Sigma^2 \sqrt{K_{\rm g} - a^2 z^2_{\rm g}}} \Bigg) \nonumber \\
 &- v^r_{\rm g} \Bigg(\frac{a^3 z^3_{\rm g} \mathcal Z(z_{\rm g}) \big(\Sigma L_{z \rm g} -2 r_{\rm g} \mathcal Z(z_{\rm g})\big)\sqrt{K_{\rm g} + r^2_{\rm g}}}{K_{\rm g}\Delta^2\Sigma^2 (1 - z^2_{\rm g}) \sqrt{K_{\rm g} - a^2 z^2_{\rm g}}} \Bigg) + (v^r_{\rm g})^2 v^z_{\rm g} \frac{a r_{\rm g} }{\Delta^2 \Sigma \sqrt{K_{\rm g} + r^2_{\rm g}}\sqrt{K_{\rm g} - a^2 z^2_{\rm g}}}   \nonumber \\
 & - (v^r_{\rm g})^2 \frac{\dd v^r_{\rm g}}{\dd r_{\rm g}} \frac{a r_{\rm g} z_{\rm g} \sqrt{K_{\rm g}-a^2 z^2_{\rm g}}}{K_{\rm g} \Delta^2 \Sigma \sqrt{K_{\rm g} + r^2_{\rm g}}} + (v^r_{\rm g})^3\frac{a z_{\rm g} r_{\rm g} \big(\Sigma(r_{\rm g} - 1) + r_{\rm g} \Delta\big)\sqrt{K_{\rm g} - a^2 z^2_{\rm g}}}{K_{\rm g}\Delta^3\Sigma^2\sqrt{K_{\rm g} + r^2_{\rm g}}} \, ,
\end{align}

\begin{align}
 \mathcal S^z_{(3)(1)} &= -v^z_{\rm g} \Bigg(\frac{a L_{z \rm g} r_{\rm g} z^2_{\rm g} \mathcal Z(z_{\rm g})\sqrt{K_{\rm g} + r^2_{\rm g}} }{K_{\rm g} (1 - z^2_{\rm g})^2 \Sigma\sqrt{K_{\rm g} - a^2 z^2_{\rm g}}} +  (v^r_{\rm g})^2\frac{a r^3_{\rm g}\sqrt{K_{\rm g} - a^2 z^2_{\rm g}}}{K_{\rm g} \Delta \Sigma^2\sqrt{K_{\rm g} + r^2_{\rm g}}} \Bigg) + (v^z_{\rm g})^2 v^r_{\rm g}\frac{a z_{\rm g}}{(1-z^2_{\rm g}) \Sigma \sqrt{K_{\rm g} + r^2_{\rm g}}\sqrt{K_{\rm g} - a^2 z^2_{\rm g}}}   \nonumber \\
 &-  v^z_{\rm g} \Bigg(\frac{r^3_{\rm g}\mathcal R(r_{\rm g}) \big( 2r_{\rm g} \mathcal Z(z_{\rm g}) - L_{z \rm g}\Sigma \big)\sqrt{K_{\rm g} - a^2 z^2_{\rm g}}}{K_{\rm g}(1 - z^2_{\rm g}) \Delta \Sigma^2  \sqrt{K_{\rm g} + r^2_{\rm g}}} \Bigg) - (v^z_{\rm g})^2 \frac{a z_{\rm g} r_{\rm g} \sqrt{K_{\rm g} + r^2_{\rm g}}}{K_{\rm g}(1-z^2_{\rm g})\Sigma \sqrt{K_{\rm g} - a^2 z^2_{\rm g}}} \nonumber \\
 &- (v^z_{\rm g})^3\frac{a r_{\rm g} z^2_{\rm g} \big(\Sigma - a^2 (1-z^2_{\rm g})\big)\sqrt{K_{\rm g} + r^2_{\rm g}}}{K_{\rm g} (1-z^2_{\rm g})^2\Sigma^2\sqrt{K_{\rm g} - a^2 z^2_{\rm g}}} \, ,
\end{align}
\end{widetext}

The projections of the Christofel symbols' components ${\Upgamma_{\mu\nu}}^t$ and  ${\Upgamma_{\mu\nu}}^\phi$ onto the Marck's tetrad are defined as
\begin{align}
 &\mathcal \Upgamma^t_\parallel \coloneqq {\Upgamma_{\mu\nu}}^t \tilde e^\mu_{~(1)} \tilde e^\nu_{~(2)} s^{(1)(2)} =  2 s_\parallel {\Upgamma_{(1)(2)}}^t  \, , \\
& \mathcal \Upgamma^t_\perp \coloneqq  {\Upgamma_{\mu\nu}}^t \Big( \tilde e^\mu_{~(2)} e^\nu_{~(3)} s^{(2)(3)} + e^\mu_{~(3)} \tilde e^\nu_{~(1)} s^{(3)(1)} \Big)  \nonumber \\
 &=  2 s_\perp \Big({\Upgamma_{(2)(3)}}^t\cos \psi_{\rm p} + {\Upgamma_{(3)(1)}}^t \sin \psi_{\rm p}\Big)   \, , 
\end{align}
The definition for the projections $\Upgamma^\phi_\parallel , \Upgamma^\phi_\perp$ are obtained by simply replacing $t \to \phi$ in the previous expressions. Before presenting these projections, we defined the following auxiliaries functions:
\begin{align}
    \mathcal Q_1 &= a^2 \Lzrd \Sigma + a^3 z^4_{\rm g} \mathcal R(r_{\rm g}) + r^3_{\rm g} (r_{\rm g} -2)\mathcal Z(z_{\rm g}) \nonumber \\
    & +K_{\rm g} \big(L_{z \rm g}\Sigma - 2 r_{\rm g} \mathcal Z(z_{\rm g}) \big) \, , \\
    \mathcal Q_2 &= 2 a r_{\rm g} (K_{\rm g} + r^2_{\rm g}) \mathcal Z(z_{\rm g}) - E_{\rm g} K_{\rm g}(r^2_{\rm g} + a^2)\Sigma  \nonumber \\
    & - a \Lzrd(r^2_{\rm g} + a^2)\Sigma  \, .
\end{align}

The components ${\Upgamma_{(1)(2)}}^t$ and ${\Upgamma_{(1)(2)}}^\phi $ are given by
\begin{widetext}
\begin{align}
     {\Upgamma_{(1)(2)}}^t &= - 2\mathcal Q_1 \Bigg(\frac{a r_{\rm g} \big( \mathcal R (r_{\rm g}) (K_{\rm g} -a^2 z^2_{\rm g}) (a^2 (r^2_{\rm g} -a^2 z^2_{\rm g}) +2 r^4_{\rm g} + r^2_{\rm g} \Sigma) + 2 a^3 z^2_{\rm g} (K_{\rm g} + r^2_{\rm g}) \Delta \mathcal Z(z_{\rm g}) \big) }{\sqrt{K_{\rm g}} \Delta^2 \Sigma^4(K_{\rm g} + r^2_{\rm g})(K_{\rm g} - a^2 z^2_{\rm g})}  \Bigg) \nonumber \\
    & - 2\mathcal Q_2 \Bigg(\frac{r_{\rm g}\big( \mathcal R(r_{\rm g}) (r^2_{\rm g} + a^2)(r^2_{\rm g} - a^2 z^2_{\rm g})(K_{\rm g} - a^2 z^2_{\rm g}) + 2 a^3 z^2_{\rm g} (K_{\rm g} + r^2_{\rm g}) \Delta \mathcal Z (z_{\rm g}) \big)}{\sqrt{K_{\rm g}} \Delta^2 \Sigma^4(K_{\rm g} + r^2_{\rm g})(K_{\rm g} - a^2 z^2_{\rm g})} \Bigg) \nonumber \\
    & - v^r_{\rm g} v^z_{\rm g} \frac{8 a^2 r^2_{\rm g} z_{\rm g}}{\sqrt{K_{\rm g}}\Delta \Sigma^3} -(v^r_{\rm g})^2 \frac{2r_{\rm g}(r^2_{\rm g} - a^2)(K_{\rm g} - a^2 z^2_{\rm g})}{\sqrt{K_{\rm g}}\Delta^2 \Sigma^2 (K_{\rm g} + r^2_{\rm g})}  \, ,
\end{align}

\begin{align}
     {\Upgamma_{(1)(2)}}^\phi &=- 2\mathcal Q_1 \Bigg(\frac{a z^2_{\rm g} \mathcal Z(z_{\rm g})\big(2 a^2 r(1 - z^2_{\rm g})+\Sigma^2\big)}{\sqrt{K_{\rm g}}(1-z^2_{\rm g})^2\Delta\Sigma^4(K_{\rm g} - a^2 z^2_{\rm g})} \Bigg) - 2\mathcal Q_1 \Bigg(\frac{r_{\rm g} \mathcal R(r_{\rm g})\big(r^2_{\rm g}(r^2_{\rm g} + a^2) + r_{\rm g} \Sigma (r_{\rm g} -\Sigma) - a^4 z^2_{\rm g} (1-z^2_{\rm g})\big)}{\sqrt{K_{\rm g}}(1-z^2_{\rm g})\Delta^2 \Sigma^4(K_{\rm g} + r^2_{\rm g})}  \Bigg) \nonumber \\
    &- 2\mathcal Q_2 \Bigg(\frac{a r_{\rm g} \big(\mathcal R(r_{\rm g})(r^2_{\rm g} - a^2 z^2_{\rm g})(1 - z^2_{\rm g})(K_{\rm g} - a^2 z^2_{\rm g}) + 2 a z^2_{\rm g} (K_{\rm g} + r^2_{\rm g}) \Delta \mathcal Z(z_{\rm g}) \big)}{\sqrt{K_{\rm g}} (1 - z^2_{\rm g})\Delta^2 \Sigma^4(K_{\rm g} + r^2_{\rm g})(K_{\rm g} - a^2 z^2_{\rm g})} \Bigg) \nonumber \\
    & + v^r_{\rm g} v^z_{\rm g} \frac{4a r_{\rm g} z_{\rm g} (\Sigma - 2 r_{\rm g}) }{\sqrt{K}(1 - z^2_{\rm g})\Delta \Sigma^3} - (v^r_{\rm g})^2 \frac{2a r_{\rm g} (r_{\rm g} - 1)(K_{\rm g} - a^2 z^2_{\rm g})}{\sqrt{K_{\rm g}}\Delta^2 \Sigma^2 (K_{\rm g} + r^2_{\rm g})} - (v^z_{\rm g})^2\frac{2a z^2_{\rm g}(K_{\rm g} + r^2_{\rm g})}{\sqrt{K_{\rm g}}(1 - z^2_{\rm g})^2\Sigma^2(K_{\rm g} - a^2 z^2_{\rm g})} \, ,
\end{align}
   
\end{widetext}

The components ${\Upgamma_{(2)(3)}}^t$ and ${\Upgamma_{(2)(3)}}^\phi $ are given by
\begin{widetext}
\begin{align}
    {\Upgamma_{(2)(3)}}^t &= \frac{2a^2 z_{\rm g} \mathcal Q_1}{\sqrt{K_{\rm g}}\Delta \Sigma^4\sqrt{K_{\rm g} + r^2_{\rm g}}\sqrt{K_{\rm g} -a^2 z^2_{\rm g}}} \bigg[ \frac{\mathcal R(r_{\rm g})}{\Delta}\Big( 3 r^4_{\rm g} -a^4 z^2_{\rm g} + a^2 r^2_{\rm g}(1 + z^2_{\rm g}) \Big) - 2 a r^2_{\rm g} \mathcal Z(z_{\rm g}) \bigg] \nonumber \\
    & + \frac{2a z_{\rm g} \mathcal Q_2}{\sqrt{K_{\rm g}}\Delta \Sigma^4\sqrt{K_{\rm g} + r^2_{\rm g}}\sqrt{K_{\rm g} -a^2 z^2_{\rm g}}} \bigg[  \frac{\mathcal R(r_{\rm g})}{\Delta}(r^2_{\rm g} + a^2)(r^2_{\rm g} -a^2 z^2_{\rm g}) - 2 a r^2_{\rm g} \mathcal Z(z_{\rm g})  \bigg]  \nonumber \\
    & + v^r_{\rm g} v^z_{\rm g} \frac{4 r_{\rm g} \big( 2 a^3 r^2 z^2_{\rm g} - a K_{\rm g} (r^2_{\rm g} - a^2 z^2_{\rm g}) \big)}{\sqrt{K_{\rm g}}\Delta \Sigma^3 \sqrt{K_{\rm g} + r^2_{\rm g}}\sqrt{K_{\rm g} - a^2 z^2_{\rm g}}} + (v^r_{\rm g})^2\frac{2a z_{\rm g} (r^2_{\rm g} - a^2)\sqrt{K_{\rm g} - a^2 z^2_{\rm g}}}{\sqrt{K_{\rm g}}\Delta^2 \Sigma^2 \sqrt{K_{\rm g} + r^2_{\rm g}}} \, ,
\end{align}

\begin{align}
    {\Upgamma_{(2)(3)}}^\phi &= \frac{z_{\rm g} \mathcal Q_1}{\sqrt{K_{\rm g}}(1-z^2_{\rm g})\Delta \Sigma^4\sqrt{K_{\rm g} + r^2_{\rm g}}\sqrt{K_{\rm g} -a^2 z^2_{\rm g}}} \bigg[ a \frac{\mathcal R(r_{\rm g})}{\Delta}\big(a^2(r^2_{\rm g} - a^2 z^2_{\rm g}) + 2 r^4_{\rm g} + \Sigma ( a^2 z^2_{\rm g} - r_{\rm g} \Sigma)  \big)  -2 a^2 r^2_{\rm g} \mathcal Z(z_{\rm g}) \nonumber \\
    & \phantom{=\frac{z_{\rm g} \mathcal Q_1}{\sqrt{K_{\rm g}}(1-z^2_{\rm g})\Delta \Sigma^4\sqrt{K_{\rm g} + r^2_{\rm g}}\sqrt{K_{\rm g} -a^2 z^2_{\rm g}}}}  -\frac{r_{\rm g} \mathcal Z(z_{\rm g}) \Sigma^2}{1-z^2_{\rm g}}   \bigg] \nonumber \\
    & -\frac{a z_{\rm g} \mathcal Q_2}{\sqrt{K_{\rm g}}(1-z^2_{\rm g})\Delta^2 \Sigma^4\sqrt{K_{\rm g} + r^2_{\rm g}}\sqrt{K_{\rm g} -a^2 z^2_{\rm g}}} \bigg[ 2 L_{z \rm g} r^4_{\rm g} +(r^2 + a^2) (3 E_{\rm g} r^2_{\rm g} -a \Lzrd)a z^2_{\rm g} - a \mathcal R(r_{\rm g})(3 r^2 + a^2 z^4_{\rm g}) \nonumber \\
    & \phantom{-\frac{a z_{\rm g} \mathcal Q_2}{\sqrt{K_{\rm g}}(1-z^2_{\rm g})\Delta^2 \Sigma^4\sqrt{K_{\rm g} + r^2_{\rm g}}\sqrt{K_{\rm g} -a^2 z^2_{\rm g}}}}-4 r^3_{\rm g} \mathcal Z(z_{\rm g}) \bigg] \nonumber \\
    & + v^r_{\rm g} v^z_{\rm g}\frac{(\Sigma - 2r_{\rm g})\big(K_{\rm g}(r^2_{\rm g} - a^2 z^2_{\rm g}) - 2 a^2 z^2_{\rm g} r^2_{\rm g} \big)}{\sqrt{K_{\rm g}}(1 - z^2_{\rm g})\Delta \Sigma^3 \sqrt{K_{\rm g} + r^2_{\rm g}}\sqrt{K_{\rm g} - a^2 z^2_{\rm g}}}  +  (v^r_{\rm g})^2\frac{a^2 z_{\rm g} (r_{\rm g} -1)\sqrt{K_{\rm g} - a^2 z^2_{\rm g}}}{\sqrt{K_{\rm g}}\Delta^2 \Sigma^2 \sqrt{K_{\rm g} + r^2_{\rm g}}} \nonumber \\
    &- (v^z_{\rm g})^2\frac{r_{\rm g} z_{\rm g} \sqrt{K_{\rm g} + r^2}}{\sqrt{K_{\rm g}}(1 - z^2_{\rm g})^2\Sigma^2 \sqrt{K_{\rm g} - a^2 z^2_{\rm g}}}  \, ,
\end{align} 
\end{widetext}

The components ${\Upgamma_{(3)(1)}}^t$ and ${\Upgamma_{(3)(1)}}^\phi $ are given by
\begin{align}
 {\Upgamma_{(3)(1)}}^t &= v^r_{\rm g} \frac{4 a^2 r_{\rm g} z_{\rm g} \mathcal Z(z_{\rm g})}{\Delta \Sigma^2\sqrt{K_{\rm g}+r^2_{\rm g}}\sqrt{K_{\rm g} - a^2 z^2_{\rm g}}} \nonumber \\
  &- v^z_{\rm g} \frac{4 a r^2_{\rm g} \mathcal R(r_{\rm g})}{\Delta \Sigma^2\sqrt{K_{\rm g}+r^2_{\rm g}}\sqrt{K_{\rm g} - a^2 z^2_{\rm g}}} \, ,
\end{align}

\begin{align}
 {\Upgamma_{(3)(1)}}^\phi & = v^z_{\rm g} \frac{2r_{\rm g}\mathcal R(r_{\rm g}) \big(\Sigma - 2 r_{\rm g}\big)}{(1-z^2_{\rm g})\Delta \Sigma^2\sqrt{K_{\rm g}+r^2_{\rm g}}\sqrt{K_{\rm g} - a^2 z^2_{\rm g}}}  \nonumber\\
  &- v^r_{\rm g} \frac{2a z_{\rm g} \mathcal Z(z_{\rm g}) \big( \Sigma - 2 r_{\rm g}\big)}{(1-z^2_{\rm g})\Delta \Sigma^2\sqrt{K_{\rm g}+r^2_{\rm g}}\sqrt{K_{\rm g} - a^2 z^2_{\rm g}}} \, ,
\end{align}

\newpage
\section{Explicit expressions for the shifts to the turning points and their averages} \label{app:turnp}

The radial shifts to the turning points are given by the following expressions:
\begin{align}
   r_{1 \rm s}(z_{\rm g},\psi_{\rm p}) &= s_\parallel r_{1{\rm s}\parallel}(z_{\rm g}) + s_\perp r_{1{\rm s}\perp}(z_{\rm g},\psi_{\rm p}) \, , \\
   r_{2 \rm s}(z_{\rm g},\psi_{\rm p}) &= s_\parallel r_{2{\rm s}\parallel}(z_{\rm g}) + s_\perp r_{2{\rm s}\perp}(z_{\rm g},\psi_{\rm p})  \,.
\end{align}
where
\begin{widetext}
\begin{align}
 r_{1{\rm s}\parallel}(z_{\rm g}) &= \frac{\mathcal R(r_{1\rm g})}{K_{\rm g} + r^2_{1 \rm g}}\bigg(\frac{2 \sqrt{K_{\rm g}} \Delta(r_{1\rm g})}{(r_{1\rm g}- r_{2\rm g})Y^2_{r \rm g}(r_{1\rm g})} -\frac{r_{1\rm g}(K_{\rm g} -a^2 z^2_{\rm g})}{\sqrt{K_{\rm g}}\Sigma(r_{1\rm g})} \bigg) + \frac{2 a \Delta(r_{1\rm g})\sgn(\Lzrd)}{(r_{1\rm g}- r_{2\rm g})Y^2_{r \rm g}(r_{1\rm g})} + \nonumber \\
 &+ \frac{1}{(r_{1\rm g}- r_{2\rm g})Y^2_{r \rm g}(r_{1\rm g})} \displaystyle\sum^{3}_{i=1} \frac{\partial R_{\rm g}(r_{1\rm g})}{\partial C_{i\rm g}} C_{i s}\, , \\
 r_{2{\rm s}\parallel}(z_{\rm g}) &= -\frac{\mathcal R(r_{2\rm g})}{K_{\rm g} + r^2_{2 \rm g}}\bigg(\frac{2 \sqrt{K_{\rm g}} \Delta(r_{2\rm g})}{(r_{1\rm g}- r_{2\rm g})Y^2_{r \rm g}(r_{2\rm g})} + \frac{r_{2\rm g}(K_{\rm g} -a^2 z^2_{\rm g})}{\sqrt{K_{\rm g}}\Sigma(r_{2\rm g})} \bigg) - \frac{2 a \Delta(r_{2\rm g})\sgn(\Lzrd)}{(r_{1\rm g} - r_{2\rm g})Y^2_{r \rm g}(r_{2\rm g})} \nonumber \\
 &- \frac{1}{(r_{1\rm g}- r_{2\rm g})Y^2_{r \rm g}(r_{2\rm g})} \displaystyle \sum^{3}_{i=1} \frac{\partial R_{\rm g}(r_{2\rm g})}{\partial C_{i\rm g}} C_{i s}\,. 
\end{align}
\end{widetext}
with $\Delta(r_{i \rm g}) = r^2_{i \rm g} -2 r_{\rm g} + a^2$ for $i =1,2$, and $\Sigma(r_{i \rm g}) = r^2_{i \rm g} + a^2 z^2_{\rm g}$ for $i = 1,2$.
The orthogonal component is the same for any choice of the fiducial geodesic, and it is given by
\begin{equation}
     r_{1{\rm s}\perp}(z_{\rm g},\psi_{\rm p}) = \frac{a z_{\rm g}\mathcal R(r_{1\rm g})\sqrt{K_{\rm g} -a^2 z^2_{\rm g}}}{\sqrt{K_{\rm g}}\sqrt{K_{\rm g} + r^2_{1 \rm g}}\Sigma(r_{1\rm g})} \cos \psi_{\rm p} \, ,
\end{equation}
while $r_{2{\rm s}\perp}(z_{\rm g},\psi_{\rm p})$ is obtained by substituting $r_{1\rm g} \to r_{2\rm g}$ in  $r_{1{\rm s}\perp}(z_{\rm g},\psi_{\rm p})$.

The polar shifts to the turning points are given by the following expressions:
\begin{align}
   z_{1 \rm s}(r_{\rm g},\psi_{\rm p}) &= s_\parallel z_{1{\rm s}\parallel}(r_{\rm g}) + s_\perp z_{1{\rm s}\perp}(r_{\rm g},\psi_{\rm p}) \,.
\end{align}
where
\begin{widetext}
\begin{align}
 z_{1{\rm s}\parallel}(r_{\rm g}) &= \frac{a\sqrt{K_{\rm g}}(1 - z^2_{1\rm g})\mathcal Z(z_{1\rm g})}{z_{1\rm g}(K_{\rm g} -a^2 z^2_{1\rm g})Y^2_{z\rm g}(z_{1\rm g})} - \frac{a z_{1\rm g}(K_{\rm g} + r^2_{\rm g})\mathcal Z(z_{1\rm g})}{\sqrt{K_{\rm g}}(K_{\rm g} -a^2 z^2_{1 \rm g})\Sigma(z_{1\rm g})} - \frac{a(1 - z^2_{1 \rm g})\sgn(\Lzrd)}{z_{1\rm g}Y^2_{z \rm g}(z_{1\rm g})} \nonumber \\
 &+\frac{1}{2 z_{1\rm g}Y^2_{z \rm g}(z_{1\rm g})} \displaystyle \sum^{3}_{i=1} \frac{\partial Z_{\rm g}(z_{1\rm g})}{\partial C_{i\rm g}}  C_{i\rm s}\,.
\end{align}
\end{widetext}
The orthogonal component is the same in `any parametrizations, and it is given by
\begin{equation}
     z_{1{\rm s}\perp}(r_{\rm g},\psi_{\rm p}) = -\frac{r_{\rm g} \mathcal Z(z_{1\rm g})\sqrt{K_{\rm g}+r^2_{\rm g}}}{\sqrt{K_{\rm g}}\Sigma(z_{1\rm g})\sqrt{K_{\rm g}- a^2 z^2_{1 \rm g}}} \cos \psi_{\rm p} \, .
\end{equation}
The shift $z_{2s \perp}$ is obtained by replacing $z_{1\rm g}\to -z_{1\rm g}$, which gives
\begin{equation}
     z_{2s \parallel} = -z_{1s \parallel} \qquad z_{2s \perp} = z_{1s \perp}
\end{equation}

\subsection{Averages to the turning points}
The averages of the turning points are the following
\begin{widetext}
\begin{align}
    \langle r_{1\rm s} \rangle &= \frac{\mathcal R(r_{1\rm g})\Big(r^2_{1 \rm g}\mathrm{K}(k_{z \rm g}) - (K_{\rm g} + r^2_{1 \rm g})\Uppi\big(\gamma(r_{1\rm g})|k_{z \rm g} \big)\Big)}{\sqrt{K_{\rm g}}r_{1\rm g}(K_{\rm g} + r^2_{1 \rm g})\mathrm{K}(k_{z \rm g})} + \frac{2 \Delta(r_{1\rm g})\Big(\sqrt{K_{\rm g}}\mathcal R(r_{1\rm g}) + a (K_{\rm g} + r^2_{1 \rm g}) \sgn(\Lzrd) \Big)}{(K_{\rm g}+r^2_{1 \rm g})(r_{1\rm g} - r_{2\rm g})Y^2_{r \rm g}(r_{1\rm g})}  \nonumber \\
    & + \frac{1}{(r_{1\rm g}- r_{2\rm g})Y^2_{r \rm g}(r_{1\rm g})} \displaystyle \sum^{3}_{i=1} \frac{\partial R_{\rm g}(r_{1\rm g})}{\partial C_{i\rm g}} C_{i s} \, , \\
    \langle r_{2 \rm s} \rangle &= \frac{\mathcal R(r_{2\rm g})\Big(r^2_{2 \rm g}\mathrm{K}(k_{z \rm g}) - (K_{\rm g} + r^2_{2 \rm g})\Uppi\big(\gamma(r_{2\rm g})|k_{z \rm g} \big)\Big)}{\sqrt{K_{\rm g}}r_{2\rm g}(K_{\rm g} + r^2_{2 \rm g})\mathrm{K}(k_{z \rm g})} - \frac{2 \Delta(r_{2\rm g})\Big(\sqrt{K_{\rm g}}\mathcal R(r_{2\rm g}) + a (K_{\rm g} + r^2_{2 \rm g}) \sgn(\Lzrd) \Big)}{(K_{\rm g}+r^2_{2 \rm g})(r_{1\rm g} - r_{2\rm g})Y^2_{r \rm g}(r_{2\rm g})}   \nonumber \\
    & - \frac{1}{(r_{1\rm g}- r_{2\rm g})Y^2_{r \rm g}(r_{2\rm g})} \displaystyle \sum^{3}_{i=1} \frac{\partial R_{\rm g}(r_{2\rm g})}{\partial C_{i\rm g}} C_{i s} \, , \\ 
\end{align}   
\begin{align}
    \langle z_{1 \rm s} \rangle &= \mathcal Z(z_{1\rm g})\frac{(r_{2\rm g} -r_{3\rm g})\Re\!\Big((i a^2 z^2_{1 \rm g} + a z_{1\rm g}(r_{2\rm g}+r_{3\rm g}) -i r_{2\rm g}r_{3\rm g}) \Uppi\Big(\frac{(r_{3\rm g} - i a z_{1\rm g})(r_{1\rm g}- r_{2\rm g})}{(r_{2\rm g} - i a z_{1\rm g})(r_{1\rm g}- r_{3\rm g})} | k_{r \rm g} \Big)  + \text{c.c.} \Big)}{2 \sqrt{K_{\rm g}} \mathrm{K}(k_{r \rm g}) (r^2_{3g} + a^2 z^2_{1 \rm g})(r^2_{2 \rm g} + a^2 z^2_{1 \rm g})} \nonumber \\
    & - \frac{a z_{1\rm g}\mathcal Z(z_{1\rm g})(r^2_{3g} + K_{\rm g})}{\sqrt{K_{\rm g}}(r^2_{3g} + a^2 z^2_{1 \rm g})(K_{\rm g}- a^2 z^2_{1 \rm g})} +\frac{a(1- z^2_{1 \rm g})\Big(\sqrt{K_{\rm g}}\mathcal Z(z_{1\rm g}) -(K_{\rm g} - a^2 z^2_{1 \rm g}) \sgn(\Lzrd) \Big)}{z_{1\rm g}(K_{\rm g} - a^2 z^2_{1 \rm g})Y^2_{z \rm g}(z_{1\rm g})} \nonumber \\
    & +\frac{1}{2 z_{1\rm g}Y^2_{z \rm g}(z_{1\rm g})} \displaystyle \sum^{3}_{i=1} \frac{\partial Z_{\rm g}(z_{1\rm g})}{\partial C_{i\rm g}}  C_{i\rm s} \, ,
\end{align}   
\end{widetext}
with $\text{c.c}$ denoting ``complex" conjugation. It is easy to check using the shifts~\eqref{eq:DH_constant_of_motion} that in the ``DH'' parametrization the averages $\langle r^\text{DH}_{1\rm s} \rangle = \langle r^\text{DH}_{2\rm s} \rangle = \langle z^\text{DH}_{1\rm s} \rangle =0$.
\section{Explicit expressions for the shifts to the radial and polar frequencies} \label{app:rad_pol_freq_shifts}

The corrections to the radial and polar frequencies can be computed as
\begin{equation}
   \Upsilon_{y \rm s} \equiv  \frac{\Upsilon_{y\rm g}}{(2\pi)^2} \int_{(0,2\pi]^2} \frac{\dd^2 w}{Y_{y \rm g}(w_y)} \bigg(  Y_{y{\rm s}\parallel}(\vec w) + y_{\rm s} \totder{Y_{y \rm g}}{y_{\rm g}} \bigg)  
\end{equation}
since the trajectories and velocities proportional to $s_\perp$ depend on either $\cos \psi_{\rm p}$ or $\sin \psi_{\rm p}$. Moreover, it was shown in Ref.~\cite{Witzany:2019nml} that the terms $\mathrm h_r$ and $\mathrm h_z$ average to zero
\begin{equation}
    \langle \mathrm h_r(\wmean) \rangle = \langle \mathrm h_z(\wmean) \rangle = 0
\end{equation}
Thus, it is convenient to compute the averages in terms of the anomalies 
\begin{align}
   \Upsilon_{r \rm s} & = \frac{\Upsilon^2_{r \rm g} \Upsilon_{z\rm g}}{(2\pi)^2} \int_{(0,2\pi]^2} \frac{1}{Y^2_{r \rm g}(\chi_r) Y_{z\rm g}(\chi_z)} \bigg( Y_{y{\rm s}\parallel}(\vec w) \nonumber \\
   &+ r_{{\rm s}\parallel} \totder{Y_{r\rm g}}{r_{\rm g}} \bigg)\dd^2 \chi \, ,
\end{align}
\begin{align}
   \Upsilon_{z\rm s} & = \frac{\Upsilon_{r\rm g} \Upsilon^2_{z \rm g}}{(2\pi)^2} \int_{(0,2\pi]^2} \frac{1}{Y_{r\rm g}(\chi_r) Y^2_{z \rm g}(\chi_z)} \bigg( Y_{y{\rm s}\parallel}(\vec w) \nonumber \\
   &+ z_{{\rm s}\parallel} \totder{Y_{z\rm g}}{z_{\rm g}} \bigg)\dd^2 \chi \, ,
\end{align}
where $\vec \chi = (\chi_r, \chi_z)$. The radial frequency shifts can be reduced to the following one-dimensional integrals of Legendre elliptic integrals.
\begin{align}
    \Upsilon_{r\rm s} &= \frac{\Upsilon^2_{r \rm g} \Upsilon_{z\rm g}}{(2\pi)^2}\int_{(0,2\pi]}\big( \mathcal Y^{(1)}_r(\chi_r) + \mathcal Y^{(2)}_r(\chi_r) \big)\dd \chi_r \nonumber \\
    &+ \frac{\Upsilon^2_{r \rm g}}{2\pi} \int_{(0,2\pi]}\mathcal Y^{(3)}_r(\chi_r)\dd \chi_r  \, ,
\end{align}
where
\begin{widetext}
\begin{align}
    \mathcal Y^{(1)}_r(\chi_r) &= \frac{4\mathcal R(r_{\rm g}) \big(\mathsf K(k_{z \rm g})r^2_{\rm g} - (K_{\rm R} + r^2_{\rm g}) \Uppi(\gamma(r_{\rm g})|k_{z \rm g}) \big) \big(K_{\rm g}(r_{\rm g}- 1) + 2 a E_{\rm g} \Lzrd r_{\rm g} + a^2 r_{\rm g} + r^2_{\rm g} (2(1 - E^2_{\rm g})r_{\rm g} -3) \big)}{\sqrt{K_{\rm g}}z_{2\rm g} r_{\rm g} (K_{\rm g} + r^2_{\rm g})(r_{1\rm g} - r_{\rm g})(r_{\rm g} - r_{2\rm g})Y^3_{r \rm g}(r_{\rm g})} \nonumber \\
    & + \frac{2\pi}{\Upsilon_{z\rm g}}\frac{1}{Y_{r\rm g}(r_{\rm g})}\bigg(\frac{\Delta \big(\sqrt{K_{\rm g}} \mathcal R(r_{\rm g}) + a (K_{\rm g} + r^2_{\rm g}) \sgn(\Lzrd)\big)}{(K_{\rm g} + r^2_{\rm g})(r_{1\rm g} - r_{\rm g})(r_{\rm g} - r_{2\rm g})Y^2_{r\rm g}(r_{\rm g})} - \frac{\langle r_{1 \rm s}\rangle}{2(r_{1\rm g} - r_{\rm g})} + \frac{\langle r_{2 \rm s}\rangle}{2(r_{\rm g} - r_{2\rm g})} \bigg) \nonumber \\
    & + \frac{2\pi}{\Upsilon_{z\rm g}}\frac{1}{2(r_{1\rm g} - r_{\rm g})(r_{\rm g} - r_{2\rm g})Y^3_{r\rm g}(r_{\rm g})} \displaystyle \sum^{3}_{i=1} \frac{\partial R_{\rm g}(r_{\rm g})}{\partial C_{i\rm g}} C_{i s} \, ,
\end{align}
\begin{align}
    \mathcal Y^{(2)}_r(\chi_r) &= -\frac{4E_{\rm g} (K_{\rm g} - r^2_{\rm g})\mathsf K(k_{z \rm g})}{\sqrt{K_{\rm g}}z_{2\rm g}(K_{\rm g} + r^2_{\rm g})Y_{r\rm g}(r_{\rm g})} + \frac{4\mathcal R(r_{\rm g})}{\sqrt{K_{\rm g}}z_{2\rm g}Y_{r\rm g}(r_{\rm g})} \bigg(\frac{1}{r^2_{\rm g}} \Uppi(\gamma(r_{\rm g})|k_{z \rm g}) - \frac{z^2_{2 \rm g} \mathsf E(k_{z \rm g})}{\Sigma(z_{1\rm g})\big((1 - E^2_{\rm g})r^2_{\rm g} + z^2_{2 \rm g}\big)}  \nonumber \\
    & + \frac{\mathsf K(k_{z \rm g})}{\Sigma(z_{1\rm g})} - \frac{(1 - E^2_{\rm g})\big(3r^2_{\rm g} + 2 a^2 z^2_{1 \rm g} \big)\Uppi(\gamma(r_{\rm g})|k_{z \rm g})}{\Sigma(z_{1\rm g})\big((1 - E^2_{\rm g})r^2_{\rm g} + z^2_{2 \rm g}\big)} - \frac{ (2 r_{\rm g}^2 + a^2 z_{1\rm g}^2) z_{2\rm g}^2\Uppi(\gamma(r_{\rm g})| k_{z \rm g})}{r^2_{\rm g} \Sigma(z_{1\rm g})\big((1 - E^2_{\rm g})r^2_{\rm g} + z^2_{2 \rm g}\big)}\bigg)
\end{align}
\end{widetext}

\begin{align}
    \mathcal Y^{(3)}_r(\chi_r) &= \frac{1}{Y^2_{r \rm g}(r_{\rm g})}\totder{Y_{r\rm g}}{r_{\rm g}} \bigg(\frac{\langle r_{1 \rm s} \rangle + \langle r_{2 \rm s} \rangle}{2} \nonumber \\
    & + \frac{\langle r_{1 \rm s} \rangle - \langle r_{2 \rm s} \rangle}{2} \sin \chi_r\bigg)
\end{align}

The polar frequency shifts can be reduced to the following one-dimensional integrals of Jacobi elliptic integrals.
\begin{align}
    \Upsilon_{z \rm s} &= \frac{\Upsilon^2_{z \rm g}}{2\pi}\int_{(0,2\pi]} \mathcal Y^{(1)}_z (\chi_z)\dd \chi_z \nonumber \\
    &+ \frac{\Upsilon^2_{z \rm g}\Upsilon_{r\rm g}}{(2\pi)^2}\int_{(0,2\pi]}\mathcal Y^{(2)}_z(\chi_r)\dd \chi_r + \frac{\Upsilon^2_{z \rm g}}{2\pi}\mathcal Y^{(3)}_z   \, ,
\end{align}
where 
\begin{widetext}
    \begin{align}
     \mathcal Y^{(1)}_z(\chi_z) &= -\frac{\langle z_{1\rm s} \rangle}{z_{1\rm g}\cos^2(\chi_z)Y_{z\rm g}(z_{\rm g})} 
     + \frac{a (1 - z^2_{\rm g})}{z^2_{1 \rm g}\cos^2(\chi_z)Y^3_{z \rm g}(z_{\rm g})} \bigg(\frac{\sqrt{K_{\rm g}}\mathcal Z(z_{\rm g})}{K_{\rm g} - a^2 z^2_{\rm g}} -\sgn(\Lzrd) \bigg) \nonumber \\
     & + \frac{\mathcal Z(z_{\rm g}) \big(a^2 (1- E^2_{\rm g})(2z^2_{\rm g} - z^2_{1 \rm g}) - z^2_{2 \rm g}\big)}{\sqrt{K_{\rm g}}(r^2_{2 \rm g} + a^2 z^2_{\rm g})(r^2_{3g} + a^2 z^2_{\rm g})}\frac{\sin(\chi_z)}{\cos^2(\chi_z)Y^3_{z \rm g}(z_{\rm g})}\bigg( a\frac{K_{\rm g} + r^2_{3g}}{K_{\rm g} - a^2 z^2_{\rm g}} (r^2_{2 \rm g} + a^2 z^2_{g})\sin(\chi_z) \nonumber \\
     & + \frac{(r_{2\rm g} - r_{3\rm g})}{2 z_{1\rm g}\mathsf K(k_{r \rm g})} \Re\!\bigg(\!\big(i r_{2\rm g}r_{3\rm g} - a(r_{2\rm g} - r_{3\rm g})z_{\rm g} - i a^2 z^2_{\rm g})\Uppi\bigg(\!\!\left.\frac{(r_{3\rm g} - i a z_{1\rm g})(r_{1\rm g} - r_{2\rm g})}{(r_{2\rm g} - i a z_{1\rm g})(r_{1\rm g}- r_{3\rm g})}\right \rvert k_{r \rm g} \bigg)  + \text{c.c.} \bigg) \bigg) \nonumber \\
     &  + \frac{1}{2z^2_{1\rm g}\cos^2(\chi_z)Y^3_{z\rm g}(z_{\rm g})} \displaystyle \sum^{3}_{i=1} \frac{\partial Z_{\rm g}(z_{\rm g})}{\partial C_{i\rm g}} C_{i s} \, ,
    \end{align}    
\begin{align}
   \mathcal Y^{(2)}_z(\chi_r) &= -\frac{8 E_{\rm g}}{\sqrt{K_{\rm g}}z_{2\rm g}Y_{r\rm g}(r_{\rm g})}\Uppi\bigg(\!\!\left.\frac{a^2 z^2_{1 \rm g}}{K_{\rm g}} \right \rvert k_{z \rm g} \!\bigg) + \frac{8 E_{\rm g} \Uppi(\gamma(r_{\rm g})|k_{z \rm g})}{\sqrt{K_{\rm g}}z_{2\rm g}Y_{r\rm g}(r_{\rm g})} + \frac{4\mathcal R(r_{\rm g})}{\sqrt{K_{\rm g}}z_{2\rm g}Y_{r\rm g}(r_{\rm g})}\bigg(\frac{1}{r^2_{\rm g}}\Uppi(\gamma(r_{\rm g})|k_{z \rm g}) + \frac{\mathsf K(k_{z \rm g})}{\Sigma(z_{1\rm g})}   \nonumber \\ 
    & - \frac{z^2_{2 \rm g}\mathsf E(k_{z\rm g})}{\Sigma(z_{1\rm g})\big((1 - E^2_{\rm g})r^2_{\rm g} + z^2_{2 \rm g}\big)}  - \frac{(1 - E^2_{\rm g})\big(3r^2_{\rm g} + 2 a^2 z^2_{1 \rm g} \big)\Uppi(\gamma(r_{\rm g})| k_{z \rm g})}{\Sigma(z_{1\rm g})\big((1 - E^2_{\rm g})r^2_{\rm g} + z^2_{2 \rm g}\big)}  - \frac{z^2_{2 \rm g}(2 r^2_{\rm g} + a^2 z^2_{1 \rm g})\Uppi(\gamma(r_{\rm g})| k_{z \rm g})}{r^2_{\rm g}\Sigma(z_{1\rm g})\big((1 - E^2_{\rm g})r^2_{\rm g} + z^2_{2 \rm g}\big)} \bigg)
\end{align}
\end{widetext}

\begin{align}
    \mathcal Y^{(3)}_z &= - 4\langle z_{1 \rm s} \rangle \frac{z^2_{2 \rm g} \mathsf E (k_{z \rm g}) - Y^2_{z \rm g}(z_{1\rm g}) \mathsf K(k_{z \rm g})}{z_{1\rm g} z_{2\rm g} Y^2_{z \rm g}(z_{1\rm g})}
\end{align}

\newpage
\section{Derivation of the shifts to the coordinate time and azimuthal frequency} \label{app:time_azi_freq_shifts}
\label{app:upstphi}

If a function $f(r, z)$ is additively separable in $r$ and $z$, i.e it can be written as
\begin{equation}
f(r, z) = f_{r\rm g}(r) + f_{z\rm g}(z) + q \delta f(\wmean)\,,
\end{equation}
then the average of $\langle \delta f(\wmean) \rangle$ can be computed without knowing the analytic expressions for the functions $\xi_r$ and $\xi_z$ as  shown in Ref. \cite{Witzany:2019nml}. We recast the derivation of this fact in the notation of our paper here.  

Using Eqs~\eqref{eq:spin_radial_trajectory},~\eqref{eq:spin_polar_trajectory} and~\eqref{eq:spin_differential},  we first notice that
\begin{align}
\langle \delta f(\wmean) \rangle &= \langle f_{\rm s}(\wmean) \rangle + \left\langle \totder{f_{r\rm g}}{r_{\rm g}}  r_{\rm s}(\wmean)\! \right\rangle + \left \langle \totder{f_{z\rm g}}{z_{\rm g}}z_{\rm s}(\wmean) \! \right \rangle  \nonumber \\
&- \left\langle \totder{f_{r\rm g}}{r_{\rm g}} \totder{r_{\rm g}}{\chi_r} \totder{\chi_r}{w_r} \xi_r (\wmean)\! \right\rangle  \nonumber \\
&- \left \langle \totder{f_{z\rm g}}{z_{\rm g}} \totder{z_{\rm g}}{\chi_z} \totder{\chi_z}{w_z} \xi_z(\wmean) \! \right \rangle + \displaystyle \sum^3_{i =1} \left\langle\frac{\partial f_{\rm g}}{\partial C_{i \rm g}} \right\rangle C_{i\rm s} \, ,
\end{align}
Using the chain rule, we get
\begin{align}
\langle \delta f(\wmean) \rangle &= \langle f_{\rm s}(\wmean) \rangle + \left\langle \totder{f_{r\rm g}}{r_{\rm g}}  r_{\rm s}(\wmean)\! \right\rangle + \left \langle \totder{f_{z\rm g}}{z_{\rm g}}z_{\rm s}(\wmean) \! \right \rangle  \nonumber \\
&- \left\langle \totder{f_{r\rm g}}{w_r} \xi_r (\wmean)\! \right\rangle - \left \langle \totder{f_{z\rm g}}{w_z} \xi_z(\wmean) \! \right \rangle \nonumber \\
& + \displaystyle \sum^3_{i =1} \left\langle\frac{\partial f_{\rm g}}{\partial C_{i \rm g}} \right\rangle C_{i\rm s}  \, ,
\end{align}
After integrating by parts in $w_r$ and $w_z$ and discarding boundary terms, $\langle \delta f(\wmean) \rangle$ becomes 
\begin{align}
\langle \delta f(\wmean) \rangle &= \langle f_{\rm s}(\wmean) \rangle + \left\langle \totder{f_{r\rm g}}{r_{\rm g}}  r_{\rm s}(\wmean)\! \right\rangle + \left \langle \totder{f_{z\rm g}}{z_{\rm g}}z_{\rm s}(\wmean) \! \right \rangle  \nonumber \\
&+ \left\langle f_{r\rm g}\frac{\partial \xi_r}{\partial w_r} \! \right\rangle + \left \langle f_{z\rm g} \frac{\partial\xi_z}{\partial w_z} \! \right \rangle + \displaystyle \sum^3_{i =1} \left\langle\frac{\partial f_{\rm g}}{\partial C_{i \rm g}} \right\rangle C_{i\rm s} \, ,
\end{align}
By construction, the shift vectors fulfill ( cf. also Chapter 6 in \cite{arnold2006celestial})
\begin{equation}
    \frac{\partial \xi_y}{\partial w_r} \Upsilon_{r\rm g} + \frac{\partial \xi_y}{\partial w_z} \Upsilon_{z\rm g} = \Upsilon_{y\rm g} \bigg( \!\left \langle \frac{1}{Y_{y \rm g}}\totder{\delta\chi_y}{\lambda} \right \rangle - \frac{1}{Y_{y \rm g}}\totder{\delta \chi_y}{\lambda}  \bigg)
\end{equation}
for $y =r,z$. Using integration by parts and the periodicity of $\xi_r$ and $\xi_z$, it is easy to show that
\begin{equation}
    \left \langle f_{r\rm g}(w_r) \frac{\partial \xi_r}{\partial w_z} \right \rangle = \left \langle f_{z\rm g}(w_z)\frac{\partial \xi_z}{\partial w_r} \right \rangle = 0
\end{equation}
Thus, we can finally write
\begin{align}
\langle \delta f(\wmean) \rangle &= \langle f_{\rm s}(\wmean) \rangle + \left\langle \totder{f_{r\rm g}}{r_{\rm g}} r_{\rm s}(\wmean)\! \right\rangle + \left \langle \totder{f_{z\rm g}}{z_{\rm g}} z_{\rm s}(\wmean) \! \right \rangle  \nonumber \\
&+\left \langle \frac{1}{Y_{r\rm g}}\totder{\delta\chi_r}{\lambda} \right \rangle \langle f_{r\rm g}\rangle - \left \langle \frac{f_{r\rm g}}{Y_{r\rm g}}\totder{\delta\chi_r}{\lambda} \right \rangle  \nonumber \\
&+\left \langle \frac{1}{Y_{z\rm g}}\totder{\delta\chi_z}{\lambda} \right \rangle \langle f_{z\rm g}\rangle - \left \langle \frac{f_{z\rm g}}{Y_{z\rm g}}\totder{\delta\chi_z}{\lambda} \right \rangle \nonumber \\
& + \displaystyle \sum^3_{i =1} \left\langle\frac{\partial f_{\rm g}}{\partial C_{i \rm g}} \right\rangle C_{i\rm s} \, , \label{eq:separable_avg}
\end{align}
Notice that $\langle \delta f(\wmean) \rangle$ does not require an explicit knowledge of $\vec \xi$. Armed with Eq.~\eqref{eq:separable_avg}, we can compute $ \Upsilon_{t \rm s}$ and $ \Upsilon_{\phi \rm s}$ as
\begin{align}
  \Upsilon_{t\rm s} &= \Upsilon_{r\rm s}\frac{\Upsilon^t_{r\rm g}}{\Upsilon_{r\rm g}} - \left \langle\frac{\delta Y_r}{Y_{r\rm g}}V^t_{r\rm g} \right \rangle + \left \langle \frac{\partial V^t_{r\rm g}}{\partial r_{\rm g}} r_{\rm s} \! \right \rangle  \nonumber\\
                & + \Upsilon_{z\rm s}\frac{\Upsilon^t_{z\rm g}}{\Upsilon_{z\rm g}} - \left \langle\frac{\delta Y_z}{Y_{z\rm g}} V^t_{z\rm g} \!\right \rangle + \left \langle\frac{\partial V^t_{z\rm g}}{\partial z_{\rm g}} z_{\rm s} \!\right \rangle  \nonumber\\
                & + \left\langle\frac{\Sigma}{2} \Upgamma^t_\parallel \right\rangle + \displaystyle \sum^2_{i =1} \left(\left \langle\frac{\partial V^t_{r\rm g}}{\partial C_{i \rm g}} \right \rangle + \left \langle \frac{\partial V^t_{z\rm g}}{\partial C_{i \rm g}} \right \rangle \right) \! C_{i\rm s} \, ,\\
  \Upsilon_{\phi\rm s} &= \Upsilon_{r\rm s}\frac{\Upsilon^\phi_{r\rm g}}{\Upsilon_{r\rm g}} - \left \langle\frac{\delta Y_r}{Y_{r\rm g}}V^\phi_{r\rm g} \right \rangle + \left \langle\frac{\partial V^\phi_{r\rm g}}{\partial r_{\rm g}} r_{\rm s} \! \right \rangle   \nonumber\\
                & + \Upsilon_{z\rm s}\frac{\Upsilon^\phi_{z\rm g}}{\Upsilon_{z\rm g}} - \left \langle\frac{\delta Y_z}{Y_{z\rm g}} V^\phi_{z\rm g} \right \rangle + \left \langle \frac{\partial V^\phi_{z\rm g}}{\partial z_{\rm g}} z_{\rm s} \! \right \rangle \nonumber\\
                & + \left\langle\frac{\Sigma}{2} \Upgamma^\phi_\parallel \right\rangle + \displaystyle \sum^2_{i =1} \left(\left \langle\frac{\partial V^\phi_{r\rm g}}{\partial C_{i \rm g}} \right \rangle + \left \langle \frac{\partial V^\phi_{z\rm g}}{\partial C_{i \rm g}} \right \rangle \right)\! C_{i\rm s}\, .
\end{align}
These results agree with Eqs. (57) and (58) of~\cite{Witzany:2019nml}. The explicit expressions for the frequency shifts are included in a companion notebook.
We notice that the averages  $\langle \Sigma/2 \Upgamma^t_\parallel \rangle$ and $\langle \Sigma/2 \Upgamma^\phi_\parallel \rangle$ are the same  in both the ``FC'' and ``DH'' parametrizations, whereas
\begin{align}
   \left \langle \frac{\partial V^t_{r\rm g}}{\partial r_{\rm g}} r^{\text{DH}}_{\rm s} \! \right \rangle = \left \langle \frac{\partial V^t_{z\rm g}}{\partial z_{\rm g}} z^{\text{DH}}_{\rm s} \! \right \rangle = 0 \, , \\
   \left \langle \frac{\partial V^\phi_{r\rm g}}{\partial r_{\rm g}} r^{\text{DH}}_{\rm s} \! \right \rangle = \left \langle \frac{\partial V^\phi_{z\rm g}}{\partial z_{\rm g}} z^{\text{DH}}_{\rm s} \! \right \rangle = 0 \, . 
\end{align}

\section{Transformations between orbital corrections in the DH and FC parametrizations} \label{app:DHmapFC}
Thanks to Eq.~\eqref{eq:DH_constant_of_motion}, we can map the radial and polar frequency shifts in the DH parametrization with their ``FC'' counterparts in the following way:
\begin{equation}    
\Upsilon^\text{DH}_{y \rm s} = \Upsilon^\text{FC}_{y \rm s} - \frac{\partial \Upsilon_{y\rm g}}{\partial r_{1\rm g}} \langle r_{1\rm s} \rangle - \frac{\partial \Upsilon_{y\rm g}}{\partial r_{2\rm g}}\langle r_{2\rm s} \rangle - \frac{\partial \Upsilon_{y\rm g}}{ \partial z_{1\rm g}}\langle z_{1\rm s}\rangle \\
\end{equation}
for $y = r,z$. Te maps between the frequency shifts $\Upsilon^\text{FC}_{t\rm s}$ and $\Upsilon^\text{FC}_{\phi\rm s}$, and the corresponding expressions obtained with the DH method are given as
\begin{align}    
\Upsilon^\text{DH}_{t\rm s} &= \Upsilon^\text{FC}_{t\rm s} - \delta \Upsilon^t_r - \delta \Upsilon^t_z \, , \\
\Upsilon^\text{DH}_{\phi\rm s} &= \Upsilon^\text{FC}_{\phi\rm s} - \delta \Upsilon^\phi_r - \delta \Upsilon^\phi_z \, , 
\end{align}
where constants shifts $\delta \Upsilon^t_r, \delta \Upsilon^t_z$ and $\delta \Upsilon^\phi_r, \delta \Upsilon^\phi_z$ are defined as
\begin{align}
    &\delta \Upsilon^t_r = \frac{\partial \Upsilon^t_{r\rm g}}{\partial r_{1\rm g}} \langle r^\text{FC}_{1\rm s} \rangle + \frac{\partial\Upsilon^t_{r\rm g}}{\partial r_{2\rm g}}\langle r^\text{FC}_{2\rm s} \rangle + \frac{\partial\Upsilon^t_{r\rm g}}{\partial z_{1\rm g}}\langle z^\text{FC}_{1\rm s}\rangle \, , \\
    &\delta \Upsilon^t_z = \frac{\partial\Upsilon^t_{z\rm g}}{\partial r_{1\rm g}} \langle r^\text{FC}_{1\rm s} \rangle + \frac{\partial\Upsilon^t_{z\rm g}}{\partial r_{2\rm g}}\langle r^\text{FC}_{2\rm s} \rangle + \frac{\partial\Upsilon^t_{z\rm g}}{\partial z_{1\rm g}}\langle z^\text{FC}_{1\rm s}\rangle \, .
\end{align}
while the analogous terms for $\delta \Upsilon^\phi_r$ and $\delta \Upsilon^\phi_z$ are defined by substituting $t \to \phi$ in the previous equations. The expressions for the geodesic frequencies terms $\Upsilon^t_{r\rm g}$, $\Upsilon^t_{z\rm g}$, $\Upsilon^\phi_{r\rm g}$ and $\Upsilon^\phi_{z\rm g}$ are given in Appendix~\ref{app:geo_freq}.

Moreover, the radial and polar spin corrections to the trajectories in the DH method are related to $\delta r(\wmean)$ and $\delta z(\wmean)$ by
\begin{align}
    \delta r^\text{DH}_\parallel &=\delta r^\text{FC}_\parallel + \eta_r(r_{\rm g})\, , \\
    \delta z^\text{DH}_\parallel &=\delta z^\text{FC}_\parallel + \eta_z(z_{\rm g}) \, ,
\end{align}
with shift functions $\eta_r(r_{\rm g})$ and $\eta_z(z_{\rm g})$ defined as
\begin{align}
   & \eta_r(r_{\rm g}) = - \frac{\partial r_{\rm g}}{\partial r_{1\rm g}}\langle r^\text{FC}_{1\rm s}\rangle -\frac{\partial r_{\rm g}}{\partial r_{2\rm g}}\langle r^\text{FC}_{2\rm s} \rangle +\displaystyle \sum^3_{i =1} \frac{\partial r_{\rm g}}{\partial C_{i\rm g}} C^\text{DH}_{i\rm s} \, , \\
   & \eta_z(z_{\rm g}) = - \frac{\partial z_{\rm g}}{\partial z_{1\rm g}}\langle z^\text{FC}_{1\rm s}\rangle + \displaystyle \sum^3_{i =1}\frac{\partial z_{\rm g}}{\partial C_{i\rm g}} C^\text{DH}_{i\rm s} \, ,
\end{align}
Similar relations apply for the radial and polar velocity corrections
\begin{align}
   \bigg(\totder{\delta r^\text{DH}_\parallel}{\lambda}\bigg) &= \bigg(\totder{\delta r^\text{FC}_\parallel}{\lambda}\bigg) + \eta_r(r_{\rm g})\frac{\partial R_{\rm g}}{\partial r_{\rm g}} + \displaystyle \sum^3_{i =1} \frac{\partial R_{\rm g}}{\partial C_{i\rm g}} C^\text{DH}_{i\rm s} \, ,\\
    \bigg(\totder{\delta z^\text{DH}_\parallel}{\lambda}\bigg) &= \bigg(\totder{\delta z^\text{FC}_\parallel}{\lambda}\bigg) + \eta_z(z_{\rm g})\frac{\partial Z_{\rm g}}{\partial z_{\rm g}} + \displaystyle \sum^3_{i =1} \frac{\partial Z_{\rm g}}{\partial C_{i\rm g}} C^{\text{DH}}_{i\rm s} \, ,
\end{align}
and for the coordinate time and azimuthal velocity corrections
\begin{equation}
   \bigg(\totder{\delta t^\text{DH}_\parallel}{\lambda}\bigg) = \bigg(\totder{\delta t^\text{FC}_\parallel}{\lambda}\bigg) + \eta^t_{r}(r_{\rm g}) + \eta^t_{z}(z_{\rm g})  \, ,
\end{equation}
with the functions $ \eta^t_{r}(r_{\rm g})$ and  $\eta^t_{z}(z_{\rm g})$ given by 
\begin{align}
    \eta^t_{r}(r_{\rm g}) &= \eta_r(r_{\rm g})\frac{\partial V^t_{r\rm g}}{\partial r_{\rm g}} + \displaystyle \sum^3_{i =1} \frac{\partial V^t_{r \rm g}}{\partial C_{i\rm g}} C^\text{DH}_{i\rm s} \, , \\
    \eta^t_{z}(z_{\rm g}) &= \eta_z(z_{\rm g})\frac{\partial V^t_{z\rm g}}{\partial z_{\rm g}} + \displaystyle \sum^3_{i =1} \frac{\partial V^t_{z\rm g}}{\partial C_{i\rm g}} C^\text{DH}_{i\rm s} \, . 
\end{align}
The equivalent map between $\dd \delta\phi^\text{DH}_\parallel/\dd\lambda$ and  $\dd \delta \phi^\text{FC}_\parallel/\dd\lambda$ can be obtained by substituting $t \to \phi$ in the previous expressions.

Finally, the following relations between the purely oscillatory terms $\osc(\delta t^\text{DH}_\parallel)$ and $\osc(\delta\phi^\text{DH}_\parallel)$, and their respective counterparts $\osc(\delta t^\text{FC}_\parallel)$ and $\osc(\delta\phi^\text{FC}_\parallel)$ hold
\begin{widetext}
\begin{align}
    \osc(\delta t^\text{DH}_\parallel) & = \osc(\delta t^\text{FC}_\parallel) + \sum_{n \neq 0} \frac{ie^{-i n w_r}}{n \Upsilon_{r\rm g}}\Big((\eta^t_r)_n +\frac{\delta \Upsilon^t_r}{\Upsilon_{r\rm g}}(V^t_{r\rm g})_n \Big) + \sum_{k \neq 0} \frac{ie^{-i k w_z}}{k \Upsilon_{z\rm g}}\Big((\eta^t_z)_k +\frac{\delta \Upsilon^t_z}{\Upsilon_{z\rm g}} (V^t_{z\rm g})_k\Big)  \, ,  \\
    \osc(\delta \phi^\text{DH}_\parallel) &= \osc(\delta \phi^\text{FC}_\parallel) + \sum_{n \neq 0} \frac{ie^{-i n w_r}}{n \Upsilon_{r\rm g}}\Big((\eta^t_r)_n +\frac{\delta \Upsilon^\phi_r}{\Upsilon_{r\rm g}}(V^\phi_{r\rm g})_n \Big) + \sum_{k \neq 0} \frac{ie^{-i k w_z}}{k \Upsilon_{z\rm g}}\Big((\eta^t_z)_k +\frac{\delta \Upsilon^\phi_z}{\Upsilon_{z\rm g}} (V^\phi_{z\rm g})_k\Big) \, .
\end{align}
\end{widetext}

\section{Source term} \label{app:source_term}
This Appendix presents the explicit expressions for the coefficients $A_{i\rm g}$,  $A_{i{\rm s}\parallel}$, $A_{i{\rm s}\perp}$, $B_{i+1{\rm s}\parallel}$, $B_{i+1{\rm s}\perp}$ for $i=0,1,2$ that appears in partial amplitudes and their spin corrections shown in Section~\ref{sec:asymptotic_amplitudes}. These coefficients contain the projection of the stress-energy tensor $T^{\mu \nu}$ 
into Kinnersley tetrad $\big(\lambda^\mu_{~0},\lambda^\mu_{~1}, \lambda^\mu_{~2}, \lambda^\mu_{~3}\big) = (l^\mu, n^\mu, m^\mu, \bar m^\mu)$:
\begin{align}
      l^\mu &= \Big(\frac{r^2+a^2}{\Delta},1,0,\frac{a}{\Delta}\Big) \, , \\
      n^\mu &= \frac{1}{2\Sigma}(r^2+a^2,-\Delta,0,a) \, , \\
      m^\mu &= \frac{\sqrt{1-z^2}}{\sqrt{2}\bar\zeta}\Big(ia,0,-1,\frac{i}{1-z^2}\Big) \, , \\
 \bar m^\mu &= \frac{\sqrt{1-z^2}}{\sqrt{2}\zeta}\Big(-ia,0,-1,-\frac{i}{1-z^2} \Big) \, ,
\end{align}
with $\zeta = r - i a z$, and $\bar \zeta = r + i a z$. In  particular, the projection of the geodesic proper-time 4-velocities $ v^\mu_{\rm g}/\Sigma$ over the Kinnerlsey tetrad reads
\begin{align}
      v^{\rm g}_l &= \frac{v^r_{\rm g}}{\Delta} - \frac{v^t_{\rm g} - a(1 - z^2_{\rm g})v^\phi_{\rm g}}{\Sigma}  \, , \\
      v^{\rm g}_n &= -\frac{\Delta}{2\Sigma^2} \bigg( v^t_{\rm g} - a (1-z^2_{\rm g}) v^\phi_{\rm g} +\frac{\Sigma}{\Delta} v^r_{\rm g}\bigg)  \, , \\
      v^{\rm g}_m &= -\frac{\sqrt{1-z^2_{\rm g}}}{\sqrt{2}\Sigma}\bigg( \frac{\zeta v^z_{\rm g}}{1-z^2_{\rm g}} + \frac{i}{\bar \zeta}\big( a v^t_{\rm g} - (r^2_{\rm g} + a^2) v^\phi_{\rm g} \big) \bigg) \, , \\
     v^{\rm g}_{\bar m} &= -\frac{\sqrt{1-z^2_{\rm g}}}{\sqrt{2}\Sigma}\bigg( \frac{\bar\zeta v^z_{\rm g}}{1-z^2_{\rm g}} - \frac{i}{\zeta}\big( a v^t_{\rm g} - (r^2_{\rm g} + a^2) v^\phi_{\rm g} \big) \bigg) \, , 
\end{align}

\subsection{Auxiliary functions proportional to $s_\parallel$} \label{app:auxiliary_func_parallel}
\subsubsection{Projections of the velocities} 
We define the following projections
\begin{align}
    v^{\rm s \parallel}_l &= \frac{\delta v^r_\parallel}{\Delta} - \frac{\delta v^t_\parallel- a(1 - z^2_{\rm g})\delta v^\phi_\parallel}{\Sigma}  \, , \\
    v^{\rm s \parallel}_n &= -\frac{\Delta}{2\Sigma^2} \bigg(\delta v^t_\parallel - a (1-z^2_{\rm g}) \delta v^\phi_\parallel +\frac{\Sigma}{\Delta} \delta v^r_\parallel\bigg)  \, , \\
    v^{\rm s \parallel}_m &= -\frac{\sqrt{1-z^2_{\rm g}}}{\sqrt{2}\Sigma}\bigg( \frac{\zeta \delta v^z_\parallel}{1-z^2_{\rm g}} + \frac{i}{\bar \zeta}\big( a \delta v^t - (r^2_{\rm g} + a^2) \delta v^\phi_\parallel \big) \bigg) \, , \\
    v^{\rm s \parallel}_{\bar m} &= -\frac{\sqrt{1-z^2_{\rm g}}}{\sqrt{2}\Sigma}\bigg( \frac{\bar\zeta \delta v^z_\parallel}{1-z^2_{\rm g}} - \frac{i}{\zeta}\big( a \delta v^t_\parallel - (r^2_{\rm g} + a^2) \delta v^\phi_\parallel \big) \bigg) \, , 
\end{align}
where $\delta v^t_\parallel$, $\delta v^r_\parallel$, $\delta v^z_\parallel$ and $\delta v^\phi_\parallel$ are the Mino-time velocities corrections given by Eqs.~\eqref{eq:1st-order-linMPDeq} due to $s_\parallel$.

\subsubsection{Projections of the spin-tensor}
We also use the following projections of the spin-tensor $S^{\mu \nu}$ over the Kinnersley tetrad
\begin{align}
    S_{l n \parallel} &= - \frac{r_{\rm g}(K_{\rm g} - a^2 z^2_{\rm g})}{\sqrt{K_{\rm g}} \Sigma} \, , \qquad S_{m \bar m \parallel} = \frac{i a z_{\rm g}(K_{\rm g} + r^2_{\rm g})}{\sqrt{K_{\rm g}} \Sigma} \, , \\
    S_{n m \parallel} &= \frac{\zeta}{\sqrt{K_{\rm g}}} v^{\rm g}_m v^{\rm g}_n \, ,\qquad S_{l \bar m \parallel} = -\frac{\zeta}{\sqrt{K_{\rm g}}} v^{\rm g}_l v^{\rm g}_{\bar m} \, ,
\end{align}
and the following partial projections:
\begin{align}
    i(\omega S^t_{~n \parallel} - m S^\phi_{~n \parallel}) &= \frac{a \omega (1 - z^2_{\rm g}) -m}{\sqrt{2(1-z^2_{\rm g})}\Sigma}\big(\zeta S_{n \bar m \parallel} - \bar \zeta S_{n m \parallel} \big) \nonumber \\
    &-\frac{i \mathcal K(r_{\rm g})}{2 \Sigma} S_{l n \parallel} \, , \\
    i(\omega S^t_{~\bar m \parallel} - m S^\phi_{~\bar m \parallel}) &= -i \mathcal K(r_{\rm g})\bigg(\frac{S_{n \bar m \parallel}}{\Delta} + \frac{S_{l \bar m \parallel}}{2\Sigma} \bigg) \nonumber \\
    & + \frac{a \omega (1 - z^2_{\rm g}) - m}{\sqrt{2(1 - z^2_{\rm g}) \zeta}} S_{m \bar m \parallel} \, .
\end{align}
\begin{align}
    S^r_{~n \parallel} &= \frac{\Delta}{2 \Sigma} S_{l n \parallel} \, , \\
    S^r_{~\bar m \parallel} &= \frac{\Delta}{2 \Sigma} S_{l \bar m \parallel}  - S_{n \bar m\parallel} \, , \\
    S^z_{~n \parallel} &= \frac{\sqrt{1- z^2_{\rm g}}}{\sqrt{2}\Sigma} \big( S_{n \bar m \parallel} \zeta + S_{n m \parallel} \bar \zeta \big) \, , \\
    S^z_{~\bar m \parallel} &= - \frac{\sqrt{1 - z^2_{\rm g}}}{\sqrt{2}\zeta} S_{m \bar m \parallel} \, , 
\end{align}

Finally, we define the following terms
\begin{align}
    C^{\text{con}}_{n n \parallel} &= S^{cd}v^{\rm g}_{n}\gamma_{ndc} + S^c_{~n}\gamma_{ndc}v^d_{\rm g} \, , \label{eq:spin_tensor_conn_nn}\\
    C^{\text{con}}_{n \bar m \parallel} &= S^{cd}v^{\rm g}_{(\bar m}\gamma_{n)dc} + S^c_{~~(n}\gamma_{\bar m)dc}v^d_{\rm g} \, , \label{eq:spin_tensor_conn_nmb}\\
    C^{\text{con}}_{\bar m \bar m \parallel} &= S^{cd}v^{\rm g}_{\bar m}\gamma_{\bar mdc} + S^{c}_{~\bar m}\gamma_{\bar mdc}v^{d}_{\rm g} \, , \label{eq:spin_tensor_conn_mbmb}
\end{align}
where the Ricci rotation coefficients are defined as
\begin{equation}
    \gamma_{adc} = \lambda_{a\mu;\rho} \lambda^\mu_{~d} \lambda^\nu_{~c} 
\end{equation}
while the terms appearing in Eqs.~\eqref{eq:spin_tensor_conn_nn}, ~\eqref{eq:spin_tensor_conn_nmb},~\eqref{eq:spin_tensor_conn_mbmb} are given in Eqs. (B1a), (B1b), (B1c), (B1d), (B1e) of Appendix B of Ref.~\cite{Skoupy:2023lih}. 

\subsubsection{Additional auxiliary functions}
We present here the last set of auxiliary functions need to define the source term, which are given in terms of the previous shown functions:
\begin{align}
    A^{\rm s \parallel}_{n n} &= 2 v^{\rm g}_n v^{\rm s \parallel}_n + C^{\text{con}}_{n n \parallel} \, , \\
    A^{\rm s \parallel}_{n \bar m} &=  v^{\rm s \parallel}_n v^{\rm g}_{\bar m} + v^{\rm s \parallel}_{\bar m} v^{\rm g}_n + C^{\text{con}}_{n \bar m \parallel} \, , \\
    A^{\rm s \parallel}_{\bar m \bar m} &= 2 v^{\rm g}_{\bar m} v^{\rm s \parallel}_{\bar m} + C^{\text{con}}_{\bar m \bar m \parallel} \, , 
\end{align}
\begin{align}
    A^r_{n n \parallel} &= S^r_{~n \parallel} v^{\rm g}_n \, , \\
    A^r_{n \bar m \parallel} &= \frac{1}{2}\Big(S^r_{~n \parallel} v^{\rm g}_{\bar m} + S^r_{~\bar m \parallel} v^{\rm g}_n  \Big) \, , \\
    A^r_{\bar m \bar m \parallel} &= S^r_{~n \parallel} v^{\rm g}_{\bar m} v^{\rm g}_{\bar m} \, , \\
\end{align}
\begin{align}
    A^z_{n n \parallel} &= S^z_{~n \parallel} v^{\rm g}_n \, , \\
    A^z_{n \bar m \parallel} &= \frac{1}{2}\Big(S^z_{~n \parallel} v^{\rm g}_{\bar m} + S^z_{~\bar m \parallel} v^{\rm g}_n  \Big) \, , \\
    A^z_{\bar m \bar m \parallel} &= S^z_{~n \parallel} v^{\rm g}_{\bar m} v^{\rm g}_{\bar m} \, , 
\end{align}
\begin{align}
    A^t_{n n \parallel} + A^\phi_{n n \parallel} &= i\big( \omega S^t_{~n \parallel} - m S^\phi_{~n \parallel} \big) v^{\rm g}_n \, , \\
    A^t_{n \bar m \parallel} + A^\phi_{n \bar m \parallel} &= \frac{i}{2}\big( \omega S^t_{~n \parallel} - m S^\phi_{~n \parallel} \big)v^{\rm g}_{\bar m}   \nonumber \\
    & + \frac{i}{2}\big( \omega S^t_{~\bar m \parallel} - m S^\phi_{~\bar m \parallel}\big) v^{\rm g}_n  \, , \\
    A^t_{\bar m \bar m \parallel} + A^\phi_{\bar m \bar m \parallel} &= i\big( i\omega S^t_{~\bar m \parallel} - m S^\phi_{~\bar m \parallel} \big) \, . 
\end{align}

\subsection{Auxiliary functions proportional to $s_\perp$}
We first present the following complete projections of the spin tensor onto the Kinnerlsey tetrad in the $s_\perp$ sector.
\begin{align}
    S_{l n \perp} &= a z_{\rm g} \frac{\sqrt{K_{\rm g} + r^2_{\rm g}}\sqrt{K_{\rm g} - a^2 z^2_{\rm g}}}{\sqrt{K_{\rm g}}\Sigma} \cos(\psi_p) \, , \\
    S_{m \bar m \perp} &= i r_{\rm g} \frac{\sqrt{K_{\rm g} + r^2_{\rm g}}\sqrt{K_{\rm g} - a^2 z^2_{\rm g}}}{\sqrt{K_{\rm g}}\Sigma}\cos(\psi_p) \, , \\
    S_{n m \perp} &= \zeta v^{\rm g}_m v^{\rm g}_n \frac{(K_{\rm g} - i a z_{\rm g} r_{\rm g})\cos(\psi_p) + \sqrt{K_{\rm g}}\bar \zeta \sin(\psi_p)}{i\sqrt{K_{\rm g}}\sqrt{K_{\rm g} + r^2_{\rm g}}\sqrt{K_{\rm g} - a^2 z^2_{\rm g}}} \, ,\\ 
   S_{l \bar m \perp} &= i \zeta v^{\rm g}_l v^{\rm g}_{\bar m}\frac{(K_{\rm g} - i a z_{\rm g} r_{\rm g})\cos(\psi_p) - \sqrt{K_{\rm g}}\bar \zeta \sin(\psi_p)}{\sqrt{K_{\rm g}}\sqrt{K_{\rm g} + r^2_{\rm g}}\sqrt{K_{\rm g} - a^2 z^2_{\rm g}}} \, ,
\end{align}
All the remaining auxiliary functions proportional to $s_\perp$ can be obtained by simply replacing $s_\parallel$ with $s_\perp$ in the expressions shown in Appendix~\ref{app:auxiliary_func_parallel}.

\subsection{Leading order coefficients}
We first define the following auxiliary functions
\begin{align}
   C^{\rm g}_{\bar m \bar m} &= (v^{\rm g}_{\bar m})^2 \, , \,\quad C^{\rm g}_{n\bar m} = v^{\rm g}_n v^{\rm g}_{\bar m} \, , \,\quad  C^{\rm g}_{nn} = (v^{\rm g}_n)^2 \, . 
\end{align}
The coefficients $A^{(i)}_{\rm g}$ for $i = 1,2,3$ are given by
\begin{align}
   A_{2\rm g} &= C^{\rm g}_{\bar m \bar m} f^{(2)}_{\bar m \bar m} \, , \\
   A_{1\rm g} &= C^{\rm g}_{\bar m \bar m} f^{(1)}_{\bar m \bar m} + C^{\rm g}_{n \bar m} f^{(1)}_{n\bar m} \, , \\
   A_{0\rm g} &= C^{\rm g}_{\bar m \bar m} f^{(0)}_{\bar m \bar m} + C^{\rm g}_{\bar m n} f^{(0)}_{\bar m n} + C^{\rm g}_{nn} f^{(0)}_{nn}  \, , 
\end{align}
where
\begin{align}
    f^{(2)\rm g}_{\bar m \bar m} &= -\frac{\zeta^2}{\bar \zeta^2} S^{\rm g}_{\ell m \vec \kappa}(z) \, , \\
    f^{(1) \rm g}_{\bar m \bar m} &= -\frac{2 \zeta^2}{\bar\zeta^2}\bigg(\frac{1}{\zeta} + i \frac{\mathcal K(r_{\rm g})}{\Delta} \bigg) S^{\rm g}_{\ell m \vec \kappa}(z) \, , \\
    f^{(0) \rm g}_{\bar m \bar m} &=  \frac{\zeta^2}{\bar \zeta^2} \bigg(i \partial_r \bigg(\frac{\mathcal K(r_{\rm g})}{\Delta}\bigg) -\frac{2i}{\zeta} \frac{\mathcal K(\hat r_{\rm g})}{\Delta} \bigg)S^{\rm g}_{\ell m \vec \kappa}(z) \nonumber \\
    & + \frac{\zeta^2}{\bar \zeta^2} \bigg(\frac{\mathcal K(r_{\rm g})}{\Delta}\bigg)^{\!\! 2} S^{\rm g}_{\ell m \vec \kappa}(z) \, , 
\end{align}
\begin{align}
    f^{(1)\rm g}_{n\bar m} &= \frac{2 \sqrt{2} \zeta^2}{\bar \zeta \Delta}\bigg( \mathcal L^\dagger_2 +  a \sqrt{1 - z^2_{\rm g}}\frac{2 a z_{\rm g}}{\Sigma}\bigg) S^{\rm g}_{\ell m \vec \kappa}(z) \, , \\
    f^{(0)\rm g}_{n\bar m} &= \frac{2\sqrt{2}\zeta^2}{\bar \zeta \Delta}\bigg(\frac{i \mathcal K(r_{\rm g})}{\Delta} + \frac{2 r_{\rm g}}{\Sigma} \bigg) \mathcal L^\dagger_2 S^{\rm g}_{\ell m \vec \kappa}(z) \nonumber \\
    & + \frac{2\sqrt{2}\zeta^2}{\bar \zeta \Delta^2} a \sqrt{1 - z^2_{\rm g}} \mathcal K(r_{\rm g}) \frac{2 i a z_{\rm g}}{\Sigma} S^{\rm g}_{\ell m \vec \kappa}(z) \, , \\
    f^{(0)\rm g}_{nn} &= -\frac{2 \zeta^2}{\Delta^2}\bigg( \mathcal L^\dagger_1 \mathcal L^\dagger_2 - \frac{2 i a}{\zeta}\sqrt{1 - z^2_{\rm g}} \mathcal L^\dagger_2 \bigg)S^{\rm g}_{\ell m \vec \kappa}(z)  \, , 
\end{align}
with
\begin{align}
    \mathcal L^\dagger_2 &= - \sqrt{1 - z^2_{\rm g}} \bigg(\partial_z - \frac{m -2 z_{\rm g}}{1 - z^2_{\rm g}} + a \omega \bigg) \, , \\
    \mathcal L^\dagger_1 &= - \sqrt{1 - z^2_{\rm g}} \bigg(\partial_z - \frac{m - z_{\rm g}}{1 - z^2_{\rm g}} + a \omega \bigg) \, ,
\end{align}

\subsection{Coefficients proportional to $s_\parallel$}
The coefficients $B_{3\rm s \parallel}$, $B_{2\rm s \parallel}$ and $B_{1\rm s \parallel}$ are given by
\begin{align}
   B_{3\rm s \parallel} &= - A^r_{\bar m \bar m\parallel} f^{(2)\rm g}_{\bar m \bar m} \, , \\  
   B_{2\rm s \parallel} &= - A^r_{n \bar m \parallel} f^{(1)\rm g}_{n \bar m} - A^r_{\bar m \bar m \parallel} f^{(1)\rm g}_{n \bar m} \, , \\
   B_{1\rm s \parallel} &= - A^r_{\bar m \bar m \parallel} f^{(0)\rm g}_{n \bar m} - A^r_{n \bar m \parallel} f^{(0)\rm g}_{n \bar m} - A^r_{n n \parallel} f^{(0) \rm g}_{n n}  \, , 
\end{align}
while coefficients $A_{2\rm s \parallel}$, $A_{1\rm s \parallel}$ and $A_{0\rm s \parallel}$ are given by
\begin{widetext}
\begin{align}
   A_{2\rm s \parallel} &= A^r_{\bar m \bar m \parallel} \partial_r f^{(2)\rm g}_{\bar m \bar m}  + A^z_{\bar m \bar m \parallel} \partial_z f^{(2)\rm g}_{\bar m \bar m} + \big(A^t_{\bar m \bar m \parallel} + A^\phi_{\bar m \bar m \parallel} + A^{\rm s \parallel}_{\bar m \bar m}\big)f^{(2)\rm g}_{\bar m \bar m} + C_{\bar m \bar m}f^{(2)\rm s \parallel}_{\bar m \bar m}  \, , \\
   A_{1\rm s \parallel} &= A^r_{\bar m \bar m \parallel} \partial_r f^{(1)\rm g}_{\bar m \bar m} + A^r_{n \bar m \parallel} \partial_r f^{(1)\rm g}_{n \bar m} + A^z_{\bar m \bar m \parallel} \partial_z f^{(1)\rm g}_{\bar m \bar m} + A^z_{n \bar m} \partial_z f^{(1)\rm g}_{n \bar m} + \big(A^t_{\bar m \bar m \parallel} + A^\phi_{\bar m \bar m \parallel} + A^{\rm s \parallel}_{\bar m \bar m}\big)f^{(1)\rm g}_{\bar m \bar m}   \nonumber \\
   & + \big(A^t_{n \bar m \parallel} + A^\phi_{n \bar m \parallel} + A^{\rm s\parallel}_{n \bar m}\big)f^{(1)\rm g}_{n \bar m} + C_{\bar m \bar m}f^{(1)\rm s \parallel}_{\bar m \bar m} + C_{n \bar m}f^{(1) \rm s \parallel}_{n \bar m}  \, , \\
   A_{0\rm s \parallel} &= A^r_{\bar m \bar m \parallel} \partial_r f^{(0) \rm g}_{\bar m \bar m} + A^r_{n \bar m \parallel} \partial_r f^{(0)\rm g}_{n \bar m} + A^r_{n n \parallel} \partial_r f^{(0)\rm g}_{n n} + A^z_{\bar m \bar m \parallel} \partial_z f^{(0) \rm g}_{\bar m \bar m} + A^z_{n \bar m \parallel} \partial_z f^{(0) \rm g}_{n \bar m} + A^z_{n n \parallel} \partial_z f^{(0) \rm g}_{n n} + \nonumber \\
   & + \big(A^t_{\bar m \bar m \parallel} + A^\phi_{\bar m \bar m \parallel} +  A^{\rm s \parallel}_{\bar m \bar m}\big)f^{(0) \rm g}_{\bar m \bar m} + \big(A^t_{n \bar m \parallel} + A^\phi_{n \bar m \parallel} +  A^{\rm s \parallel}_{n \bar m}\big)f^{(0) \rm g}_{n \bar m} + \big(A^t_{n n \parallel} + A^\phi_{n n \parallel} +  A^{\rm s \parallel}_{n n}\big)f^{(0) \rm g}_{n n} + \nonumber \\
   & + C_{\bar m \bar m}f^{(0) \rm s \parallel}_{\bar m \bar m} + C_{n \bar m}f^{(0) \rm s \parallel}_{n \bar m} + C_{n n}f^{(0) \rm s \parallel} _{n n} \, ,
\end{align}
\end{widetext}
where the functions $f^{(i)\rm s \parallel}_{\bar m \bar m}$ are obtained by replacing $S^{\rm g}_{\ell m \kappa}$ with $\delta S_{\ell m \kappa}$ in, respectively, $f^{(i)\rm g}_{\bar m \bar m}$ for $i=0,1,2$. The same also applies for $f^{(1)\rm s \parallel}_{n \bar m}$, $f^{(0)\rm s \parallel}_{n \bar m}$, $f^{(0)\rm s \parallel}_{n n}$.

\subsection{Coefficients proportional to $s_\perp$}
The coefficients $B_{3\rm s \parallel}$, $B_{2\rm s \parallel}$ and $B_{1\rm s \parallel}$ are given by
\begin{align}
   B_{3\rm s \perp} &= - A^r_{\bar m \bar m \perp} f^{(2)\rm g}_{\bar m \bar m} \, , \\  
   B_{2\rm s \perp} &= - A^r_{n \bar m \perp} f^{(1)\rm g}_{n \bar m} - A^r_{\bar m \bar m \perp} f^{(1)\rm g}_{n \bar m} \, , \\
   B_{1\rm s \perp} &= - A^r_{\bar m \bar m \perp} f^{(0)\rm g}_{n \bar m} - A^r_{n \bar m \perp} f^{(0) \rm g}_{n \bar m} - A^r_{n n \perp} f^{(0)\rm g}_{n n}  \, , 
\end{align}
while coefficients $A_{2\rm s \perp}$, $A_{1\rm s \perp}$ and $A_{0\rm s \perp}$ are given by
\begin{widetext}
\begin{align}
   A_{2\rm s \perp} &= A^r_{\bar m \bar m \perp} \partial_r f^{(2)\rm g}_{\bar m \bar m} + A^z_{\bar m \bar m \perp} \partial_z f^{(2)\rm g}_{\bar m \bar m} + \big(A^t_{\bar m \bar m \perp} + A^\phi_{\bar m \bar m \perp} + A^{\rm s \perp}_{\bar m \bar m}\big)f^{(2)\rm g}_{\bar m \bar m}  \, , \\  
   A_{1\rm s \perp} &= A^r_{\bar m \bar m \perp} \partial_r f^{(1)\rm g}_{\bar m \bar m} + A^r_{n \bar m \perp} \partial_r f^{(1)\rm g}_{n \bar m} + A^z_{\bar m \bar m \perp} \partial_z f^{(1)\rm g}_{\bar m \bar m} + A^z_{n \bar m} \partial_z f^{(1)\rm g}_{n \bar m} + \big(A^t_{\bar m \bar m \perp} + A^\phi_{\bar m \bar m \perp} + A^{\rm s \perp}_{\bar m \bar m}\big)f^{(1)\rm g}_{\bar m \bar m}   \nonumber \\
   & + \big(A^t_{n \bar m \perp} + A^\phi_{n \bar m \perp} + A^{\rm s\perp}_{n \bar m}\big)f^{(1)\rm g}_{n \bar m}  \, , \\
   A_{0\rm s \perp} &= A^r_{\bar m \bar m \perp} \partial_r f^{(0) \rm g}_{\bar m \bar m} + A^r_{n \bar m \perp} \partial_r f^{(0)\rm g}_{n \bar m} + A^r_{n n \perp} \partial_r f^{(0)\rm g}_{n n} + A^z_{\bar m \bar m \perp} \partial_z f^{(0) \rm g}_{\bar m \bar m} + A^z_{n \bar m \perp} \partial_z f^{(0) \rm g}_{n \bar m} + A^z_{n n \perp} \partial_z f^{(0) \rm g}_{n n} + \nonumber \\
   & + \big(A^t_{\bar m \bar m \perp} + A^\phi_{\bar m \bar m \perp} +  A^{\rm s \perp}_{\bar m \bar m}\big)f^{(0) \rm g}_{\bar m \bar m} + \big(A^t_{n \bar m \perp} + A^\phi_{n \bar m \perp} +  A^{\rm s \perp}_{n \bar m}\big)f^{(0) \rm g}_{n \bar m} + \big(A^t_{n n \parallel} + A^\phi_{n n \perp} +  A^{\rm s \perp}_{n n}\big)f^{(0) \rm g}_{n n} \, .
\end{align}
\end{widetext}

\section{Data tables}\label{app:data}

In this appendix we present a few data tables for future reference.

We start with the shifts to the frequencies due to the spin of the secondary.  Table \ref{tbl:freq_FC} lists the Mino-time shifts to frequencies for three different sets of orbital parameters $\{a, p,e,x\}$ when the constants of motion are kept fixed (``FC'' scheme). Specifically, we consider an equatorial orbit with $\{a,p,e,x\}=\{0.9,4,1/2,1\}$ and two generic orbits $\{a,p_{\rm g},e_{\rm g},x_{\rm g}\}=\{0.9,4,0.3,1/\sqrt{2}\}$, $\{a,p_{\rm g},e_{\rm g},x_{\rm g}\}=\{0.9,4,1/2,\sqrt{3}/2\}$.
 
\begin{table*}[h!]
    \begin{tabular}{| c|c|c|c|c|}
        \hline
        $\{a,p_{\rm g},e_{\rm g},x_{\rm g}\}$ &
        $\Upsilon_{r \rm s}$&$\Upsilon_{z \rm s}$&$\Upsilon_{t \rm s}$& $\Upsilon_{\phi \rm s}$\\
         \hline
        $\{0.9,4,1/2,1\}$ &
        -0.8763992&-0.5081304 &-15.21017 & 6.001243 $\times 10^{-2}$\\
        \hline
        $\{0.9,4,1/2,\sqrt{3}/2\}$ &
        -1.700439 &-0.6362760 & -28.43601& 0.5795217\\
        \hline 
        $\{0.9,4,0.3,1/\sqrt{2}\}$ &
        -5.040008&-0.7785672 &-61.47855 & 2.148015\\
        \hline
    \end{tabular}
    \caption{Corrections to the Mino-time frequencies due to the secondary spin for specific values of $\{a,p_{\rm g},e_{\rm g},x_{\rm g}\}$ in the ``FC'' scheme. } \label{tbl:freq_FC}
\end{table*}
 Similarly, Table~\ref{tbl:freq_DH} shows the Mino-time shifts to frequencies for the same orbits when the turning points are kept fixed (``DH'' scheme). Note that we have explicitly checked that the results for the two parametrizations agree, when appropriately mapped to each other. Finally, Table~\ref{tbl:shift_constant_DH} presents the shits to the constants of motion for the ``DH'' parametrization.
\begin{table*}[h!]
     \begin{tabular}{|c|c|c|c|c|}
        \hline
        $\{a,p_{\rm g},e_{\rm g},x_{\rm g}\}$ &$\Upsilon_{r \rm s}$&$\Upsilon_{z  \rm s}$&$\Upsilon_{t  \rm s}$& $\Upsilon_{\phi  \rm s}$ \\
         \hline
        $\{0.9,4,1/2,1\}$ & 0.4969736 &-3.297780 $\times 10^{-3}$  &1.156303  & -0.4746781 \\
        \hline
        $\{0.9,4,1/2,\sqrt{3}/2\}$ &0.5560860
         &-0.05904496 &1.843115 & -0.5419529\\
        \hline 
        $\{0.9,4,0.3,1/\sqrt{2}\}$ &0.6855624
        & -0.1362528& 0.4337835 & -0.5707891 \\
        \hline
    \end{tabular}
    \caption{Corrections to the Mino-time frequencies due to the secondary spin for specific values of $\{a,p_{\rm g},e_{\rm g},x_{\rm g}\}$ in the ``DH'' scheme.. These results are in agreement with~\cite{Drummond:2022efc,Drummond:2022xej}.} \label{tbl:freq_DH}
\end{table*}

\begin{table*}[h!]
    \begin{tabular}{|c|c|c|c|}
        \hline
        $\{a,p_{\rm g},e_{\rm g},x_{\rm g}\}$ & $E_{\rm s}$& $J_{z\rm s}$& $K_{\rm s}$\\
         \hline
        $\{0.9,4,1/2,1\}$ &-1.107603 $\times 10^{-2}$ &0.507194& 1.689610 \\
        \hline
        $\{0.9,4,1/2,\sqrt{3}/2\}$ &-1.120130 $\times 10^{-2}$& 0.2645343 & 2.595226\\
        \hline 
        $\{0.9,4,0.3,1/\sqrt{2}\}$ & -1.697615 $\times 10^{-2}$ & 5.358741 $\times 10^{-2}$& 3.413985\\
        \hline
    \end{tabular}
    \caption{Corrections to the constants of motion due to the secondary spin for specific values of $\{a,p_{\rm g},e_{\rm g},x_{\rm g}\}$ in the ``DH'' scheme. These results are in agreement with~\cite{Drummond:2022efc,Drummond:2022xej}.} \label{tbl:shift_constant_DH}
\end{table*}

 
Furthermore, in Table~\ref{tbl:amplParFC},~\ref{tbl:amplParDH},~\ref{tbl:amplOrtPlus} and \ref{tbl:amplOrtMinus} we list the $l$, $m$, $n$, $k$ modes of the shift of the amplitudes for a generic orbit with orbital parameters $\{a,p_{\rm g},e_{\rm g},x_{\rm g}\}=\{0.9,12,1/5,\sqrt{3}/2\}$ in the ``DH'' and ``FC'' parametrization, as well as for the orthogonal part of the spin for $j=+1$ and $j=-1$. In addition, in Table~\ref{tbl:flux}, we list the $m$ modes of the linear in spin part of the energy flux and for the angular momentum flux for a specific orbit in the two parametrizations of interest. All the numerical values presented in these tables were obtained using 12 Fourier modes in the spin corrections to the orbits.
\begin{table*}[h]
    \begin{tabular}{| c |c|c|c|c|c|c|c|}
        \hline
        $m$&$l$&$n$&$k$& $\Re(\delta Z^\infty_{ l m n k0})$ &$\Im(\delta Z^\infty_{ l m n k0})$ & $\Re(\delta Z^H_{l m n k0})$ & $\Im(\delta Z^H_{ l m n k0})$ \\
        \hline
        1 &2&0&1&$3.749 \times 10^{-4}$ & $-8.306 \times 10^{-5}$ &$5.982  \times 10^{-4}$&$8.752 \times 10^{-5}$\\
        1 &2&1&1&$-9.537 \times 10^{-4}$ & $2.597 \times 10^{-4}$ &$-1.177 \times 10^{-3}$&$-2.125 \times 10^{-4}$\\
        1 &2&2&1&$-1.295\times 10^{-3}$ & $3.954 \times 10^{-4}$ &$-1.379  \times 10^{-3}$&$-2.923 \times 10^{-4}$\\
        1 &2&3&1&$-9.286\times 10^{-4}$ & $3.043 \times 10^{-4}$ &$-8.892  \times 10^{-4}$&$-2.1725 \times 10^{-4}$\\
        1 &3&0&2&$-6.343\times 10^{-5}$ & $-1.643 \times 10^{-4}$ &$3.944  \times 10^{-5}$&$2.721 \times 10^{-5}$\\
        1 &3&1&2&$6.314\times 10^{-5}$ & $1.374 \times 10^{-4}$ &$-3.067  \times 10^{-5}$&$-2.002 \times 10^{-5}$\\
        1 &3&2&2&$1.614\times 10^{-4}$ & $3.122\times 10^{-4}$ &$-6.732  \times 10^{-5}$&$-4.164 \times 10^{-5}$\\
        1 &3&3&2&$1.588\times 10^{-4}$ & $2.7895\times 10^{-4}$ &$-5.965  \times 10^{-5}$&$-3.475 \times 10^{-5}$\\
        2&2&0&0&$-7.625\times 10^{-4}$ & $1.708\times 10^{-4}$ &$-1.111  \times 10^{-3}$&$-2.842 \times 10^{-4}$\\
        2&2&1&0&$1.9605\times 10^{-3}$ & $-5.393\times 10^{-4}$ &$2.153  \times 10^{-3}$&$6.614 \times 10^{-4}$\\
        2&2&2&0&$2.747\times 10^{-3}$ & $-8.462\times 10^{-4}$ &$2.520 \times 10^{-3}$&$8.925 \times 10^{-4}$\\
        2&2&3&0&$2.025\times 10^{-3}$ & $-6.702\times 10^{-4}$ &$1.615 \times 10^{-3}$&$6.498 \times 10^{-4}$\\
        2&3&0&1&$1.635\times 10^{-4}$ & $4.194\times 10^{-4}$ &$-7.113 \times 10^{-5}$&$-1.3925 \times 10^{-4}$\\
        2&3&1&1&$-1.586\times 10^{-4}$ & $-3.417\times 10^{-4}$ &$5.186 \times 10^{-5}$&$1.01275 \times 10^{-4}$\\
        2&3&2&1&$-4.205\times 10^{-4}$ & $-8.059\times 10^{-4}$ &$1.1397 \times 10^{-4}$&$2.2098 \times 10^{-4}$\\
        2&3&3&1&$-4.231\times 10^{-4}$ & $-7.362\times 10^{-4}$ &$9.968 \times 10^{-5}$&$1.91 \times 10^{-4}$\\
        \hline  
    \end{tabular}
    \caption{Real and imaginary parts of the spin parallel amplitude corrections ($j=0$) in the ``FC'' parametrization for given $l$, $m$, $n$ and $k$ of a generic orbit with $\{a,p_{\rm g},e_{\rm g},x_{\rm g}\}=\{0.9,12,1/5,\sqrt{3}/2\}$. The numbers are rounded to the third decimal place.}\label{tbl:amplParFC}
\end{table*}
\begin{table*}[h!]
    \begin{tabular}{| c |c|c|c|c|c|c|c|}
        \hline
        $m$&$l$&$n$&$k$& $\Re(\delta Z^\infty_{ l m n k0})$ &$\Im(\delta Z^\infty_{ l m n k0})$ & $\Re(\delta Z^H_{l m n k0})$ & $\Im(\delta Z^H_{ l m n k0})$ \\
        \hline
        1 &2&0&1&$4.896 \times 10^{-6}$ & $-1.602 \times 10^{-6}$ &$-5.272  \times 10^{-6}$&$2.782 \times 10^{-7}$\\
        1 &2&1&1& $9.951 \times 10^{-6}$ &$-2.78456 \times 10^{-6}$ &$3.859 \times 10^{-6}$&$8.739 \times 10^{-7}$\\
        1 &2&2&1 &$7.303 \times 10^{-6}$ &$-2.147 \times 10^{-6}$ &$3.741 \times 10^{-6}$ & $7.806\times 10^{-7}$ \\
        1 &2&3&1&$3.601 \times 10^{-6}$ &$-1.123 \times 10^{-6}$ &$1.774 \times 10^{-6}$ & $3.884 \times 10^{-7}$ \\
        1 &3&0&2&$-9.459 \times 10^{-6}$ & $1.302\times 10^{-6}$ &$-5.877 \times 10^{-7}$ & $-4.197 \times 10^{-7}$ \\
        1 &3&1&2&$-8.880 \times10^{-7}$ & $-1.793 \times 10^{-6}$ &$-3.2098 \times 10^{-7}$ & $-2.175 \times 10^{-7}$ \\
        1 &3&2&2&$-9.857 \times 10^{-7}$ & $-1.947 \times 10^{-6}$ &$-1.106 \times 10^{-7}$ & $-6.4897 \times 10^{-8}$ \\
        1 &3&3&2&$-6.657 \times 10^{-7}$ & $-1.239 \times 10^{-6}$ &$-3.735 \times  10^{-8}$ & $-1.485 \times 10^{-8}$ \\
        2&2&0&0&$-1.989 \times 10^{-5}$ & $5.899 \times 10^{-6}$ &$-7.564 \times 10^{-6}$ & $-3.698 \times 10^{-6}$ \\
        2&2&1&0&$-3.653 \times 10^{-5}$ & $1.047 \times 10^{-5}$ &$-2.873 \times 10^{-5}$ & $-9.511 \times 10^{-6}$ \\
        2&2&2&0&$-2.824 \times 10^{-5}$ & $8.643 \times 10^{-6}$ &$-2.135 \times 10^{-5}$ & $-7.584 \times 10^{-6}$\\
        2&2&3&0&$-1.530 \times 10^{-5}$ & $4.973 \times 10^{-6}$ &$-1.060 \times 10^{-5}$ & $-4.141 \times 10^{-6}$\\
        2&3&0&1&$8.242 \times 10^{-7}$ & $9.229 \times 10^{-7}$ &$6.845 \times 10^{-7}$ & $1.338 \times 10^{-6}$\\
        2&3&1&1&$3.873 \times 10^{-6}$ & $7.747 \times 10^{-6}$ &$-1.289 \times 10^{-7}$ & $-2.494 \times 10^{-7}$\\
        2&3&2&1&$4.309 \times 10^{-6}$ & $8.210 \times 10^{-6}$ &$-4.252 \times 10^{-7}$ & $-8.281 \times 10^{-7}$\\
        2&3&3&1 &$3.064\times 10^{-6}$ & $5.447 \times 10^{-6}$ &$-3.322 \times 10^{-7}$ & $-6.455 \times 10^{-7}$\\
        \hline
    \end{tabular}
    \caption{Real and imaginary parts of the spin parallel amplitude corrections ($j=0$) in the ``DH'' parametrization for given $l$, $m$, $n$ and $k$ of a generic orbit with $\{a,p_{\rm g},e_{\rm g},x_{\rm g}\}=\{0.9,12,1/5,\sqrt{3}/2\}$.} \label{tbl:amplParDH}
\end{table*}

\begin{table*}[h]
    \begin{tabular}{| c |c|c|c|c|c|c|c|}
        \hline
        $m$&$l$&$n$&$k$& $\Re(\delta Z^\infty_{ l m n k1})$ &$\Im(\delta Z^\infty_{ l m n k1})$ & $\Re(\delta Z^H_{l m n k1})$ & $\Im(\delta Z^H_{ l m n k1})$ \\
        \hline
        1&2&0&1&$-8.973\times 10^{-5}$ &$ 1.022\times 10^{-4}$ &$-3.191 \times 10^{-5}$ &$ 2.435 \times 10^{-5}$\\
        1&2&1&1&$-1.215\times 10^{-4}$ &$ 1.533\times 10^{-4}$ &$-3.870 \times 10^{-5}$ &$ 2.381 \times 10^{-5}$\\
        1&2&2&1&$-8.842\times 10^{-5}$ &$ 1.203\times 10^{-4}$ &$-2.606 \times 10^{-5}$ &$ 1.222 \times 10^{-5}$\\
        1&2&3&1&$-4.878\times 10^{-5}$ &$ 7.004\times 10^{-5}$ &$-1.374 \times 10^{-5}$ &$ 4.298 \times 10^{-6}$\\
        1&3&0&2&$ 3.778\times 10^{-5}$ &$ 3.689\times 10^{-5}$ &$-2.940 \times 10^{-6}$ &$ 2.103 \times 10^{-6}$\\
        1&3&1&2&$ 7.837\times 10^{-5}$ &$ 7.378\times 10^{-5}$ &$-5.344 \times 10^{-6}$ &$ 3.934 \times 10^{-6}$\\
        1&3&2&2&$ 7.324\times 10^{-5}$ &$ 6.664\times 10^{-5}$ &$-4.305 \times 10^{-6}$ &$ 3.511 \times 10^{-6}$\\
        1&3&3&2&$ 4.878\times 10^{-5}$ &$ 4.330\times 10^{-5}$ &$-2.510 \times 10^{-6}$ &$ 2.280 \times 10^{-6}$\\
        2&2&0&0&$-2.082\times 10^{-5}$ &$ 1.763\times 10^{-5}$ &$-7.317 \times 10^{-6}$ &$-1.596 \times 10^{-5}$\\
        2&2&1&0&$-2.660\times 10^{-5}$ &$ 1.573\times 10^{-5}$ &$-4.034 \times 10^{-6}$ &$-1.993 \times 10^{-5}$\\
        2&2&2&0&$-2.074\times 10^{-5}$ &$ 5.921\times 10^{-6}$ &$-4.454 \times 10^{-7}$ &$-1.363 \times 10^{-5}$\\
        2&2&3&0&$-1.278\times 10^{-5}$ &$-1.101\times 10^{-7}$ &$ 8.899 \times 10^{-7}$ &$-7.147 \times 10^{-6}$\\
        2&3&0&1&$-2.964\times 10^{-5}$ &$-1.541\times 10^{-5}$ &$ 2.787 \times 10^{-6}$ &$ 1.253 \times 10^{-7}$\\
        2&3&1&1&$-6.877\times 10^{-5}$ &$-2.835\times 10^{-5}$ &$ 5.424 \times 10^{-6}$ &$ 5.324 \times 10^{-7}$\\
        2&3&2&1&$-6.829\times 10^{-5}$ &$-2.225\times 10^{-5}$ &$ 4.592 \times 10^{-6}$ &$ 4.739 \times 10^{-7}$\\
        2&3&3&1&$-4.780\times 10^{-5}$ &$-1.203\times 10^{-5}$ &$ 2.814 \times 10^{-6}$ &$ 2.910 \times 10^{-7}$\\
        \hline  
    \end{tabular}
    \caption{Real and imaginary parts of the spin orthogonal amplitude corrections for $j=+1$ and given $l$, $m$, $n$ and $k$ of a generic orbit with $\{a,p_{\rm g},e_{\rm g},x_{\rm g}\}=\{0.9,12,1/5,\sqrt{3}/2\}$. The numbers are rounded to the third decimal place.} \label{tbl:amplOrtPlus} 
\end{table*}
\begin{table*}[h]
    \begin{tabular}{| c |c|c|c|c|c|c|c|}
        \hline
        $m$&$l$&$n$&$k$& $\Re(\delta Z^\infty_{ l m n k-1})$ &$\Im(\delta Z^\infty_{ l m n k-1})$ & $\Re(\delta Z^H_{l m n k-1})$ & $\Im(\delta Z^H_{ l m n k-1})$ \\
        \hline
        1&2&0&1&$-1.303\times 10^{-6}$ &$ 2.372\times 10^{-6}$ &$-8.611 \times 10^{-5}$ &$-1.256 \times 10^{-5}$\\
        1&2&1&1&$ 3.820\times 10^{-7}$ &$ 4.287\times 10^{-7}$ &$-4.999 \times 10^{-5}$ &$-1.095 \times 10^{-5}$\\
        1&2&2&1&$ 4.053\times 10^{-7}$ &$-6.016\times 10^{-7}$ &$-2.093 \times 10^{-5}$ &$-5.876 \times 10^{-6}$\\
        1&2&3&1&$ 9.166\times 10^{-8}$ &$-5.710\times 10^{-7}$ &$-7.564 \times 10^{-6}$ &$-2.523 \times 10^{-6}$\\
        1&3&0&2&$ 2.191\times 10^{-6}$ &$ 2.976\times 10^{-6}$ &$-8.036 \times 10^{-6}$ &$-6.23 \times 10^{-8}$\\
        1&3&1&2&$ 1.247\times 10^{-6}$ &$ 2.813\times 10^{-6}$ &$-7.166 \times 10^{-6}$ &$ 1.573 \times 10^{-7}$\\
        1&3&2&2&$ 1.374\times 10^{-7}$ &$ 1.841\times 10^{-6}$ &$-4.044 \times 10^{-6}$ &$ 2.120 \times 10^{-7}$\\
        1&3&3&2&$-2.277\times 10^{-7}$ &$ 1.049\times 10^{-6}$ &$-1.839 \times 10^{-6}$ &$ 1.540 \times 10^{-7}$\\
        2&2&0&0&$-3.413\times 10^{-6}$ &$ 2.779\times 10^{-6}$ &$-2.217 \times 10^{-6}$ &$-1.545 \times 10^{-5}$\\
        2&2&1&0&$-4.337\times 10^{-6}$ &$ 3.379\times 10^{-6}$ &$ 1.794 \times 10^{-6}$ &$-1.075 \times 10^{-5}$\\
        2&2&2&0&$-3.269\times 10^{-6}$ &$ 2.369\times 10^{-6}$ &$ 1.842 \times 10^{-6}$ &$-5.107 \times 10^{-6}$\\
        2&2&3&0&$-1.886\times 10^{-6}$ &$ 1.270\times 10^{-6}$ &$ 1.035 \times 10^{-6}$ &$-2.040 \times 10^{-6}$\\
        2&3&0&1&$-1.428\times 10^{-6}$ &$-2.521\times 10^{-7}$ &$ 3.249 \times 10^{-6}$ &$ 6.604 \times 10^{-6}$\\
        2&3&1&1&$-7.344\times 10^{-7}$ &$ 4.245\times 10^{-7}$ &$ 2.863 \times 10^{-6}$ &$ 6.178 \times 10^{-6}$\\
        2&3&2&1&$ 6.180\times 10^{-8}$ &$ 4.889\times 10^{-7}$ &$ 1.580 \times 10^{-6}$ &$ 3.6389\times 10^{-6}$\\
        2&3&3&1&$ 2.860\times 10^{-7}$ &$ 2.612\times 10^{-7}$ &$ 7.048 \times 10^{-7}$ &$ 1.716 \times 10^{-6}$\\
        \hline
    \end{tabular}
    \caption{Real and imaginary parts of the spin orthogonal amplitude corrections for $j=-1$ and given $l$, $m$, $n$ and $k$ of a generic orbit with $\{a,p_{\rm g},e_{\rm g},x_{\rm g}\}=\{0.9,12,1/5,\sqrt{3}/2\}$. The numbers are rounded to the third decimal place.} \label{tbl:amplOrtMinus} 
\end{table*}

\begin{table*}[h]
\begin{tabular}{|c| c|c|c|}
\hline
                    & $m$ &  $\delta \mathcal{F}^{E}_m$& $\delta\mathcal{F}^{J_z}_m$\\
 \hline
\multirow{2}{*}{DH} & 1& $-1.132 \times 10^{-7}$  & $-1.184 \times 10^{-6}$\\
                    & 2& $4.9545 \times 10^{-11}$ & $1.203 \times 10^{-9}$\\
\multirow{2}{*}{FC} & 1& $5.771  \times 10^{-6}$  &$3.609  \times 10^{-5}$\\
                    & 2& $2.073 \times 10^{-9}$   &$3.352\times 10^{-8}$\\
\hline
\end{tabular}
\caption{Linear in spin parts of the total energy fluxes and the angular momentum fluxes for $\{a,p_{\rm g},e_{\rm g},x_{\rm g}\}=\{0.9,12,1/5,\sqrt{3}/2\}$ for two values of $m$. The fluxes are summed over the $l, n$ and $k$ shown in Table~\ref{tbl:amplParFC} (for the ``FC'' scheme) and Table~\ref{tbl:amplParDH} (for the``DH'' scheme)} \label{tbl:flux}
\end{table*}


\bibliographystyle{utphys}
\bibliography{Ref}

@article{Grant:2024ivt,
    author = "Grant, Alexander M.",
    title = "{Flux-balance laws for spinning bodies under the gravitational self-force}",
    eprint = "2406.10343",
    archivePrefix = "arXiv",
    primaryClass = "gr-qc",
    month = "6",
    year = "2024"
}

@article{Ramond:2024ozy,
    author = "Ramond, Paul",
    title = "{On the integrability of extended test body dynamics around black holes}",
    eprint = "2402.02670v2",
    archivePrefix = "arXiv",
    primaryClass = "gr-qc",
    month = "2",
    year = "2024"
}

@article{Kubiznak:2011ay,
    author = "Kubiz{\v n}{\'a}k, David and Cariglia, Marco",
    title = "{On Integrability of spinning particle motion in higher-dimensional black hole spacetimes}",
    eprint = "1110.0495",
    archivePrefix = "arXiv",
    primaryClass = "hep-th",
    reportNumber = "DAMTP-2011-80",
    doi = "10.1103/PhysRevLett.108.051104",
    journal = "Phys. Rev. Lett.",
    volume = "108",
    pages = "051104",
    year = "2012"
}

@article{Gibbons:1993ap,
    author = "Gibbons, G. W. and Rietdijk, R. H. and van Holten, J. W.",
    title = "{SUSY in the sky}",
    eprint = "hep-th/9303112",
    archivePrefix = "arXiv",
    reportNumber = "NIKHEF-H93-04, DAMTP-R-92-43",
    doi = "10.1016/0550-3213(93)90472-2",
    journal = "Nucl. Phys. B",
    volume = "404",
    pages = "42--64",
    year = "1993"
}

@article{Hackmann:2014tga,
    author = {Hackmann, Eva and L\"ammerzahl, Claus and Obukhov, Yuri N. and Puetzfeld, Dirk and Schaffer, Isabell},
    title = "{Motion of spinning test bodies in Kerr spacetime}",
    eprint = "1408.1773",
    archivePrefix = "arXiv",
    primaryClass = "gr-qc",
    doi = "10.1103/PhysRevD.90.064035",
    journal = "Phys. Rev. D",
    volume = "90",
    number = "6",
    pages = "064035",
    year = "2014"
}

@article{Suzuki:1999si,
    author = "Suzuki, Shingo and Maeda, Kei-ichi",
    title = "{Signature of chaos in gravitational waves from a spinning particle}",
    eprint = "gr-qc/9910064",
    archivePrefix = "arXiv",
    reportNumber = "WU-AP-83-99",
    doi = "10.1103/PhysRevD.61.024005",
    journal = "Phys. Rev. D",
    volume = "61",
    pages = "024005",
    year = "2000"
}

@article{Suzuki:1996gm,
    author = "Suzuki, Shingo and Maeda, Kei-ichi",
    title = "{Chaos in Schwarzschild space-time: The motion of a spinning particle}",
    eprint = "gr-qc/9604020",
    archivePrefix = "arXiv",
    reportNumber = "WU-AP-59-96",
    doi = "10.1103/PhysRevD.55.4848",
    journal = "Phys. Rev. D",
    volume = "55",
    pages = "4848--4859",
    year = "1997"
}

@article{Han:2010tp,
    author = "Han, Wen-Biao",
    title = "{Gravitational Radiations from a Spinning Compact Object around a supermassive Kerr black hole in circular orbit}",
    eprint = "1008.3324",
    archivePrefix = "arXiv",
    primaryClass = "gr-qc",
    doi = "10.1103/PhysRevD.82.084013",
    journal = "Phys. Rev. D",
    volume = "82",
    pages = "084013",
    year = "2010"
}

@article{Skoupy:2023lih,
    author = "Skoup{\'y}, Viktor and Lukes-Gerakopoulos, Georgios and Drummond, Lisa V. and Hughes, Scott A.",
    title = "{Asymptotic gravitational-wave fluxes from a spinning test body on generic orbits around a Kerr black hole}",
    eprint = "2303.16798",
    archivePrefix = "arXiv",
    primaryClass = "gr-qc",
    doi = "10.1103/PhysRevD.108.044041",
    journal = "Phys. Rev. D",
    volume = "108",
    number = "4",
    pages = "044041",
    year = "2023"
}

@article{Skoupy:2024jsi,
    author = "Skoup\'y, Viktor and Witzany, Vojt\v{e}ch",
    title = "{Post-Newtonian expansions of extreme mass ratio inspirals of spinning bodies into Schwarzschild black holes}",
    eprint = "2406.14291",
    archivePrefix = "arXiv",
    primaryClass = "gr-qc",
    doi = "10.1103/PhysRevD.110.084061",
    journal = "Phys. Rev. D",
    volume = "110",
    number = "8",
    pages = "084061",
    year = "2024"
}

@article{Sago:2005fn,
    author = "Sago, Norichika and Tanaka, Takahiro and Hikida, Wataru and Ganz, Katsuhiko and Nakano, Hiroyuki",
    title = "{The Adiabatic evolution of orbital parameters in the Kerr spacetime}",
    eprint = "gr-qc/0511151",
    archivePrefix = "arXiv",
    doi = "10.1143/PTP.115.873",
    journal = "Prog. Theor. Phys.",
    volume = "115",
    pages = "873--907",
    year = "2006"
}

@ARTICLE{1975PhRvD..11.1387D,
       author = {{D'Eath}, P.~D.},
        title = "{Dynamics of a small black hole in a background universe}",
      journal = {\prd},
         year = 1975,
        month = mar,
       volume = {11},
        pages = {1387-1403},
          doi = {10.1103/PhysRevD.11.1387},
       adsurl = {https://ui.adsabs.harvard.edu/abs/1975PhRvD..11.1387D}
}

@article{Thorne:1984mz,
    author = "Thorne, Kip S. and Hartle, James B.",
    title = "{Laws of motion and precession for black holes and other bodies}",
    doi = "10.1103/PhysRevD.31.1815",
    journal = "Phys. Rev. D",
    volume = "31",
    pages = "1815--1837",
    year = "1984"
}

@article{Damour:2016gwp,
    author = "Damour, Thibault",
    title = "{Gravitational scattering, post-Minkowskian approximation and Effective One-Body theory}",
    eprint = "1609.00354",
    archivePrefix = "arXiv",
    primaryClass = "gr-qc",
    doi = "10.1103/PhysRevD.94.104015",
    journal = "Phys. Rev. D",
    volume = "94",
    number = "10",
    pages = "104015",
    year = "2016"
}

@article{Damour:2009sm,
    author = "Damour, Thibault",
    title = "{Gravitational Self Force in a Schwarzschild Background and the Effective One Body Formalism}",
    eprint = "0910.5533",
    archivePrefix = "arXiv",
    primaryClass = "gr-qc",
    doi = "10.1103/PhysRevD.81.024017",
    journal = "Phys. Rev. D",
    volume = "81",
    pages = "024017",
    year = "2010"
}

@article{Bini:2013zaa,
    author = "Bini, Donato and Damour, Thibault",
    title = "{Analytical determination of the two-body gravitational interaction potential at the fourth post-Newtonian approximation}",
    eprint = "1305.4884",
    archivePrefix = "arXiv",
    primaryClass = "gr-qc",
    doi = "10.1103/PhysRevD.87.121501",
    journal = "Phys. Rev. D",
    volume = "87",
    number = "12",
    pages = "121501",
    year = "2013"
}

@article{Bini:2011nhv,
    author = "Bini, Donato and Geralico, Andrea and Jantzen, Robert T.",
    title = "{Spin-geodesic deviations in the Schwarzschild spacetime}",
    eprint = "1408.4946",
    archivePrefix = "arXiv",
    primaryClass = "gr-qc",
    doi = "10.1007/s10714-010-1111-4",
    journal = "Gen. Rel. Grav.",
    volume = "43",
    pages = "959",
    year = "2011"
}

@article{Albertini:2022rfe,
    author = "Albertini, Angelica and Nagar, Alessandro and Pound, Adam and Warburton, Niels and Wardell, Barry and Durkan, Leanne and Miller, Jeremy",
    title = "{Comparing second-order gravitational self-force, numerical relativity, and effective one body waveforms from inspiralling, quasicircular, and nonspinning black hole binaries}",
    eprint = "2208.01049",
    archivePrefix = "arXiv",
    primaryClass = "gr-qc",
    doi = "10.1103/PhysRevD.106.084061",
    journal = "Phys. Rev. D",
    volume = "106",
    number = "8",
    pages = "084061",
    year = "2022"
}

@article{Antonelli:2019fmq,
    author = "Antonelli, Andrea and van de Meent, Maarten and Buonanno, Alessandra and Steinhoff, Jan and Vines, Justin",
    title = "{Quasicircular inspirals and plunges from nonspinning effective-one-body Hamiltonians with gravitational self-force information}",
    eprint = "1907.11597",
    archivePrefix = "arXiv",
    primaryClass = "gr-qc",
    doi = "10.1103/PhysRevD.101.024024",
    journal = "Phys. Rev. D",
    volume = "101",
    number = "2",
    pages = "024024",
    year = "2020"
}

@article{vandeMeent:2023ols,
    author = "van de Meent, Maarten and Buonanno, Alessandra and Mihaylov, Deyan P. and Ossokine, Serguei and Pompili, Lorenzo and Warburton, Niels and Pound, Adam and Wardell, Barry and Durkan, Leanne and Miller, Jeremy",
    title = "{Enhancing the SEOBNRv5 effective-one-body waveform model with second-order gravitational self-force fluxes}",
    eprint = "2303.18026",
    archivePrefix = "arXiv",
    primaryClass = "gr-qc",
    doi = "10.1103/PhysRevD.108.124038",
    journal = "Phys. Rev. D",
    volume = "108",
    number = "12",
    pages = "124038",
    year = "2023"
}

@article{Amaro-Seoane:2012lgq,
    author = "Amaro-Seoane, Pau",
    title = "{Relativistic dynamics and extreme mass ratio inspirals}",
    eprint = "1205.5240",
    archivePrefix = "arXiv",
    primaryClass = "astro-ph.CO",
    doi = "10.1007/s41114-018-0013-8",
    journal = "Living Rev. Rel.",
    volume = "21",
    number = "1",
    pages = "4",
    year = "2018"
}

@article{Barack:2011ed,
    author = "Barack, Leor and Sago, Norichika",
    title = "{Beyond the geodesic approximation: conservative effects of the gravitational self-force in eccentric orbits around a Schwarzschild black hole}",
    eprint = "1101.3331",
    archivePrefix = "arXiv",
    primaryClass = "gr-qc",
    doi = "10.1103/PhysRevD.83.084023",
    journal = "Phys. Rev. D",
    volume = "83",
    pages = "084023",
    year = "2011"
}

@article{Hughes:2005qb,
    author = "Hughes, Scott A. and Drasco, Steve and Flanagan, Eanna E. and Franklin, Joel",
    title = "{Gravitational radiation reaction and inspiral waveforms in the adiabatic limit}",
    eprint = "gr-qc/0504015",
    archivePrefix = "arXiv",
    doi = "10.1103/PhysRevLett.94.221101",
    journal = "Phys. Rev. Lett.",
    volume = "94",
    pages = "221101",
    year = "2005"
}

@book{baumgarte2010numerical,
  title={Numerical relativity: solving Einstein's equations on the computer},
  author={Baumgarte, Thomas W and Shapiro, Stuart L},
  year={2010},
  publisher={Cambridge University Press}
}

@article{Klein:2015hvg,
    author = "Klein, Antoine and others",
    title = "{Science with the space-based interferometer eLISA: Supermassive black hole binaries}",
    eprint = "1511.05581",
    archivePrefix = "arXiv",
    primaryClass = "gr-qc",
    doi = "10.1103/PhysRevD.93.024003",
    journal = "Phys. Rev. D",
    volume = "93",
    number = "2",
    pages = "024003",
    year = "2016"
}

@article{Colpi:2024xhw,
    author = "Colpi, Monica and others",
    title = "{LISA Definition Study Report}",
    eprint = "2402.07571",
    archivePrefix = "arXiv",
    primaryClass = "astro-ph.CO",
    month = "2",
    year = "2024"
}

@article{LISA:2017pwj,
    author = "Amaro-Seoane, Pau and others",
    collaboration = "LISA",
    title = "{Laser Interferometer Space Antenna}",
    eprint = "1702.00786",
    archivePrefix = "arXiv",
    primaryClass = "astro-ph.IM",
    month = "2",
    year = "2017"
}

@misc{gracedbO4,
  title = "{LIGO/Virgo/KAGRA Public Alerts on GraceDB}",
  howpublished = "\url{https://gracedb.ligo.org/superevents/public/O4/}",
  year = {2024}, 
  note = "[121 O4 detection candidates during the 64 weeks of operation since May 2023, as of 22 August 2024]"
}

@article{KAGRA:2013rdx,
    author = "Abbott, B. P. and others",
    collaboration = "KAGRA, LIGO Scientific, Virgo",
    title = "{Prospects for observing and localizing gravitational-wave transients with Advanced LIGO, Advanced Virgo and KAGRA}",
    eprint = "1304.0670",
    archivePrefix = "arXiv",
    primaryClass = "gr-qc",
    reportNumber = "LIGO-P1200087, VIR-0288A-12, JGW-P1808427",
    doi = "10.1007/s41114-020-00026-9",
    journal = "Living Rev. Rel.",
    volume = "19",
    pages = "1",
    year = "2016"
}

@article{Reitze:2019iox,
    author = "Reitze, David and others",
    title = "{Cosmic Explorer: The U.S. Contribution to Gravitational-Wave Astronomy beyond LIGO}",
    eprint = "1907.04833",
    archivePrefix = "arXiv",
    primaryClass = "astro-ph.IM",
    reportNumber = "LIGO-P1900316",
    journal = "Bull. Am. Astron. Soc.",
    volume = "51",
    number = "7",
    pages = "035",
    year = "2019"
}

@software{KerrGeodesics090,
   author       = {Niels Warburton and
                  Barry Wardell and
                  Oliver Long and
                  Sam Upton and
                  Philip Lynch and
                  Zachary Nasipak and
                  Leo C. Stein},
  title        = {{KerrGeodesics} 0.9.0},
  month        = jul,
  year         = 2023,
  publisher    = {Zenodo},
  version      = {0.9.0},
  doi          = {10.5281/zenodo.8108265},
  url          = {https://doi.org/10.5281/zenodo.8108265}
}

@article{Chakrabarti:2013tca,
    author = {Chakrabarti, Sayan and Delsate, T\'erence and G\"urlebeck, Norman and Steinhoff, Jan},
    title = "{I-Q relation for rapidly rotating neutron stars}",
    eprint = "1311.6509",
    archivePrefix = "arXiv",
    primaryClass = "gr-qc",
    doi = "10.1103/PhysRevLett.112.201102",
    journal = "Phys. Rev. Lett.",
    volume = "112",
    pages = "201102",
    year = "2014"
}

@article{marck1983solution,
  title={Solution to the equations of parallel transport in Kerr geometry; tidal tensor},
  author={Marck, J-A},
  journal={Proceedings of the Royal Society of London. A. Mathematical and Physical Sciences},
  volume={385},
  number={1789},
  pages={431--438},
  year={1983},
  publisher={The Royal Society London}
}

@article{tulczyjew1959motion,
  title={Motion of multipole particles in general relativity theory},
  author={Tulczyjew, W},
  journal={Acta Phys. Pol},
  volume={18},
  pages={393-409},
  year={1959},
  url = {https://www.actaphys.uj.edu.pl/fulltext?series=T&vol=18&no=5&page=393}
}

@article{Hughes:2021exa,
    author = "Hughes, Scott A. and Warburton, Niels and Khanna, Gaurav and Chua, Alvin J. K. and Katz, Michael L.",
    title = "{Adiabatic waveforms for extreme mass-ratio inspirals via multivoice decomposition in time and frequency}",
    eprint = "2102.02713",
    archivePrefix = "arXiv",
    primaryClass = "gr-qc",
    doi = "10.1103/PhysRevD.103.104014",
    journal = "Phys. Rev. D",
    volume = "103",
    number = "10",
    pages = "104014",
    year = "2021",
    note = "[Erratum: Phys.Rev.D 107, 089901 (2023)]"
}

@article{Fujita:2009bp,
    author = "Fujita, Ryuichi and Hikida, Wataru",
    title = "{Analytical solutions of bound timelike geodesic orbits in Kerr spacetime}",
    eprint = "0906.1420",
    archivePrefix = "arXiv",
    primaryClass = "gr-qc",
    doi = "10.1088/0264-9381/26/13/135002",
    journal = "Class. Quant. Grav.",
    volume = "26",
    pages = "135002",
    year = "2009"
}

@article{Pound:2021qin,
    author = "Pound, Adam and Wardell, Barry",
    title = "{Black hole perturbation theory and gravitational self-force}",
    eprint = "2101.04592",
    archivePrefix = "arXiv",
    primaryClass = "gr-qc",
    month = "1",
    year = "2021"
}

@article{Skoupy:2021asz,
    author = "Skoup\'y, Viktor and Lukes-Gerakopoulos, Georgios",
    title = "{Spinning test body orbiting around a Kerr black hole: Eccentric equatorial orbits and their asymptotic gravitational-wave fluxes}",
    eprint = "2102.04819",
    archivePrefix = "arXiv",
    primaryClass = "gr-qc",
    doi = "10.1103/PhysRevD.103.104045",
    journal = "Phys. Rev. D",
    volume = "103",
    number = "10",
    pages = "104045",
    year = "2021"
}

@article{Akcay:2019bvk,
    author = "Akcay, Sarp and Dolan, Sam R. and Kavanagh, Chris and Moxon, Jordan and Warburton, Niels and Wardell, Barry",
    title = "{Dissipation in extreme-mass ratio binaries with a spinning secondary}",
    eprint = "1912.09461",
    archivePrefix = "arXiv",
    primaryClass = "gr-qc",
    doi = "10.1103/PhysRevD.102.064013",
    journal = "Phys. Rev. D",
    volume = "102",
    number = "6",
    pages = "064013",
    year = "2020"
}

@article{Pound:2012nt,
      author         = "Pound, Adam",
      title          = "{Second-order gravitational self-force}",
      journal        = "Phys. Rev. Lett.",
      volume         = "109",
      year           = "2012",
      pages          = "051101",
      doi            = "10.1103/PhysRevLett.109.051101",
      eprint         = "1201.5089",
      archivePrefix  = "arXiv",
      primaryClass   = "gr-qc",
      SLACcitation   = "%%CITATION = ARXIV:1201.5089;%%"
}

@article{Chua:2020stf,
    author = "Chua, Alvin J. K. and Katz, Michael L. and Warburton, Niels and Hughes, Scott A.",
    title = "{Rapid generation of fully relativistic extreme-mass-ratio-inspiral waveform templates for LISA data analysis}",
    eprint = "2008.06071",
    archivePrefix = "arXiv",
    primaryClass = "gr-qc",
    doi = "10.1103/PhysRevLett.126.051102",
    journal = "Phys. Rev. Lett.",
    volume = "126",
    number = "5",
    pages = "051102",
    year = "2021"
}

@article{Katz:2021yft,
    author = "Katz, Michael L. and Chua, Alvin J. K. and Speri, Lorenzo and Warburton, Niels and Hughes, Scott A.",
    title = "{Fast extreme-mass-ratio-inspiral waveforms: New tools for millihertz gravitational-wave data analysis}",
    eprint = "2104.04582",
    archivePrefix = "arXiv",
    primaryClass = "gr-qc",
    doi = "10.1103/PhysRevD.104.064047",
    journal = "Phys. Rev. D",
    volume = "104",
    number = "6",
    pages = "064047",
    year = "2021"
}

@article{Tanaka:1996ht,
      author         = "Tanaka, Takahiro and Mino, Yasushi and Sasaki, Misao and
                        Shibata, Masaru",
      title          = "{Gravitational waves from a spinning particle in circular
                        orbits around a rotating black hole}",
      journal        = "Phys. Rev.",
      volume         = "D54",
      year           = "1996",
      pages          = "3762-3777",
      doi            = "10.1103/PhysRevD.54.3762",
      eprint         = "gr-qc/9602038",
      archivePrefix  = "arXiv",
      primaryClass   = "gr-qc",
      reportNumber   = "OU-TAP-27",
      SLACcitation   = "%%CITATION = GR-QC/9602038;%%"
}

@article{Hinderer:2008dm,
      author         = "Hinderer, Tanja and Flanagan, Eanna E.",
      title          = "{Two timescale analysis of extreme mass ratio inspirals
                        in Kerr. I. Orbital Motion}",
      journal        = "Phys. Rev.",
      volume         = "D78",
      year           = "2008",
      pages          = "064028",
      doi            = "10.1103/PhysRevD.78.064028",
      eprint         = "0805.3337",
      archivePrefix  = "arXiv",
      primaryClass   = "gr-qc",
      SLACcitation   = "%%CITATION = ARXIV:0805.3337;%%"
}

@article{Dixon:1970I,
      author         = "Dixon, W. G.",
      title          = "{Dynamics of extended bodies in general relativity. I.
                        Momentum and angular momentum}",
      journal        = "Proc. Roy. Soc. Lond.",
      volume         = "A314",
      year           = "1970",
      pages          = "499-527",
      doi            = "10.1098/rspa.1970.0020",
      SLACcitation   = "%%CITATION = PRSLA,A314,499;%%"
}

@article{Dixon:1970II,
      author         = "Dixon, W. G.",
      title          = "{Dynamics of extended bodies in general relativity. II.
                        Moments of the charge-current vector}",
      journal        = "Proc. Roy. Soc. Lond.",
      volume         = "A319",
      year           = "1970",
      pages          = "509-547",
      doi            = "10.1098/rspa.1970.0191",
      SLACcitation   = "%%CITATION = PRSLA,A319,509;%%"
}

@ARTICLE{DixonIII,
       author = {{Dixon}, W.~G.},
        title = "{Dynamics of Extended Bodies in General Relativity. III. Equations of Motion}",
      journal = {Philosophical Transactions of the Royal Society of London Series A},
         year = 1974,
        month = aug,
       volume = {277},
       number = {1264},
        pages = {59-119},
          doi = {10.1098/rsta.1974.0046},
       adsurl = {https://ui.adsabs.harvard.edu/abs/1974RSPTA.277...59D}
}

@article{Steinhoff:2009tk,
      author         = "Steinhoff, Jan and Puetzfeld, Dirk",
      title          = "{Multipolar equations of motion for extended test bodies
                        in General Relativity}",
      journal        = "Phys. Rev.",
      volume         = "D81",
      year           = "2010",
      pages          = "044019",
      doi            = "10.1103/PhysRevD.81.044019",
      eprint         = "0909.3756",
      archivePrefix  = "arXiv",
      primaryClass   = "gr-qc",
      SLACcitation   = "%%CITATION = ARXIV:0909.3756;%%"
}

@article{Kyrian:2007zz,
      author         = "Kyrian, K and Semer{\'a}k, O",
      title          = "{Spinning test particles in a Kerr field}",
      journal        = "Mon. Not. Roy. Astron. Soc.",
      volume         = "382",
      year           = "2007",
      pages          = "1922",
      doi            = "10.1111/j.1365-2966.2007.12502.x",
      SLACcitation   = "%%CITATION = MNRAA,382,1922;%%"
}

@article{Mathisson:1937zz,
      author         = "Mathisson, Myron",
      title          = "{Neue mechanik materieller systemes}",
      journal        = "Acta Phys. Polon.",
      volume         = "6",
      year           = "1937",
      pages          = "163-200",
      SLACcitation   = "%%CITATION = APPOA,6,163;%%"
}

@article{Papapetrou:1951pa,
      author         = "Papapetrou, Achille",
      title          = "{Spinning test particles in general relativity. 1.}",
      journal        = "Proc. Roy. Soc. Lond.",
      volume         = "A209",
      year           = "1951",
      pages          = "248-258",
      doi            = "10.1098/rspa.1951.0200",
      SLACcitation   = "%%CITATION = PRSLA,A209,248;%%"
}

@article{Drasco:2005kz,
      author         = "Drasco, Steve and Hughes, Scott A.",
      title          = "{Gravitational wave snapshots of generic extreme mass
                        ratio inspirals}",
      journal        = "Phys. Rev.",
      volume         = "D73",
      year           = "2006",
      number         = "2",
      pages          = "024027",
      doi            = "10.1103/PhysRevD.88.109905, 10.1103/PhysRevD.90.109905,
                        10.1103/PhysRevD.73.024027",
      note           = "[Erratum: Phys. Rev.D88,no.10,109905(2013); Erratum:
                        Phys. Rev.D90,no.10,109905(2014)]",
      eprint         = "gr-qc/0509101",
      archivePrefix  = "arXiv",
      primaryClass   = "gr-qc",
      SLACcitation   = "%%CITATION = GR-QC/0509101;%%"
}

@article{Babak:2017tow,
      author         = "Babak, Stanislav and Gair, Jonathan and Sesana, Alberto
                        and Barausse, Enrico and Sopuerta, Carlos F. and Berry,
                        Christopher P. L. and Berti, Emanuele and Amaro-Seoane,
                        Pau and Petiteau, Antoine and Klein, Antoine",
      title          = "{Science with the space-based interferometer LISA. V:
                        Extreme mass-ratio inspirals}",
      journal        = "Phys. Rev.",
      volume         = "D95",
      year           = "2017",
      number         = "10",
      pages          = "103012",
      doi            = "10.1103/PhysRevD.95.103012",
      eprint         = "1703.09722",
      archivePrefix  = "arXiv",
      primaryClass   = "gr-qc",
      SLACcitation   = "%%CITATION = ARXIV:1703.09722;%%"
}

@article{Lindblom:2008cm,
      author         = "Lindblom, Lee and Owen, Benjamin J. and Brown, Duncan A.",
      title          = "{Model Waveform Accuracy Standards for Gravitational Wave
                        Data Analysis}",
      journal        = "Phys. Rev.",
      volume         = "D78",
      year           = "2008",
      pages          = "124020",
      doi            = "10.1103/PhysRevD.78.124020",
      eprint         = "0809.3844",
      archivePrefix  = "arXiv",
      primaryClass   = "gr-qc",
      SLACcitation   = "%%CITATION = ARXIV:0809.3844;%%"
}

@article{Carter:1968rr,
      author         = "Carter, Brandon",
      title          = "{Global structure of the Kerr family of gravitational
                        fields}",
      journal        = "Phys. Rev.",
      volume         = "174",
      year           = "1968",
      pages          = "1559-1571",
      doi            = "10.1103/PhysRev.174.1559",
      SLACcitation   = "%%CITATION = PHRVA,174,1559;%%"
}

@article{Fujita:2009us,
      author         = "Fujita, Ryuichi and Hikida, Wataru and Tagoshi, Hideyuki",
      title          = "{An Efficient Numerical Method for Computing
                        Gravitational Waves Induced by a Particle Moving on
                        Eccentric Inclined Orbits around a Kerr Black Hole}",
      journal        = "Prog. Theor. Phys.",
      volume         = "121",
      year           = "2009",
      pages          = "843-874",
      doi            = "10.1143/PTP.121.843",
      eprint         = "0904.3810",
      archivePrefix  = "arXiv",
      primaryClass   = "gr-qc",
      SLACcitation   = "%%CITATION = ARXIV:0904.3810;%%"
}

@article{Teukolsky:1972my,
      author         = "Teukolsky, S. A.",
      title          = "{Rotating black holes - separable wave equations for
                        gravitational and electromagnetic perturbations}",
      journal        = "Phys. Rev. Lett.",
      volume         = "29",
      year           = "1972",
      pages          = "1114-1118",
      doi            = "10.1103/PhysRevLett.29.1114",
      reportNumber   = "OAP-291",
      SLACcitation   = "%%CITATION = PRLTA,29,1114;%%"
}

@article{Teukolsky:1973ha,
    author = "Teukolsky, Saul A.",
    title = "{Perturbations of a rotating black hole. 1. Fundamental equations for gravitational electromagnetic and neutrino field perturbations}",
    doi = "10.1086/152444",
    journal = "Astrophys. J.",
    volume = "185",
    pages = "635--647",
    year = "1973"
}

@misc{xAct,
  title = {{xAct: Efficient tensor computer algebra for the Wolfram Language}},
  howpublished = {(\href{http://www.xact.es/}{xact.es})},
}

@misc{BHPToolkit,
  title = {{Black Hole Perturbation Toolkit}},
  howpublished = {(\href{http://bhptoolkit.org/}{bhptoolkit.org})},
}

@software{BHPToolkitKerrGeodesics,
  author       = {Niels Warburton and
                  Barry Wardell and
                  Oliver Long and
                  Sam Upton and
                  Philip Lynch and
                  Zachary Nasipak and
                  Leo C. Stein},
  title        = {KerrGeodesics},
  month        = jul,
  year         = 2023,
  publisher    = {Zenodo},
  version      = {0.9.0},
  doi          = {10.5281/zenodo.8108265},
  url          = {https://doi.org/10.5281/zenodo.8108265}
}

@software{BHPToolkitSpinWeightedSpheroidalHarmonicsv0.3.0,
  author       = {Barry Wardell and
                  Niels Warburton and
                  Kwinten Fransen and
                  Sam Upton},
  title        = {SpinWeightedSpheroidalHarmonics},
  month        = jun,
  year         = 2023,
  publisher    = {Zenodo},
  version      = {0.3.0},
  doi          = {10.5281/zenodo.8091168},
  url          = {https://doi.org/10.5281/zenodo.8091168}
}

@software{BHPToolkitTeukolsky,
  author       = {Wardell, Barry and
                  Warburton, Niels and
                  Cunningham, Kevin and
                  Durkan, Leanne and
                  Leather, Benjamin and
                  Nasipak, Zachary and
                  Kavanagh, Chris and
                  Torres, Theo and
                  Ottewill, Adrian and
                  Casals, Marc},
  title        = {Teukolsky},
  month        = oct,
  year         = 2023,
  publisher    = {Zenodo},
  version      = {1.0.4},
  doi          = {10.5281/zenodo.10040501},
  url          = {https://doi.org/10.5281/zenodo.10040501}
}

@article{Piovano:2020zin,
    author = "Piovano, Gabriel Andres and Maselli, Andrea and Pani, Paolo",
    title = "{Extreme mass ratio inspirals with spinning secondary: a detailed study of equatorial circular motion}",
    eprint = "2004.02654",
    archivePrefix = "arXiv",
    primaryClass = "gr-qc",
    doi = "10.1103/PhysRevD.102.024041",
    journal = "Phys. Rev. D",
    volume = "102",
    number = "2",
    pages = "024041",
    year = "2020"
}

@article{Piovano:2021iwv,
    author = "Piovano, Gabriel Andres and Brito, Richard and Maselli, Andrea and Pani, Paolo",
    title = "{Assessing the detectability of the secondary spin in extreme mass-ratio inspirals with fully relativistic numerical waveforms}",
    eprint = "2105.07083",
    archivePrefix = "arXiv",
    primaryClass = "gr-qc",
    doi = "10.1103/PhysRevD.104.124019",
    journal = "Phys. Rev. D",
    volume = "104",
    number = "12",
    pages = "124019",
    year = "2021"
}

@article{Lukes-Gerakopoulos:2017vkj,
      author         = "Lukes-Gerakopoulos, Georgios and Harms, Enno and
                        Bernuzzi, Sebastiano and Nagar, Alessandro",
      title          = "{Spinning test-body orbiting around a Kerr black hole:
                        circular dynamics and gravitational-wave fluxes}",
      journal        = "Phys. Rev.",
      volume         = "D96",
      year           = "2017",
      number         = "6",
      pages          = "064051",
      doi            = "10.1103/PhysRevD.96.064051",
      eprint         = "1707.07537",
      archivePrefix  = "arXiv",
      primaryClass   = "gr-qc",
      SLACcitation   = "%%CITATION = ARXIV:1707.07537;%%"
}

@article{Harms:2016ctx,
      author         = "Harms, Enno and Lukes-Gerakopoulos, Georgios and
                        Bernuzzi, Sebastiano and Nagar, Alessandro",
      title          = "{Spinning test body orbiting around a Schwarzschild black
                        hole: Circular dynamics and gravitational-wave fluxes}",
      journal        = "Phys. Rev.",
      volume         = "D94",
      year           = "2016",
      number         = "10",
      pages          = "104010",
      doi            = "10.1103/PhysRevD.100.129902, 10.1103/PhysRevD.94.104010",
      eprint         = "1609.00356",
      archivePrefix  = "arXiv",
      primaryClass   = "gr-qc",
      SLACcitation   = "%%CITATION = ARXIV:1609.00356;%%"
}

@article{Semerak:1999qc,
      author         = "Semer{\'a}k, O.",
      title          = "{Spinning test particles in a Kerr field. 1.}",
      journal        = "Mon. Not. Roy. Astron. Soc.",
      volume         = "308",
      year           = "1999",
      pages          = "863-875",
      doi            = "10.1046/j.1365-8711.1999.02754.x",
      SLACcitation   = "%%CITATION = MNRAA,308,863;%%"
}

@article{Warburton:2017sxk,
      author         = "Warburton, Niels and Osburn, Thomas and Evans, Charles
                        R.",
      title          = "{Evolution of small-mass-ratio binaries with a spinning
                        secondary}",
      journal        = "Phys. Rev.",
      volume         = "D96",
      year           = "2017",
      number         = "8",
      pages          = "084057",
      doi            = "10.1103/PhysRevD.96.084057",
      eprint         = "1708.03720",
      archivePrefix  = "arXiv",
      primaryClass   = "gr-qc",
      SLACcitation   = "%%CITATION = ARXIV:1708.03720;%%"
}

@article{Ehlers:1977,
       author = "Ehlers, J{\"u}rgen and Rudolph, Ekkart",
        title = "{Dynamics of extended bodies in general relativity center-of-mass description and quasirigidity}",
      journal = {General Relativity and Gravitation},
         year = "1977",
        month = "Mar",
       volume = {8},
       number = {3},
        pages = {197-217},
          doi = {10.1007/BF00763547},
       adsurl = {https://ui.adsabs.harvard.edu/abs/1977GReGr...8..197E},
}

@article{Harms:2015ixa,
      author         = "Harms, Enno and Lukes-Gerakopoulos, Georgios and
                        Bernuzzi, Sebastiano and Nagar, Alessandro",
      title          = "{Asymptotic gravitational wave fluxes from a spinning
                        particle in circular equatorial orbits around a rotating
                        black hole}",
      journal        = "Phys. Rev.",
      volume         = "D93",
      year           = "2016",
      number         = "4",
      pages          = "044015",
      doi            = "10.1103/PhysRevD.100.129901, 10.1103/PhysRevD.93.044015",
      note           = "[Addendum: Phys. Rev.D100,no.12,129901(2019)]",
      eprint         = "1510.05548",
      archivePrefix  = "arXiv",
      primaryClass   = "gr-qc",
      SLACcitation   = "%%CITATION = ARXIV:1510.05548;%%"
}

@article{Witzany:2018ahb,
      author         = "Witzany, Vojt\u{e}ch and Steinhoff, Jan and
                        Lukes-Gerakopoulos, Georgios",
      title          = "{Hamiltonians and canonical coordinates for spinning
                        particles in curved space-time}",
      journal        = "Class. Quant. Grav.",
      volume         = "36",
      year           = "2019",
      number         = "7",
      pages          = "075003",
      doi            = "10.1088/1361-6382/ab002f",
      eprint         = "1808.06582",
      archivePrefix  = "arXiv",
      primaryClass   = "gr-qc",
      SLACcitation   = "%%CITATION = ARXIV:1808.06582;%%"
}

@article{Barack:2018yvs,
      author         = "Barack, Leor and Pound, Adam",
      title          = "{Self-force and radiation reaction in general
                        relativity}",
      journal        = "Rept. Prog. Phys.",
      volume         = "82",
      year           = "2019",
      number         = "1",
      pages          = "016904",
      doi            = "10.1088/1361-6633/aae552",
      eprint         = "1805.10385",
      archivePrefix  = "arXiv",
      primaryClass   = "gr-qc",
      SLACcitation   = "%%CITATION = ARXIV:1805.10385;%%"
}

@article{Ruangsri:2015cvg,
      author         = "Ruangsri, Uchupol and Vigeland, Sarah J. and Hughes,
                        Scott A.",
      title          = "{Gyroscopes orbiting black holes: A frequency-domain
                        approach to precession and spin-curvature coupling for
                        spinning bodies on generic Kerr orbits}",
      journal        = "Phys. Rev.",
      volume         = "D94",
      year           = "2016",
      number         = "4",
      pages          = "044008",
      doi            = "10.1103/PhysRevD.94.044008",
      eprint         = "1512.00376",
      archivePrefix  = "arXiv",
      primaryClass   = "gr-qc",
      SLACcitation   = "%%CITATION = ARXIV:1512.00376;%%"
}

@article{Bini:2014soa,
    author = "Bini, D. and Geralico, A.",
    title = "{Spin-geodesic deviations in the Kerr spacetime}",
    eprint = "1408.4952",
    archivePrefix = "arXiv",
    primaryClass = "gr-qc",
    doi = "10.1103/PhysRevD.84.104012",
    journal = "Phys. Rev. D",
    volume = "84",
    pages = "104012",
    year = "2011"
}

@article{Witzany:2019nml,
      author         = "Witzany, Vojt\u{e}ch",
      title          = "{Hamilton-Jacobi equation for spinning particles near
                        black holes}",
      journal        = "Phys. Rev.",
      volume         = "D100",
      year           = "2019",
      number         = "10",
      pages          = "104030",
      doi            = "10.1103/PhysRevD.100.104030",
      eprint         = "1903.03651",
      archivePrefix  = "arXiv",
      primaryClass   = "gr-qc",
      SLACcitation   = "%%CITATION = ARXIV:1903.03651;%%"
}

@article{Zelenka:2019nyp,
      author         = "Zelenka, Ond\u{r}ej and Lukes-Gerakopoulos, Georgios and
                        Witzany, Vojtu\u{e}ch and Kop\'{a}\u{c}ek, Ondu\u{r}ej",
      title          = "{Growth of resonances and chaos for a spinning test
                        particle in the Schwarzschild background}",
      journal        = "Phys. Rev.",
      volume         = "D101",
      year           = "2020",
      number         = "2",
      pages          = "024037",
      doi            = "10.1103/PhysRevD.101.024037",
      eprint         = "1911.00414",
      archivePrefix  = "arXiv",
      primaryClass   = "gr-qc",
      SLACcitation   = "%%CITATION = ARXIV:1911.00414;%%"
}

@article{Zenginoglu:2011jz,
    author = "Zengino\u{g}lu, An{\i}l",
    title = "{A Geometric framework for black hole perturbations}",
    eprint = "1102.2451",
    archivePrefix = "arXiv",
    primaryClass = "gr-qc",
    doi = "10.1103/PhysRevD.83.127502",
    journal = "Phys. Rev. D",
    volume = "83",
    pages = "127502",
    year = "2011"
}

@article{Maggiore:2019uih,
    author = "Maggiore, Michele and others",
    title = "{Science Case for the Einstein Telescope}",
    eprint = "1912.02622",
    archivePrefix = "arXiv",
    primaryClass = "astro-ph.CO",
    doi = "10.1088/1475-7516/2020/03/050",
    journal = "JCAP",
    volume = "03",
    pages = "050",
    year = "2020"
}

@article{Penrose:1973um,
    author = "Penrose, R.",
    title = "{Naked singularities}",
    doi = "10.1111/j.1749-6632.1973.tb41447.x",
    journal = "Annals N. Y. Acad. Sci.",
    volume = "224",
    pages = "125--134",
    year = "1973"
}

@article{Walker:1970un,
    author = "Walker, M. and Penrose, R.",
    title = "{On quadratic first integrals of the geodesic equations for type [22] spacetimes}",
    doi = "10.1007/BF01649445",
    journal = "Commun. Math. Phys.",
    volume = "18",
    pages = "265--274",
    year = "1970"
}

@article{Ruiz:2007yx,
    author = "Ruiz, Milton and Takahashi, Ryoji and Alcubierre, Miguel and Nunez, Dario",
    title = "{Multipole expansions for energy and momenta carried by gravitational waves}",
    eprint = "0707.4654",
    archivePrefix = "arXiv",
    primaryClass = "gr-qc",
    doi = "10.1007/s10714-007-0570-8",
    journal = "Gen. Rel. Grav.",
    volume = "40",
    pages = "2467",
    year = "2008"
}

@article{Hirata:2010xn,
    author = "Hirata, Christopher M",
    title = "{Resonant recoil in extreme mass ratio binary black hole mergers}",
    eprint = "1011.4987",
    archivePrefix = "arXiv",
    primaryClass = "gr-qc",
    doi = "10.1103/PhysRevD.83.104024",
    journal = "Phys. Rev. D",
    volume = "83",
    pages = "104024",
    year = "2011"
}

@article{vandeMeent:2014raa,
    author = "van de Meent, Maarten",
    title = "{Resonantly enhanced kicks from equatorial small mass-ratio inspirals}",
    eprint = "1406.2594",
    archivePrefix = "arXiv",
    primaryClass = "gr-qc",
    doi = "10.1103/PhysRevD.90.044027",
    journal = "Phys. Rev. D",
    volume = "90",
    number = "4",
    pages = "044027",
    year = "2014"
}

@article{Rudiger:1981,
 ISSN = {00804630},
 URL = {http://www.jstor.org/stable/2990231},
 author = {R. R{\"u}diger},
 journal = {Proceedings of the Royal Society of London. Series A, Mathematical and Physical Sciences},
 number = {1761},
 pages = {185--193},
 publisher = {The Royal Society},
 title = {Conserved Quantities of Spinning Test Particles in General Relativity. I},
 volume = {375},
 year = {1981}
}

@article{Rudiger:1983,
 ISSN = {00804630},
 URL = {http://www.jstor.org/stable/2397480},
 author = {R. R{\"u}diger},
 journal = {Proceedings of the Royal Society of London. Series A, Mathematical and Physical Sciences},
 number = {1788},
 pages = {229--239},
 publisher = {The Royal Society},
 title = {Conserved Quantities of Spinning Test Particles in General Relativity. II},
 volume = {385},
 year = {1983}
}

@article{Mathews:2021rod,
    author = "Mathews, Josh and Pound, Adam and Wardell, Barry",
    title = "{Self-force calculations with a spinning secondary}",
    eprint = "2112.13069",
    archivePrefix = "arXiv",
    primaryClass = "gr-qc",
    doi = "10.1103/PhysRevD.105.084031",
    journal = "Phys. Rev. D",
    volume = "105",
    number = "8",
    pages = "084031",
    year = "2022"
}

@article{Skoupy:2022adh,
    author = "Skoup\'y, Viktor and Lukes-Gerakopoulos, Georgios",
    title = "{Adiabatic equatorial inspirals of a spinning body into a Kerr black hole}",
    eprint = "2201.07044",
    archivePrefix = "arXiv",
    primaryClass = "gr-qc",
    doi = "10.1103/PhysRevD.105.084033",
    journal = "Phys. Rev. D",
    volume = "105",
    number = "8",
    pages = "084033",
    year = "2022"
}

@article{Warburton:2021kwk,
    author = "Warburton, Niels and Pound, Adam and Wardell, Barry and Miller, Jeremy and Durkan, Leanne",
    title = "{Gravitational-Wave Energy Flux for Compact Binaries through Second Order in the Mass Ratio}",
    eprint = "2107.01298",
    archivePrefix = "arXiv",
    primaryClass = "gr-qc",
    doi = "10.1103/PhysRevLett.127.151102",
    journal = "Phys. Rev. Lett.",
    volume = "127",
    number = "15",
    pages = "151102",
    year = "2021"
}

@article{Wardell:2021fyy,
    author = "Wardell, Barry and Pound, Adam and Warburton, Niels and Miller, Jeremy and Durkan, Leanne and Le Tiec, Alexandre",
    title = "{Gravitational Waveforms for Compact Binaries from Second-Order Self-Force Theory}",
    eprint = "2112.12265",
    archivePrefix = "arXiv",
    primaryClass = "gr-qc",
    doi = "10.1103/PhysRevLett.130.241402",
    journal = "Phys. Rev. Lett.",
    volume = "130",
    number = "24",
    pages = "241402",
    year = "2023"
}

@article{PhysRevD.67.084027,
  title = {Perturbative approach to an orbital evolution around a supermassive black hole},
  author = {Mino, Yasushi},
  journal = {Phys. Rev. D},
  volume = {67},
  issue = {8},
  pages = {084027},
  numpages = {17},
  year = {2003},
  month = {Apr},
  publisher = {American Physical Society},
  doi = {10.1103/PhysRevD.67.084027},
  url = {https://link.aps.org/doi/10.1103/PhysRevD.67.084027}
}

@book{arnold2006celestial,
  title={Mathematical aspects of classical and celestial mechanics},
  author={Arnold, Vladimir Igorevich and Kozlov, Valery V and Neishtadt, Anatoly I and Iacob, I},
  volume={3},
  year={2006},
  publisher={Springer}
}

@article{Blanchet:2013haa,
    author = "Blanchet, Luc",
    title = "{Gravitational Radiation from Post-Newtonian Sources and Inspiralling Compact Binaries}",
    eprint = "1310.1528",
    archivePrefix = "arXiv",
    primaryClass = "gr-qc",
    doi = "10.12942/lrr-2014-2",
    journal = "Living Rev. Rel.",
    volume = "17",
    pages = "2",
    year = "2014"
}

@article{Ossokine:2020kjp,
    author = "Ossokine, Serguei and others",
    title = "{Multipolar Effective-One-Body Waveforms for Precessing Binary Black Holes: Construction and Validation}",
    eprint = "2004.09442",
    archivePrefix = "arXiv",
    primaryClass = "gr-qc",
    doi = "10.1103/PhysRevD.102.044055",
    journal = "Phys. Rev. D",
    volume = "102",
    number = "4",
    pages = "044055",
    year = "2020"
}

@article{vandeMeent:2020xgc,
    author = "van de Meent, Maarten and Pfeiffer, Harald P.",
    title = "{Intermediate mass-ratio black hole binaries: Applicability of small mass-ratio perturbation theory}",
    eprint = "2006.12036",
    archivePrefix = "arXiv",
    primaryClass = "gr-qc",
    doi = "10.1103/PhysRevLett.125.181101",
    journal = "Phys. Rev. Lett.",
    volume = "125",
    number = "18",
    pages = "181101",
    year = "2020"
}

@article{LeTiec:2011dp,
    author = "Le Tiec, Alexandre and Barausse, Enrico and Buonanno, Alessandra",
    title = "{Gravitational Self-Force Correction to the Binding Energy of Compact Binary Systems}",
    eprint = "1111.5609",
    archivePrefix = "arXiv",
    primaryClass = "gr-qc",
    doi = "10.1103/PhysRevLett.108.131103",
    journal = "Phys. Rev. Lett.",
    volume = "108",
    pages = "131103",
    year = "2012"
}

@article{LIGOScientific:2016aoc,
    author = "Abbott, B. P. and others",
    collaboration = "LIGO Scientific, Virgo",
    title = "{Observation of Gravitational Waves from a Binary Black Hole Merger}",
    eprint = "1602.03837",
    archivePrefix = "arXiv",
    primaryClass = "gr-qc",
    reportNumber = "LIGO-P150914",
    doi = "10.1103/PhysRevLett.116.061102",
    journal = "Phys. Rev. Lett.",
    volume = "116",
    number = "6",
    pages = "061102",
    year = "2016"
}

@article{Drasco:2003ky,
    author = "Drasco, Steve and Hughes, Scott A.",
    title = "{Rotating black hole orbit functionals in the frequency domain}",
    eprint = "astro-ph/0308479",
    archivePrefix = "arXiv",
    reportNumber = "CSR-03-51",
    doi = "10.1103/PhysRevD.69.044015",
    journal = "Phys. Rev. D",
    volume = "69",
    pages = "044015",
    year = "2004"
}

@article{Mino:1996nk,
    author = "Mino, Yasushi and Sasaki, Misao and Tanaka, Takahiro",
    title = "{Gravitational radiation reaction to a particle motion}",
    eprint = "gr-qc/9606018",
    archivePrefix = "arXiv",
    reportNumber = "OU-TAP-38, KUNS-1394",
    doi = "10.1103/PhysRevD.55.3457",
    journal = "Phys. Rev. D",
    volume = "55",
    pages = "3457--3476",
    year = "1997"
}

@article{Drummond:2022xej,
    author = "Drummond, Lisa V. and Hughes, Scott A.",
    title = "{Precisely computing bound orbits of spinning bodies around black holes. I. General framework and results for nearly equatorial orbits}",
    eprint = "2201.13334",
    archivePrefix = "arXiv",
    primaryClass = "gr-qc",
    doi = "10.1103/PhysRevD.105.124040",
    journal = "Phys. Rev. D",
    volume = "105",
    number = "12",
    pages = "124040",
    year = "2022"
}

@article{Drummond:2022efc,
    author = "Drummond, Lisa V. and Hughes, Scott A.",
    title = "{Precisely computing bound orbits of spinning bodies around black holes. II. Generic orbits}",
    eprint = "2201.13335",
    archivePrefix = "arXiv",
    primaryClass = "gr-qc",
    doi = "10.1103/PhysRevD.105.124041",
    journal = "Phys. Rev. D",
    volume = "105",
    number = "12",
    pages = "124041",
    year = "2022"
}

@article{vandeMeent:2019cam,
    author = "van de Meent, Maarten",
    title = "{Analytic solutions for parallel transport along generic bound geodesics in Kerr spacetime}",
    eprint = "1906.05090",
    archivePrefix = "arXiv",
    primaryClass = "gr-qc",
    doi = "10.1088/1361-6382/ab79d5",
    journal = "Class. Quant. Grav.",
    volume = "37",
    number = "14",
    pages = "145007",
    year = "2020"
}

@article{vandeMeent:2017bcc,
    author = "van de Meent, Maarten",
    title = "{Gravitational self-force on generic bound geodesics in Kerr spacetime}",
    eprint = "1711.09607",
    archivePrefix = "arXiv",
    primaryClass = "gr-qc",
    doi = "10.1103/PhysRevD.97.104033",
    journal = "Phys. Rev. D",
    volume = "97",
    number = "10",
    pages = "104033",
    year = "2018"
}

@article{Schmidt:2002qk,
    author = "Schmidt, Wolfram",
    title = "{Celestial mechanics in Kerr space-time}",
    eprint = "gr-qc/0202090",
    archivePrefix = "arXiv",
    doi = "10.1088/0264-9381/19/10/314",
    journal = "Class. Quant. Grav.",
    volume = "19",
    pages = "2743",
    year = "2002"
}

@article{LISAConsortiumWaveformWorkingGroup:2023arg,
    author = "Afshordi, Niayesh and others",
    collaboration = "LISA Consortium Waveform Working Group",
    title = "{Waveform Modelling for the Laser Interferometer Space Antenna}",
    eprint = "2311.01300",
    archivePrefix = "arXiv",
    primaryClass = "gr-qc",
    month = "11",
    year = "2023"
}

@article{Witzany:2023bmq,
    author = "Witzany, Vojt\v{e}ch and Piovano, Gabriel Andres",
    title = "{Analytic Solutions for the Motion of Spinning Particles near Spherically Symmetric Black Holes and Exotic Compact Objects}",
    eprint = "2308.00021",
    archivePrefix = "arXiv",
    primaryClass = "gr-qc",
    doi = "10.1103/PhysRevLett.132.171401",
    journal = "Phys. Rev. Lett.",
    volume = "132",
    number = "17",
    pages = "171401",
    year = "2024"
}

@article{Drummond:2023wqc,
    author = "Drummond, Lisa V. and Lynch, Philip and Hanselman, Alexandra G. and Becker, Devin R. and Hughes, Scott A.",
    title = "{Extreme mass-ratio inspiral and waveforms for a spinning body into a Kerr black hole via osculating geodesics and near-identity transformations}",
    eprint = "2310.08438",
    archivePrefix = "arXiv",
    primaryClass = "gr-qc",
    doi = "10.1103/PhysRevD.109.064030",
    journal = "Phys. Rev. D",
    volume = "109",
    number = "6",
    pages = "064030",
    year = "2024"
}

@article{Warburton:2013yj,
    author = "Warburton, Niels and Barack, Leor and Sago, Norichika",
    title = "{Isofrequency pairing of geodesic orbits in Kerr geometry}",
    eprint = "1301.3918",
    archivePrefix = "arXiv",
    primaryClass = "gr-qc",
    doi = "10.1103/PhysRevD.87.084012",
    journal = "Phys. Rev. D",
    volume = "87",
    number = "8",
    pages = "084012",
    year = "2013"
}

@article{Burke:2023lno,
    author = "Burke, Ollie and Piovano, Gabriel Andres and Warburton, Niels and Lynch, Philip and Speri, Lorenzo and Kavanagh, Chris and Wardell, Barry and Pound, Adam and Durkan, Leanne and Miller, Jeremy",
    title = "{Assessing the importance of first postadiabatic terms for small-mass-ratio binaries}",
    eprint = "2310.08927",
    archivePrefix = "arXiv",
    primaryClass = "gr-qc",
    doi = "10.1103/PhysRevD.109.124048",
    journal = "Phys. Rev. D",
    volume = "109",
    number = "12",
    pages = "124048",
    year = "2024"
}

@article{Miller:2020bft,
    author = "Miller, Jeremy and Pound, Adam",
    title = "{Two-timescale evolution of extreme-mass-ratio inspirals: waveform generation scheme for quasicircular orbits in Schwarzschild spacetime}",
    eprint = "2006.11263",
    archivePrefix = "arXiv",
    primaryClass = "gr-qc",
    doi = "10.1103/PhysRevD.103.064048",
    journal = "Phys. Rev. D",
    volume = "103",
    number = "6",
    pages = "064048",
    year = "2021"
}

@article{Pratten:2020ceb,
    author = "Pratten, Geraint and others",
    title = "{Computationally efficient models for the dominant and subdominant harmonic modes of precessing binary black holes}",
    eprint = "2004.06503",
    archivePrefix = "arXiv",
    primaryClass = "gr-qc",
    doi = "10.1103/PhysRevD.103.104056",
    journal = "Phys. Rev. D",
    volume = "103",
    number = "10",
    pages = "104056",
    year = "2021"
}

@article{Scheel:2014ina,
    author = "Scheel, Mark A. and Giesler, Matthew and Hemberger, Daniel A. and Lovelace, Geoffrey and Kuper, Kevin and Boyle, Michael and Szil\'agyi, B. and Kidder, Lawrence E.",
    title = "{Improved methods for simulating nearly extremal binary black holes}",
    eprint = "1412.1803",
    archivePrefix = "arXiv",
    primaryClass = "gr-qc",
    doi = "10.1088/0264-9381/32/10/105009",
    journal = "Class. Quant. Grav.",
    volume = "32",
    number = "10",
    pages = "105009",
    year = "2015"
}

@article{Varma:2019csw,
    author = "Varma, Vijay and Field, Scott E. and Scheel, Mark A. and Blackman, Jonathan and Gerosa, Davide and Stein, Leo C. and Kidder, Lawrence E. and Pfeiffer, Harald P.",
    title = "{Surrogate models for precessing binary black hole simulations with unequal masses}",
    eprint = "1905.09300",
    archivePrefix = "arXiv",
    primaryClass = "gr-qc",
    doi = "10.1103/PhysRevResearch.1.033015",
    journal = "Phys. Rev. Research.",
    volume = "1",
    pages = "033015",
    year = "2019"
}

@article{Blackman:2017pcm,
    author = "Blackman, Jonathan and Field, Scott E. and Scheel, Mark A. and Galley, Chad R. and Ott, Christian D. and Boyle, Michael and Kidder, Lawrence E. and Pfeiffer, Harald P. and Szil\'agyi, B\'ela",
    title = "{Numerical relativity waveform surrogate model for generically precessing binary black hole mergers}",
    eprint = "1705.07089",
    archivePrefix = "arXiv",
    primaryClass = "gr-qc",
    reportNumber = "YITP-17-44",
    doi = "10.1103/PhysRevD.96.024058",
    journal = "Phys. Rev. D",
    volume = "96",
    number = "2",
    pages = "024058",
    year = "2017"
}

@article{Ramos-Buades:2023ehm,
    author = "Ramos-Buades, Antoni and Buonanno, Alessandra and Estell\'es, H\'ector and Khalil, Mohammed and Mihaylov, Deyan P. and Ossokine, Serguei and Pompili, Lorenzo and Shiferaw, Mahlet",
    title = "{Next generation of accurate and efficient multipolar precessing-spin effective-one-body waveforms for binary black holes}",
    eprint = "2303.18046",
    archivePrefix = "arXiv",
    primaryClass = "gr-qc",
    doi = "10.1103/PhysRevD.108.124037",
    journal = "Phys. Rev. D",
    volume = "108",
    number = "12",
    pages = "124037",
    year = "2023"
}

@article{SkoupySpherInprep,
    author = "Skoupý, V and Piovano, G and Witzany, V",
    title = "{Adiabatic spherical inspirals of a spinning body into a Kerr black hole}",
    eprint = "",
    archivePrefix = "arXiv",
    primaryClass = "gr-qc",
    doi = "",
    journal = "In preparation",
    year = "2024"
}

@article{Boyle:2019kee,
    author = "Boyle, Michael and others",
    title = "{The SXS Collaboration catalog of binary black hole simulations}",
    eprint = "1904.04831",
    archivePrefix = "arXiv",
    primaryClass = "gr-qc",
    doi = "10.1088/1361-6382/ab34e2",
    journal = "Class. Quant. Grav.",
    volume = "36",
    number = "19",
    pages = "195006",
    year = "2019"
}

@article{Compere:2023alp,
    author = "Comp\`ere, Geoffrey and Druart, Adrien and Vines, Justin",
    title = "{Generalized Carter constant for quadrupolar test bodies in Kerr spacetime}",
    eprint = "2302.14549",
    archivePrefix = "arXiv",
    primaryClass = "gr-qc",
    doi = "10.21468/SciPostPhys.15.6.226",
    journal = "SciPost Phys.",
    volume = "15",
    number = "6",
    pages = "226",
    year = "2023"
}

@software{repoHJproject,
  author       = {Piovano, Gabriel Andres and
                  Pantelidou, Christiana and
                  Mac Uilliam, Jake and
                  Witzany, Vojtěch},
  title        = {gabriel-andres-piovano/Spinning-Body-Hamilton-
                   Jacobi: Spinning-Body-Hamilton-Jacobi 0.1.0
                  },
  month        = dec,
  year         = 2024,
  publisher    = {Zenodo},
  version      = {0.1.0},
  doi          = {10.5281/zenodo.14281747},
  url          = {https://doi.org/10.5281/zenodo.14281747}
}

\end{document}